\documentclass[12pt]{article}

\usepackage[margin=1in]{geometry}

\usepackage{babel}
\usepackage[T1]{fontenc}
\usepackage{lmodern}

\usepackage{authblk}
\usepackage{amsthm,amsmath,amsfonts,amssymb}
\usepackage{graphicx,color,xcolor}
\usepackage{caption,subcaption}
\usepackage{multicol,longtable,multirow,booktabs}
\usepackage{hyperref}
\hypersetup{colorlinks=true,urlcolor=blue,citecolor=blue,allcolors=blue}
\usepackage{comment,enumitem}
\usepackage[title]{appendix}
\usepackage[authoryear]{natbib}
\theoremstyle{definition}
\numberwithin{equation}{section}  

\newtheorem{lemma}{Lemma}[section]
\newtheorem{theorem}{Theorem}[section]
\newtheorem{corollary}{Corollary}[section]
\newtheorem{remark}{Remark}[section]

\providecommand{\keywords}[1]{ \noindent\small\textbf{\textit{Keywords:}} #1 }

\def\E{\mathbb{E}}
\def\R{\mathbb{R}}

\def\bbtheta{\boldsymbol{\theta}}

\def\bbxi{\boldsymbol{\xi}}

\def\bbeta{\boldsymbol{\eta}}

\newcommand{\fracpartial}[2]{\frac{\partial #1}{\partial  #2}}
\newcommand{\bb}[1]{\boldsymbol{#1}}
\newcommand{\bbhat}[1]{\widehat{\boldsymbol{#1}}}

\newcommand{\bbtilde}[1]{\widetilde{\boldsymbol{#1}}}

\newcommand{\diag}[1]{\text{\normalfont Diag}\left( #1 \right)}
\newcommand{\trace}[1]{\text{\normalfont Trace}\left( #1 \right)}
\newcommand{\rank}[1]{\text{\normalfont Rank}\left( #1 \right)}

\newcommand{\tr}{^{\intercal}}
\newcommand{\normdist}{\mathcal{N}}

\newcommand{\var}{\mathrm{Var}}

\newcommand{\dash}{^{\prime}}
\newcommand{\ddash}{^{\prime\prime}}
\DeclareMathOperator*{\argmin}{arg\,min}

\title{Robust Principal Component Analysis using Density Power Divergence}
\author[1]{Subhrajyoty Roy}
\author[1]{Ayanendranath Basu}
\author[1]{Abhik Ghosh}
\affil[1]{Indian Statistical Institute, Kolkata}
\affil[ ]{\small\textit{subhrajyoty\_r@isical.ac.in,ayanbasu@isical.ac.in,abhik.ghosh@isical.ac.in}}
\date{}

\begin{document}
\maketitle

\begin{abstract}
    Principal component analysis (PCA) is a widely employed statistical tool used primarily for dimensionality reduction. However, it is known to be adversely affected by the presence of outlying observations in the sample, which is quite common. Robust PCA methods using M-estimators have theoretical benefits, but their robustness drop substantially for high dimensional data. On the other end of the spectrum, robust PCA algorithms solving principal component pursuit or similar optimization problems have high breakdown, but lack theoretical richness and demand high computational power compared to the M-estimators. We introduce a novel robust PCA estimator based on the minimum density power divergence estimator. This combines the theoretical strength of the M-estimators and the minimum divergence estimators with a high breakdown guarantee regardless of data dimension. We present a computationally efficient algorithm for this estimate. Our theoretical findings are supported by extensive simulations and comparisons with existing robust PCA methods. We also showcase the proposed algorithm's applicability on two benchmark datasets and a credit card transactions dataset for fraud detection. 
\end{abstract}
    


\section{Introduction}\label{sec:intro}

The classical problem of finding the principal components aims to approximate the covariance structure of a high dimensional sample of many features by the covariance structure of a lower dimensional sample of ``principal components'', obtained as linear combinations of the original feature variables. Mathematically, starting with an independent and identically distributed (i.i.d.) sample $\bb{X}_1, \bb{X}_2, \dots \bb{X}_n$, where each $\bb{X}_i \in \R^p$, and a scale measure $S_n(y_1, \dots y_n)$ to measure the dispersion in a univariate sample $\{ y_1, \dots y_n \}$, the first eigenvector associated with the principal components is defined as the unit length vector maximizing the function
\begin{equation}
    \bb{v} \rightarrow S_n(\bb{v}\tr \bb{X}_1, \dots \bb{v}\tr \bb{X}_n); \ \bb{v} \in \R^p.
    \label{eqn:pc-function}
\end{equation}
\noindent Similarly, assuming that the first $(k-1)$ eigenvectors $\bbhat{v}_1, \bbhat{v}_2, \dots \bbhat{v}_{k-1}$ has already been found, one can obtain the subsequent $k$-th eigenvector as the unit vector maximizing the same function given in Eq.~\eqref{eqn:pc-function}, but under the set of restrictions $\bb{v}\tr \bbhat{v}_i = 0$ for all $i = 1, \dots (k-1)$. The corresponding eigenvalues are defined as the maximum values of the scale function, i.e.,
\begin{equation*}
    \widehat{\lambda}_k = S_n(\bbhat{v}_k\tr\bb{X}_1, \dots \bbhat{v}_k\tr\bb{X}_n).
\end{equation*}
\noindent In essence, principal component analysis (PCA) takes input $n$ observations of dimension $p$, where $p$ is presumably very large, and outputs a set of pairs $\{ (\widehat{\lambda}_k, \bbhat{v}_k): k = 1, 2, \dots r\}$ where $r$ is a pre-specified number of components, generally much smaller compared to both $n$ and $p$. For each $k$, the former of the pair $\widehat{\lambda}_k$ denotes the maximum variability expressed by the $k$-th principal component, and the latter of the pair $\bbhat{v}_k$ denotes the direction along which this maximum variability can be found in the given i.i.d. sample. The $k$-th principal component is then defined by the variable obtained from projecting the observations along the $k$-th eigenvector scaled by the $k$-th eigenvalue, i.e., $\{ \widehat{\lambda}_k\bbhat{v}_k\tr \bb{X}_i : i = 1, \dots n\}$.

Since a small number of principal components can explain most of the variation present in the random sample, it is primarily used for the purpose of dimensionality reduction. PCA provides a simple method of visualizing any high-dimensional data by plotting the first two or three principal components, and subsequently one can identify potential outliers~\citep{locantore1999robust}. \cite{jolliffe2002principal} also provides an application of PCA for variable selection in the regression context. In machine learning and pattern recognition, PCA has been used abundantly for both supervised and unsupervised paradigms~\citep{vathy2013graph}. PCA has also found its applications across many disciplines ranging from multi-sensor data fusion~\citep{lock2013jive}, signal processing, image compression~\citep{bouwmans2018applications}, video event detection~\citep{roy2021new} to material and chemical sciences~\citep{smilde2005multi}. The readers are referred to see~\cite{sanguansat2012principal} and the references therein for further details on the multitude of applications of PCA.

In the classical PCA, the scale estimator $S_n(y_1, y_2, \dots y_n)$ is chosen to be the square root of the sample variance. As a result, the eigenvalues and the eigenvectors of the sample covariance matrix of $\bb{X}_1, \dots \bb{X}_n$ become the solution to the aforementioned principal components problem. It is well known that the sample covariance matrix is very sensitive to outliers, hence the principal components resulting from classical PCA also suffer from the presence of outlying observations in the data~\citep{hubert2005robpca,candes2011robust}. In the context of the high dimensional datasets pertaining to the above applications, it is very challenging to locate these outlying observations beforehand in order to discard them. Thus, any practitioner relying solely on the classical PCA to interpret multivariate data may end up with a distorted visualization of the data, false detection of outliers, and a wrong conclusion about the data. Several robustified versions of PCA have been proposed to date to provide reliable estimates of the principal components even under the presence of outlying observations~\citep{jolliffe2002principal}. A brief discussion of the existing literature in this area is provided in the following subsection.

\subsection{Existing Literature}\label{sec:literature}

Most of the early literature to derive a robust principal component analysis (RPCA) followed one of the two primary approaches. The first class of estimators estimated the principal components robustly from the eigenvalues and the eigenvectors of a robust covariance matrix of the sample. Notable among this class of estimators are those due to~\cite{maronna1976robust} and~\cite{campbell1980robust}, where the authors create affine-equivariant principal component estimates from robust M-estimators of the covariance matrix. \cite{devlin1981robust} proposed to use minimum covariance determinant (MCD) estimator and minimum volume ellipsoid (MVE) estimator~\citep{rousseeuw1985multivariate} for this purpose due to their high breakdown compared to the M-estimators.

The other approach considered robustifying PCA by using a robust scale function $S_n$ in Eq.~\eqref{eqn:pc-function}. This idea was first presented by~\cite{li1985projection} and was further developed later by~\cite{croux1996fast} where they considered the median absolute deviation about sample median as the scale function. Various theoretical properties like the influence function, asymptotic distribution and the breakdown point of this estimator have also been established in the literature~\citep{croux2000principal,croux2005high}. These estimators and their variants primarily restricted their attention to the elliptically symmetric family of distributional models, i.e., the random observations $\bb{X}_i$ for $i= 1, 2 \dots n$ were assumed to follow a density function of the form
\begin{equation}
  f(\bb{x}) \propto g\left( (\bb{x} - \bb{\mu})\tr \bb{\Sigma}^{-1} (\bb{x} - \bb{\mu}) \right),
  \label{eqn:elliptic-model}
\end{equation}
\noindent where $g: \R^+ \rightarrow \R$ is a known function governing the shape of the density. It turns out that under this model, $\E(\bb{X}_i) = \bb{\mu}$ and $\E\left( (\bb{X}_i - \bb{\mu})(\bb{X}_i - \bb{\mu})\tr \right) = \bb{\Sigma}$ (Based on the usual notation for elliptically symmetric family, the variance of $\bb{X}_i$ is $k_g\bb{\Sigma}$ where $k_g$ is a constant depending on the function $g$, but we assume that such $k_g$ is included in the dispersion matrix $\bb{\Sigma}$ itself by modifying the function $g$ appropriately). Even though these statistical RPCA approaches guarantee the highest possible asymptotic breakdown point of $1/2$, they show low asymptotic efficiency and sometimes large bias even at considerably lower levels of contaminations than their breakdowns~\citep{fishbone2023new}. 

Recent advances in the area of RPCA view the estimation of the principal components in a different light through the guise of the factor model. \cite{wright2009robust} define the RPCA problem as the problem of recovering $\bb{L}$ from the unknown decomposition of the data matrix $\bb{X} = \bb{L} + \bb{S}$, where $\bb{L}$ is a low rank matrix and $\bb{S}$ is a sparse noise component. The direct solution to this problem would consider the optimization problem 
\begin{equation}
    \min_{\bb{L}, \bb{S}} \rank{\bb{L}} + \gamma \Vert \bb{S} \Vert_0,
    \label{eqn:rpca-problem}
\end{equation}
\noindent subject to the restriction that $\Vert \bb{S} \Vert_0 \leq k$ and $\bb{X} = \bb{L} + \bb{S}$, for a predetermined value of $k$. Here, $\Vert \bb{A} \Vert_0$ denotes the $L_0$-norm of the matrix $\bb{A}$, i.e., the sum of the nonzero entries of $\bb{A}$ and $\gamma$ is a tuning parameter to control the balance between the rank of $\bb{L}$ and the sparsity of $\bb{S}$. As noted in \cite{candes2011robust}, the classical PCA seeks the best low-rank component $\bb{L}$ in terms of minimizing the usual Euclidean $L_2$ norm, i.e., it is related to the optimization problem $\min_{\bb{L}} \Vert \bb{X} - \bb{L}\Vert_2$ subject to the restriction that $\rank{\bb{L}} \leq k$. However, the problem in Eq.~\eqref{eqn:rpca-problem} is notoriously difficult to solve, hence \cite{wright2009robust} and \cite{candes2011robust} considered the convex optimization problem $\min_{\bb{L}, \bb{S}} \Vert \bb{L}\Vert_{\ast} + \gamma \Vert \bb{S}\Vert_1$ where $\Vert\bb{L}\Vert_{\ast}$ is the nuclear norm of the matrix $\bb{L}$, i.e., the sum of its singular values and $\Vert \bb{S}\Vert_1$ is the $L_1$ norm of the matrix $\bb{S}$. Various algorithmic techniques like principal component pursuit (PCP) method~\citep{candes2011robust}, augmented Lagrange multiplier (ALM) method~\citep{lin2010augmented} and alternating projection (AltProj) algorithm~\citep{cai2019accelerated} have been developed to solve this optimization problem efficiently. This new approach radically differs from the traditional statistical methods: these methods are non-parametric in nature and assume that the data matrix $\bb{X}$ is non-stochastic, rather the only source of randomness comes from the positions of the nonzero entries of the sparse matrix $\bb{S}$. The convergence and correctness guarantees of these methods are then provided based on the bounds on the entries of these matrices $\bb{L}$ and $\bb{S}$ directly. This exact decomposition is often far from practical applications as every entry of the data matrix $\bb{X}$ is subject to measurement errors. To mitigate this, \cite{zhou2010stable} considered the decomposition
\begin{equation}
    \bb{X} = \bb{L} + \bb{S} + \bb{E},
    \label{eqn:decomposition}
\end{equation}
\noindent where $\bb{E}$ is a dense perturbation matrix (such as matrix with i.i.d. mean zero and homoscadastic entries). Although such a decomposition is considered, the analysis of the algorithm still assumed $\bb{X}$ to be deterministic and considered $\Vert \bb{E}\Vert_2 \leq \delta$, a prespecified level of noise variance to maintain a high signal-to-noise ratio.

\subsection{Connection between RPCA Approaches and Our Contributions}

The two existing RPCA approaches, one based on the maximization of the scale function as in Eq.~\eqref{eqn:pc-function} and another based on the minimization of objective function Eq.~\eqref{eqn:rpca-problem} with matrix decomposition, are usually not equivalent except for the trivial cases of classical PCA. In this paper, we consider a combination of both approaches by taking the decomposition given in Eq.~\eqref{eqn:decomposition} but with stochastic modelling of the data matrix $\bb{X}$. We assume that the rows of the data matrix $\bb{X}$, namely $\bb{X}_1, \dots \bb{X}_n$ are i.i.d. observations generated from an elliptically symmetric family of distributions having a density function of the form as in Eq.~\eqref{eqn:elliptic-model}. Clearly, the sample observations can be expressed as $\bb{X}_i = \bb{\mu} + \bb{\Sigma}^{1/2} \bb{Z}_i$, for $i = 1, 2, \dots n$, where $\bb{Z}_i$ are i.i.d. random variables with $\E(\bb{Z}_i) = 0$ and $\E(\bb{Z}_i \bb{Z}_i\tr) = \bb{I}_p$, the identity matrix of order $p$. The density function of the random variable $\bb{Z}_i$ depends on the specific form of the $g$ function. Then, incorporating the eigendecomposition of $\bb{\Sigma} = \sum_{k = 1}^{p} \gamma_k \bb{v}_k \bb{v}_k\tr$ (with $\gamma_1 \geq \gamma_2 \dots \geq \gamma_p$), we can rewrite the data matrix as
\begin{equation}
  \bb{X} = \bb{1}_n\bb{\mu}\tr + \sum_{k=1}^p \sqrt{\gamma_k} \bb{Z}\bb{v}_k \bb{v}_k\tr,
  \label{eqn:scale-model-decomposition}
\end{equation}
\noindent where $\bb{1}_n$ denotes the $n$-length column vector with all elements equal to $1$ and $\bb{Z}$ is the $n\times p$ matrix whose $i$-th row is equal to $\bb{Z}_i\tr$. Denoting $\bb{u}_k = \bb{Z}\bb{v}_k/\sqrt{n}$ for $k = 1, 2, \dots p$, one can easily see that $\bb{u}_k$s form a set of orthonormal vectors in expectation, i.e., $\E(\bb{u}_k\tr \bb{u}_l) = \delta_{kl}$, the Kronecker delta function. This enables us to rewrite Eq.~\eqref{eqn:scale-model-decomposition} as
\begin{equation}
  \bb{X} = \bb{1}_n\bb{\mu}\tr + \sum_{k=1}^r \sqrt{n\gamma_k} \bb{u}_k \bb{v}_k\tr + \sum_{k=(r+1)}^p \sqrt{n\gamma_k} \bb{u}_k \bb{v}_k\tr, 
  \label{eqn:pca-model}
\end{equation}
\noindent for some prespecified rank $r$. Ignoring the location, the rest of the decomposition is $\bb{X} = \bb{L} + \bb{E}$ which is a subset of the model given in Eq.~\eqref{eqn:decomposition} without any sparse component. In the presence of outlying observations in the data matrix $\bb{X}$, the resulting error matrix $\bb{E}$ will contain occasional spikes which can be separated into the sparse component $\bb{S}$ giving rise to the decomposition in Eq.~\eqref{eqn:decomposition}. Connecting the low rank matrix $\bb{L}$ in Eq.~\eqref{eqn:decomposition} to the sum $\sum_{k=1}^r \sqrt{n\gamma_k} \bb{u}_k \bb{v}_k\tr$ in Eq.~\eqref{eqn:pca-model}, it is now evident that maximizing the scale function of Eq.~\eqref{eqn:pc-function} would result in the eigenvectors $\bb{v}_k$s which are the right singular vectors of the $\bb{L}$ matrix. This provides a connection between the two approaches when the rows of the data matrix are i.i.d. observations from an elliptically symmetric family of distributions. Thus, in this paper, we propose a fast, scalable novel robust PCA algorithm based on the popular minimum density power divergence estimation (MDPDE) approach~\citep{basu1998robust} for the aforementioned setup along with a decomposition as in Eq.~\eqref{eqn:pca-model}. The major contributions of this paper are as follows:
\begin{enumerate}
    \item We propose a novel robust PCA estimator (to be henceforth called rPCAdpd) based on the popular MDPDE, which allows balancing the robustness and efficiency in estimation by simply tuning a robustness parameter $\alpha$ and is able to work under a general decomposition model as in Eq.~\eqref{eqn:decomposition}.
    \item We propose a fast, parallelizable iterative algorithm to obtain the rPCAdpd estimate based on alternating regression; this contrasts with the existing robust PCA algorithms which do not scale well due to large matrix inversion steps.
    \item We also derive various theoretical properties such as equivariance, $\sqrt{n}$-consistency and asymptotic distribution of the proposed rPCAdpd estimator akin to the widely used robust M-estimators. There exists little literature on the theoretical behaviour of the existing PCP methods and often the asymptotic distributions of these estimators are non-Gaussian~\citep{bickel2018projection}.
    \item We also theoretically demonstrate the robustness of the proposed rPCAdpd estimator by demonstrating that its influence function is bounded, and by deriving a lower bound of its asymptotic breakdown point which is independent of the data dimension $p$ but only a function of the robustness tuning parameter $\alpha$. This ensures the scalability of the proposed rPCAdpd estimator for arbitrarily high dimensional random samples.
    \item We corroborate our theoretical findings with extensive simulations. For all the simulation setups considered, rPCAdpd performs better (and sometimes closely on par) than the existing RPCA algorithms.
    \item We also compare the performances of the existing robust PCA algorithms with the rPCAdpd for a few benchmark datasets, and demonstrate how the estimator can be used to detect fraudulent transactions for a credit card transactions dataset.
\end{enumerate}

The rest of the paper is structured as follows: our proposed rPCAdpd estimator is described in detail in Section~\ref{sec:proposed-method} when the model family is elliptically symmetric. In Section~\ref{sec:algorithm}, we derive a computationally efficient iterative technique to obtain the rPCAdpd estimator using the solution to an alternating regression problem. Section~\ref{sec:theory} describes the necessary theoretical results regarding the convergence of the algorithm, equivariance properties, consistency and asymptotic distribution of the estimator. All of these theoretical results are then corroborated by extensive simulation studies in Section~\ref{sec:simulation}, where we compare the performance of the rPCAdpd estimator with several existing robust PCA algorithms. Finally, in Section~\ref{sec:real-data-analysis}, we demonstrate the practical applicability of the proposed estimator for two popular benchmark datasets (namely the Car dataset and Octane dataset in~\cite{hubert2005robpca}) and a Credit Card Fraud Detection dataset. For streamlining the presentation, the proofs of all of the theoretical results are deferred till the Appendix.

\section{The rPCAdpd Estimator}

Before proceeding with the description of the proposed rPCAdpd estimator, we introduce some notations to be used throughout the paper unless otherwise specified. Let, for a matrix $\bb{A}$, $\diag{\bb{A}}$ denote the vector comprising the diagonal elements of $\bb{A}$. The notations $\bb{I}_n$ and $\bb{1}_n$ denote the $n \times n$-size identity matrix and $n$-length vector of $1$s respectively. The transpose, rank and the trace of a matrix $\bb{A}$ will be denoted as $\bb{A}\tr$, $\rank{\bb{A}}$ and $\trace{\bb{A}}$. For any two matrices $\bb{A}$ and $\bb{B}$, their usual matrix product will be denoted as $\bb{A}\bb{B}$ and the Kronecker product will be denoted as $\bb{A} \otimes \bb{B}$. We shall use the symbol $\Vert \bb{x}\Vert_2$ and $\Vert \bb{A}\Vert_2$ to denote the usual Euclidean norm of a vector $\bb{x}$ and the Frobenius norm of the matrix $\bb{A}$ respectively. The notation $f_{\bbtheta}(\bb{x})$ will denote a generic symbol of the probability density function of a random variable $\bb{X}$ following a distribution parametrized by $\bbtheta$ and evaluated at a point $\bb{x}$. The expectation and the covariance operator will be denoted by $\E(\cdot)$ and $\var(\cdot)$ respectively. 

\subsection{Description of the rPCAdpd Estimator}\label{sec:proposed-method}

Let $\bb{X}_1, \dots \bb{X}_n$ be a $p$-variate sample such that each of the observations $\bb{X}_i$ follows an elliptically symmetric family of distributions with a density function of the form
\begin{equation}
    f_{\bbtheta}(\bb{x}) = c_g^{-1}\det(\bb{\Sigma})^{-1/2} \exp\left[ g\left( \bb{x}\tr \sum_{k=1}^p \gamma_k^{-1}\bb{v}_k\bb{v}_k\tr \bb{x} \right) \right],
    \label{eqn:dpd-general-model-1}
\end{equation}
\noindent where $\bb{\Sigma} = \sum_{k=1}^p \gamma_k \bb{v}_k\bb{v}_k\tr$ is the eigendecomposition of the dispersion matrix. The parameter $\bb{\theta} = \left(\gamma_1, \dots \gamma_p, \bb{\eta}\right)$ in Eq.~\eqref{eqn:dpd-general-model-1} consists of the eigenvalues $\gamma_1, \dots \gamma_p$ of the dispersion matrix $\bb{\Sigma}_{p\times p}$ and the parameter $\bb{\eta}$ parametrizing the eigenvectors $\bb{v}_1, \dots \bb{v}_k$ residing in the Stiefel manifold $S_{(p-1)}^p$, i.e., the space of all $p \times p$ orthogonal matrices. Here, $g: \R^+ \rightarrow \R$ is a scalar function that parametrizes the family of distribution and is assumed to be known. For instance, the multivariate Gaussian family of distributions corresponds to $g(x) = (-x/2)$. Note that, since the principal components primarily deal with the variance structure of the data, the location parameter $\bb{\mu} = \E(\bb{X}_i)$ is a nuisance parameter, hence it is assumed to be a known constant. Without the loss of generality, we take this known location parameter equal to $\bb{0}$, otherwise, one may treat $\bb{Y}_i = \bb{X}_i - \bb{\mu}$ as the i.i.d sample under consideration. However, for all practical purposes when it is unknown, one can substitute $\bb{\mu}$ by any consistent robust estimate of the location parameter (some choices will be described later in Section~\ref{sec:choice-robust-location}). We shall show later in Section~\ref{sec:theory} that the choice of this location estimator does not affect the asymptotic properties of the robust estimator of $\bbtheta$ we will propose.

Based on the above formulation, we shall use the popular minimum density power divergence estimator (MDPDE) to estimate these parameters in $\bbtheta$. As shown in several studies~\citep{basu1998robust,ghosh2013robust}, the MDPDE is robust and highly efficient in inference and provides a smooth bridge between the efficient yet non-robust maximum likelihood estimator and the robust but less efficient minimum $L_2$ distance estimator. \cite{basu1998robust} introduced the density power divergence between two densities $g$ and $f$ as
\begin{equation}
    d_\alpha(h, f) = \int f^{1+\alpha}dx - \left( 1 + \dfrac{1}{\alpha} \right)\int f^{\alpha}h dx + \dfrac{1}{\alpha} \int h^{1+\alpha}dx, \ \alpha > 0
\end{equation}
\noindent which provides a smooth bridge between the Kullback Leibler divergence and the $L_2$ distance between $h$ and $f$ via the robustness tuning parameter $\alpha$. Given the true distribution $H$ with density $h$ and a parametric model family of distributions $\mathcal{F} = \{F_\theta: \theta \in \Theta\}$ with corresponding densities $f_\theta$, the MDPD functional $T(H)$ is defined as the value of the parameter $\theta \in \Theta$ such that $d_\alpha(h, f_\theta)$ is minimized. Using the same objective function for MDPDE and substituting the empirical measure of the sample observations instead of the true distribution $H$, our proposed estimator of robust principal components turns out to be the solution to the optimization problem
\begin{equation}
  \bbhat{\theta} = \argmin_{\bb{\theta} \in \bb{\Theta}} \int f_{\bbtheta}^{1+\alpha}(\bb{x})d\bb{x} - \left(1 + \dfrac{1}{\alpha} \right)\dfrac{1}{n}\sum_{i=1}^{n} f_{\bb{\theta}}^\alpha(\bb{X}_i),
  \label{eqn:dpd-general-model-3}
\end{equation}
\noindent where $f_{\bbtheta}(\bb{x})$ is as given in Eq.~\eqref{eqn:dpd-general-model-1} and the parameter space $\bb{\Theta} = (\R^+)^p \times S$ where $S$ is the parameter space for $\bbeta$. Combining Eq.~\eqref{eqn:dpd-general-model-1} and Eq.~\eqref{eqn:dpd-general-model-3}, we can recover MDPDE as 
\begin{equation}
    \bbhat{\theta} = \argmin_{\bb{\theta} \in \bb{\Theta}} c_g^{-\alpha} \prod_{k=1}^p \gamma_k^{-\alpha/2} \left[ \dfrac{c_{(1+\alpha)g}}{c_g}  \right.
    \left. - \left( 1 + \dfrac{1}{\alpha} \right)\dfrac{1}{n}\sum_{i=1}^n e^{ \alpha g\left( \bb{X}_i\tr \sum_{k=1}^p \gamma_k^{-1}\bb{v}_k(\bb{\eta})\bb{v}_k\tr(\bb{\eta}) \bb{X}_i  \right) } \right].
    \label{eqn:dpd-general-model-4}
\end{equation}
\noindent We refer to this as the rPCAdpd estimator of the principal components under the general elliptically symmetric family of distributions. This estimator assumes the description of the model family through the specification of the completely known function $g(\cdot)$. In particular, when $g(x) = (-x/2)$, i.e., the model family is a $p$-variate Gaussian distribution, then the corresponding optimization problem in Eq.~\eqref{eqn:dpd-general-model-4} becomes 
\begin{multline}
    \bbhat{\theta} 
    = \argmin_{\bb{\theta} \in \bb{\Theta}} (2\pi)^{-\alpha p/2} \prod_{k=1}^p \gamma_k^{-\alpha/2} \left[ (1+\alpha)^{-p/2} - \right.\\
    \left. \left( 1 + \dfrac{1}{\alpha} \right)\dfrac{1}{n}\sum_{i=1}^n e^{ -\frac{\alpha}{2} \left( \bb{X}_i \tr \sum_{k=1}^p \gamma_k^{-1}\bb{v}_k(\bbeta)\bb{v}_k\tr(\bbeta) \bb{X}_i  \right) } \right].
    \label{eqn:dpd-normal-model-4}
\end{multline}

\subsection{Algorithm for Efficient Compution of the rPCAdpd Estimator}\label{sec:algorithm}

Clearly, if the minimization given in Eq.~\eqref{eqn:dpd-general-model-4} was to be performed on the entries of the dispersion matrix to obtain a robust estimate of covariance directly, it would be difficult to restrict the optimization space to the space of all positive definite matrices. Thus, the optimization is deliberately made with respect to the eigenvectors and the eigenvalues of the dispersion matrix to ensure that the estimated dispersion matrix remains positive definite and symmetric. While it is easy to optimize the objective function in Eq.~\eqref{eqn:dpd-general-model-4} with respect to the eigenvalues, it still remains computationally expensive to solve it for the eigenvectors since one has to perform an optimization over the non-convex Steifel manifold $S_{p-1}^p$. Although there exist some efficient optimization algorithms on the Riemannian manifold as proposed by~\cite{wen2013feasible,jiang2015framework,li2020efficient}, these general-purpose optimization techniques require complicated iteration steps via Cayley transformation and curvilinear searches. To circumvent this direct optimization, we apply a procedure similar to the alternating regression approach of the rSVDdpd algorithm by~\cite{roy2021new}.

We start by assuming that the unknown location parameter $\bb{\mu}$ is already estimated using a robust consistent estimator of the location. For our purpose, we use the $L_1$-median as the location estimator; however, in Section~\ref{sec:choice-robust-location}, we shall describe some alternative choices that may be used. In the decomposition of Eq.~\eqref{eqn:decomposition}, we assume that elements of the error matrix $\bb{E}$ are independent and identically distributed. For instance, when the model densities $f_{\bbtheta}$ follow a multivariate Gaussian distribution (or multivariate $t$-distribution), the entries of $\bb{E}$ follow approximately univariate Gaussian distribution (or univariate $t$-distribution) respectively. The sparse matrix $\bb{S}$ has a few nonzero entries, which may be regarded as outlying observations in the original data matrix $\bb{X}$ at the corresponding places. This is a classic setup for robust statistical inference, hence the MDPDE approach can be directly used to tackle this estimation problem. For ease of explanation, in the following text, we develop the proposed algorithm assuming the particular model of Gaussian distribution as in Eq.~\eqref{eqn:dpd-normal-model-4}. However, the same algorithm can be modified to fit any choice of $g(\cdot)$ in Eq.~\eqref{eqn:dpd-general-model-4} using its univariate analogous distribution. 

To estimate the principal components robustly, we perform a robust singular value decomposition of the centred data matrix using an iterative algorithm rSVDdpd~\citep{roy2021new}. To illustrate the approach, we rewrite the decomposition model of Eq.~\eqref{eqn:decomposition} as 
\begin{equation}
    X_{ij} = \mu_j + \sum_{k=1}^r u_{ki}\beta_{kj} + \epsilon_{ij} 
    = \mu_j + \sum_{k=1}^r \alpha_{ki}v_{kj} + \epsilon_{ij},  \ i = 1, \dots n; j = 1, \dots p,
    \label{eqn:svd-model-each}
\end{equation}
where $\beta_{kj} = \lambda_k v_{kj}$, $\alpha_{ki} = \lambda_k u_{ki}$, $u_{ki}$ is the $i$-th coordinate of $\bb{u}_k$ and $v_{kj}$ is the $j$-th coordinate of $\bb{v}_k$. For a fixed choice of $j$ and known value of $r$ and $\bb{u}_{k}$s (for $k = 1, \dots r$), Eq.~\eqref{eqn:svd-model-each} simply denotes a linear regression problem with intercept $\mu_j$ and $r$ slope coefficients $\beta_{1j}, \dots \beta_{rj}$. Let, $\widehat{\mu}_j$ be the robust consistent estimator of $\mu_j$. Also, let $(\widehat{u}_{ki}^{(t)}, \widehat{v}_{kj}^{(t)}, \widehat{\lambda}_k^{(t)}, (\widehat{\sigma}^2)^{(t)})$ be the estimates at the $t$-th iteration of the algorithm and $\widehat{\beta}_{kj}^{(t)}$ and $\widehat{\alpha}_{ki}^{(t)}$ be defined accordingly. The iteration rule for the rSVDdpd algorithm is then defined by the system of equations
\begin{equation}
    \begin{split}
    \left(\widehat{\beta}_{1j}^{(t+1)}, \dots \widehat{\beta}_{rj}^{(t+1)}\right) 
    & = \argmin_{\beta_{1j},\dots \beta_{rj}} \dfrac{1}{n}\sum_{i=1}^n V\left( Z_{ij}; \widehat{u}_{ki}^{(t)}, \beta_{kj}, (\widehat{\sigma}^2)^{(t)} \right),\\
    \left(\widehat{\alpha}_{1i}^{(t+1)}, \dots \widehat{\alpha}_{ri}^{(t+1)}\right) 
    & = \argmin_{\alpha_{1i},\dots \alpha_{ri}} \dfrac{1}{p}\sum_{j=1}^p V\left( Z_{ij}; \alpha_{ki}, \widehat{v}_{kj}^{(t+1)}, (\widehat{\sigma}^2)^{(t)} \right),\\
    (\widehat{\sigma}^2)^{(t+1)} 
    & = \argmin_{\sigma^2} \dfrac{1}{np}\sum_{i=1}^n \sum_{j=1}^p V\left( Z_{ij}; \widehat{\alpha}_{ki}^{(t)}, \widehat{v}_{kj}^{(t+1)}, \sigma^2 \right).
    \end{split}
    \label{eqn:iteration-rule}
\end{equation}
\noindent where 
\begin{equation*}
  V(y; c, d, \sigma^2) = \dfrac{1}{(2\pi)^{\alpha/2}\sigma^{\alpha}} \left[ \dfrac{1}{\sqrt{1+\alpha}} - \left( \dfrac{1+\alpha}{\alpha} \right) \exp\left\{ -\alpha\dfrac{(y-cd)^2}{2\sigma^2} \right\} \right],  
\end{equation*}
\noindent with $\alpha$ being the robustness tuning parameter as in Eq.~\eqref{eqn:dpd-general-model-4}. In between these steps the vectors $(\widehat{\alpha}_{k1}^{(t)}, \dots \widehat{\alpha}_{kn}^{(t)})\tr$ and $(\widehat{\beta}_{k1}^{(t)}, \dots \widehat{\beta}_{kp}^{(t)})\tr$ are normalized accordingly to produce unit vectors $\bbhat{u}_k^{(t)} = ( \widehat{u}_{k1}^{(t)}, \dots,  \widehat{u}_{kn}^{(t)} )\tr$ and $\bbhat{v}_k^{(t)} = ( \widehat{v}_{k1}^{(t)}, \dots,  \widehat{v}_{kr}^{(t)} )\tr$, and the norm of the $\beta$-vector is regarded as the estimate $\widehat{\lambda}_k^{(t)}$, the $k$-th singular value at $t$-th step of the iteration.

We repeat these alternating steps until convergence. Using the converged estimates from the aforementioned rSVDdpd procedure, the unit vector $\bbhat{v}_k^{(\infty)}$ and the quantity $(\widehat{\lambda}_k^{(\infty)})^2/n$ are outputted as the $k$-th eigenvector and $k$-th eigenvalue corresponding to the principal components of the i.i.d. sample $\bb{X}_1, \dots \bb{X}_n$ respectively. We shall call this entire procedure as the robust principal component analysis using the density power divergence (rPCAdpd) algorithm.

\subsection{Choice of the Robust Location Estimator}\label{sec:choice-robust-location}
There are several choices for the robust estimators of the location for the rPCAdpd algorithm. We shall discuss only a few of these estimators which are quick and simple since the primary focus is to estimate the principal components. As we will show later in Section~\ref{sec:theory}, the asymptotic properties of the estimated principal components are free of the choice of this location estimator, as long as the location estimator is robust and asymptotically consistent.

Naturally, we may want to use the MDPDE~\citep{basu1998robust} for a normal location model family, extended to a multivariate setup. However, estimating the location parameter in this way would force us to estimate the unknown dispersion matrix $\bb{\Sigma}$ as well, which is already taken care of using the rPCAdpd algorithm. Also, as will be discussed later in Section~\ref{sec:orthogonal-equivariance}, this multivariate MDPDE does not satisfy the desirable orthogonal equivariance property, and in particular, the permutation equivariance property. So instead, we can resort to a coordinatewise MDPDE under the normal location model family. In this case, the coordinates of the estimated location vector satisfy 
\begin{equation*}
    \widehat{\mu}_j = \argmin_{\mu}\min_{\sigma}\dfrac{1}{(2\pi)^{\alpha/2}\sigma^\alpha} \left[ \dfrac{1}{\sqrt{1+\alpha}} - \left(\dfrac{1+\alpha}{\alpha} \right)\dfrac{1}{n}\sum_{i=1}^n\exp\left\{ -\alpha \dfrac{(X_{ij} - \mu)^2}{2\sigma^2} \right\} \right], \ j = 1, \dots p, 
\end{equation*} 
\noindent where $\alpha$ is the robustness parameter lying between $0$ and $1$, $X_{ij}$ is the $j$-th coordinate of $\bb{X}_i$. This coordinatewise MDPDE still retains its robustness properties while being permutation and scale equivariant, but it still does not satisfy orthogonal equivariance for general orthogonal matrices.

Alternative choices of a robust and consistent estimator of the location parameter would include the $L_1$ median~\citep{vardi2000multivariate}, coordinatewise median or any $M$-estimator for location~\citep{huber1964location}. The $L_1$ median possesses the desirable orthogonal equivariance property. Based on extensive simulation studies, we have found that $L_1$ median fits our purpose and provides a desirable balance between speed (computational advantage) and accuracy (robustness and efficiency), and hence it is chosen to be used as a robust location estimator during the rPCAdpd algorithm for all our subsequent studies.

\subsection{Choices of Hyperparameters}\label{sec:choice-hyperparameter}

The two hyperparameters associated with the rPCAdpd estimator are the rank of the $\bb{L}$ matrix, i.e., the number of significant eigenvalues or the number of principal components to output, and the robustness parameter $\alpha$ in the objective function~\eqref{eqn:dpd-general-model-3}.

To determine the rank of the matrix $\bb{L}$, we robustly estimate all the $\min(n, p)$ eigenvalues and the corresponding eigenvectors using the rPCAdpd algorithm. Subsequently, we select a rank $r \leq \min(n, p)$, ensuring that the first $r$ eigenvalues and corresponding eigenvectors can account for a proportion of variation of at least $(1 - \delta)$. Common choices for $\delta$ are typically $0.1$ or $0.25$. Thus, the rank of the matrix $\bb{L}$ is estimated as
\begin{equation*}
    \widehat{r} = \min\left\{ 1 \leq r \leq \min(n,p): \frac{\sum_{k=1}^{r} \widehat{\gamma}^{(\alpha)}_k}{\sum_{k=1}^{\min(n,p)} \widehat{\gamma}^{(\alpha)}_k  } > (1-\delta) \right\},
\end{equation*}
\noindent where $\widehat{\gamma}^{(\alpha)}_k$ is the $k$-th eigenvalue as estimated by rPCAdpd method with robustness parameter $\alpha$. Similar criteria have been used to determine the number of significant principal components by many authors~\citep{he2012incremental,xu2012robust}.

Applying the general result pertaining to the asymptotic breakdown of the MDPDE as in~\cite{roy2023breakdown}, the asymptotic breakdown of the rPCAdpd estimator turns out to be at least $\alpha/(1+\alpha)$. We discuss this in detail later in Section~\ref{sec:breakdown}. Clearly, as $\alpha$ increases to $1$, one approaches the highest possible breakdown $1/2$, by sacrificing some efficiency in estimation. On the other hand, the efficiency is most when $\alpha \rightarrow 0$, but the breakdown becomes unacceptably low for the rPCAdpd algorithm to be of any use as a robust PCA estimator in that case. Therefore, there must be a balance between robustness and efficiency with an adaptive optimal choice of $\alpha \in [0, 1]$. Since we use the rSVDdpd procedure to obtain the estimate of the singular values from which we obtain the robust estimates of the principal components, we follow the same criterion as introduced by~\cite{roy2021new}. The authors consider that the optimal choice of the robustness parameter is the minimizer of a conditional MSE criterion
\begin{multline*}
    (n + p) (\widehat{\sigma}^{(\alpha)})^2 \left( 1 + \dfrac{\alpha^2}{1 + 2\alpha} \right)^{3/2} + \dfrac{1}{r} \sum_{k=1}^r \Vert \widehat{\lambda}_k^{(\alpha)}\widehat{\boldsymbol{a}}_k^{(\alpha)} - \widehat{\lambda}_k^{(1)}\widehat{\boldsymbol{a}}_k^{(1)}\Vert_2^2 + \dfrac{1}{r}\sum_{k=1}^r \Vert \widehat{\lambda}_k^{(\alpha)}\widehat{\boldsymbol{b}}_k^{(\alpha)} - \widehat{\lambda}_k^{(1)}\widehat{\boldsymbol{b}}_k^{(1)}\Vert_2^2,
\end{multline*}
\noindent where $\widehat{\lambda}_k^{(\alpha)}, \widehat{\boldsymbol{a}}_k^{(\alpha)}, \widehat{\bb{b}}_k^{(\alpha)}$ are the estimates of $k$-th singular value and vectors as obtained by the rSVDdpd procedure with robustness parameter $\alpha$.

\section{Theoretical Properties}\label{sec:theory}

In this section, we explore various theoretical properties of the rPCAdpd estimator. First, we show the existence and the uniqueness of the estimator and that the proposed iterative algorithm converges to the estimator for any finite $n$. Next, we prove various equivariance properties, and asymptotic consistency, following which we derive the asymptotic distribution of the robust eigenvalues and eigenvectors estimated by the rPCAdpd estimator. Finally, we derive the influence function and asymptotic breakdown point of the estimator to demonstrate its robustness properties. All of these theoretical results hold for any location estimator that is robust, asymptotically consistent and equivariant under the orthogonal transformation (like $L_1$-median), used in the rPCAdpd algorithm.

\subsection{Existence of the Estimator}
We start by writing the objective function in Eq.~\eqref{eqn:dpd-general-model-4} as a function of the individual term of the parameter vector $\bbtheta$ as 
\begin{multline}
    Q(\gamma_1, \dots, \gamma_p, \bbeta) = \prod_{k=1}^{p} \gamma_k^{-\alpha/2} \left[ \dfrac{c_{(1+\alpha)g}}{c_g} \right. \\
    \left. - \left( 1 + \dfrac{1}{\alpha} \right) \dfrac{1}{n}\sum_{i=1}^n \exp\left\{ \alpha g\left( (\bb{X}_i - \bb{\widehat{\mu}})\tr\sum_{k=1}^p \gamma_k^{-1}\bb{v}_k(\bbeta)\bb{v}_k(\bbeta)\tr (\bb{X}_i - \bb{\widehat{\mu}})  \right) \right\} \right].
    \label{eqn:Q-function}
\end{multline}
\noindent where $\bb{\widehat{\mu}}$ is a robust consistent estimate of the location including those described in Section~\ref{sec:choice-robust-location}.
The following result establishes the existence of the rPCAdpd estimator.
\begin{theorem}\label{thm:existence}
    If the generating function $g: [0, \infty) \rightarrow \R$ of the elliptically symmetric family of distributions is a decreasing continuous function, then for a sufficiently large number of sample observations $n$, there exists a minimum of the objective function $Q(\cdot)$ given in Eq.~\eqref{eqn:Q-function} with probability tending to $1$. 
\end{theorem}
\noindent For instance, when the model family is $p$-variate $t$-distribution with $\nu$ degrees of freedom, then $g(x)$ turns out to be $-\frac{\nu + p}{2} \log(1 + x/\nu)$, which is a decreasing continuous function, hence the rPCAdpd estimator exists for the multivariate $t$-distribution family.

\subsection{Convergence of the Algorithm}\label{sec:convergence}
Once the existence of the rPCAdpd estimator is established, the convergence of the algorithm follows directly from the convergence of the rSVDdpd procedure as presented in~\cite{roy2021new}. Observe that, the iterations in Eq.~\eqref{eqn:iteration-rule} monotonically decrease the value of objective function $Q(\gamma_1, \dots, \gamma_p, \bbeta)$, which is also continuous in its arguments. Since Theorem~\ref{thm:existence} asserts the existence of the minimizer, it means that the sequence $Q(\widehat{\gamma}_1^{(t)}, \dots, \widehat{\gamma}_p^{(t)}, \widehat{\bbeta}^{(t)})$ (where $\widehat{\gamma}_1^{(t)}$ and $\widehat{\bbeta}^{(t)}$ denote the estimated parameters at $t$-th iteration) is bounded below. Then an application of the monotone convergence theorem combined with the uniqueness of the rSVDdpd estimator asserts the convergence of the rPCAdpd estimator.

\subsection{Orthogonal Equivariance}\label{sec:orthogonal-equivariance}

As mentioned in~\cite{rousseeuw1985multivariate}, orthogonal equivariance is one of the fundamental properties that an estimator of the principal components should possess. Let, $\bb{Y}_1, \dots \bb{Y}_n$ be a transformed sample $\bb{Y}_i = a\bb{P}\bb{X}_i + \bb{b}$ for $i = 1, 2, \dots n$, where $\bb{P}_{p\times p}$ is an orthogonal matrix, $a \in (0, \infty)$ and $\bb{b}$ is a $p$-length vector. Then, an orthogonally equivariant estimator $T_\lambda(\bb{X}_1, \dots \bb{X}_n)$ of an eigenvalue should satisfy $T_\lambda(\bb{Y}_1, \dots \bb{Y}_n) = a^2 T_\lambda(\bb{X}_1, \dots \bb{X}_n)$. Similarly, for an orthogonally equivariant estimate $T_{\bb{v}}(\bb{X}_1, \dots \bb{X}_n)$ of the corresponding eigenvector, it satisfies $T_{\bb{v}}(\bb{Y}_1, \dots \bb{Y}_n) = \bb{P} T_{\bb{v}}(\bb{X}_1, \dots \bb{X}_n)$. For any orthogonal equivariant estimate of the principal components, both of these two conditions should hold for all eigenvalues and their corresponding eigenvectors. 

Since our primary focus is on the principal components, we will assume that the robust estimator of the location parameter is orthogonally equivariant. The choice of $L_1$-median as a robust estimator of location satisfies this property. Given the orthogonal equivariance property of the location estimator, it follows that the resulting rPCAdpd estimator also satisfies the same.
\begin{theorem}\label{thm:orthogonal-equivariance}
    The rPCAdpd estimators of the eigenvalues and eigenvectors are equivariant under the transformation 
    \begin{equation}
        \bb{Y}_i = a\bb{P}\bb{X}_i + \bb{b}, \ i = 1, 2, \dots n,
        \label{eqn:equivariant-transformation}
    \end{equation}
    \noindent where $\bb{P}_{p\times p}$ is an orthogonal matrix, $a \in (0, \infty)$ and $\bb{b}$ is a $p$-length vector provided that the location estimator used in the rPCAdpd procedure also satisfy same equivariance property.
\end{theorem}
\begin{corollary}
    As in the case of the rSVDdpd estimator discussed in~\cite{roy2021new}, the rPCAdpd estimator also satisfies scale and permutation equivariance. This follows from the observation that both are special cases of the transformation mentioned in Eq.~\eqref{eqn:equivariant-transformation}. In particular, with $\bb{P} = \bb{I}_p$, we get scale equivariance. If $a = 1$ and $\bb{P}$ is a permutation matrix, then permutation equivariance follows.
\end{corollary}

\subsection{Consistency and Asymptotic Distribution}

One of the integral components of the proposed rPCAdpd estimator is the MDPDE. As shown in~\cite{basu1998robust}, the MDPDE, being an M-estimator and a minimum distance estimator, enjoys a vast set of nice asymptotic properties including consistency and asymptotic normality. In this subsection, we will investigate how these properties carry over to the special scenario of principal component estimation under elliptically symmetric models. Thus, throughout this entire subsection, unless otherwise specified, we will consider the setup that the sample observations $\bb{X}_1, \dots \bb{X}_n$ are i.i.d. random variables from a $p$-variate elliptically symmetric distribution with unknown mean $\bb{\mu}^\ast$ and unknown dispersion matrix $\bb{\Sigma}^\ast$, having density function 
\begin{equation}
    f_{\bbtheta^\ast}(\bb{x}) = c_g^{-1} \text{det}(\bb{\Sigma}^\ast)^{-1/2} e^{ g\left( (\bb{x}-\bb{\mu}^\ast)\tr (\bb{\Sigma}^\ast)^{-1} (\bb{x}-\bb{\mu}^\ast) \right) }, \ \bb{x} \in \R^p,
    \label{eqn:model-family}
\end{equation}
\noindent where $g$ is the characterizing function of the elliptically symmetric family of distributions. The covariance matrix $\bb{\Sigma}^{\ast}$ is assumed to have an eigendecomposition $\bb{\Sigma}^{\ast} = \sum_{k=1}^p \gamma_k^\ast \bb{v}_k^\ast (\bb{v}_k^\ast)\tr$ where $\gamma_k^{\ast} \geq 0$ are eigenvalues and $\bb{v}_k^{\ast}$s are the corresponding eigenvectors of the covariance matrix. We wish to estimate the parameter of interest $\bbtheta^\ast = (\gamma_1^\ast, \dots \gamma_p^\ast, \bbeta^\ast)$, comprising of the eigenvalue $\gamma_1^\ast, \dots \gamma_p^{\ast}$ and the natural parameter $\bbeta^\ast$ parametrizing the eigenvectors in the Stiefel manifold $S_{(p-1)}^p$. The location parameter $\bb{\mu}^\ast$ is a nuisance parameter in this setup.

Following the footsteps of~\cite{basu1998robust}, we consider the following quantities
\begin{equation*}
  \bbxi_{\bbtheta} = \int u_{\bbtheta}(\bb{x})f_{\bbtheta}^{(1+\alpha)}dx, \
  \bb{J}_{\bbtheta} = \int u_{\bbtheta}(\bb{x})u_{\bbtheta}(\bb{x})\tr f_{\bbtheta}^{(1+\alpha)}dx, \
  \bb{K}_{\bbtheta} = \int u_{\bbtheta}(\bb{x})u_{\bbtheta}(\bb{x})\tr f_{\bbtheta}^{(1+2\alpha)}dx - \bbxi_{\bbtheta}\bbxi_{\bbtheta}\tr,  
\end{equation*}
\noindent which are essential for obtaining different asymptotic properties of the MDPDE. Here, $f_{\bbtheta}(\bb{x})$ denotes the same family of distributions as in Eq.~\eqref{eqn:model-family} at parameter $\bbtheta$ and corresponding score function is denoted by $u_{\bbtheta}(\bb{x}) = \fracpartial{}{\bbtheta}\log(f_{\bbtheta}(\bb{x}))$. To calculate all of these quantities, we will resort to the following assumptions. 
\begin{enumerate}[label = (A\arabic*), ref = (A\arabic*)]
    \item\label{assum:0} The generating function $g(\cdot)$ for the elliptically symmetric family of distributions is thrice differentiable and the third order derivative is continuous.
    \item\label{assum:2} The true eigenvalues $\gamma_1^\ast, \dots, \gamma_p^\ast$ are distinct.
    \item\label{assum:3} The functions $s^2 g\dash(s)e^{g(s)}$, $s^4 (g\dash(s))^2 e^{g(s)}$, $s^4 g\ddash(s)e^{g(s)}$ and $s^4 g^{\prime\prime\prime}(s)e^{g(s)}$ are uniformly bounded above by some constant $M^\ast$ for any $s \geq 0$, where $g\dash(s), g\ddash(s)$ and $g^{\prime\prime\prime}(s)$ denotes the first, second and third order derivatives of $g$.
\end{enumerate}
\noindent Assumptions~\ref{assum:0} and~\ref{assum:3} are similar in spirit to the assumptions (R1) and (R2) of~\cite{ghosh2013robust}, which in turn imply the assumptions (A1)-(A5) of~\cite{basu1998robust}. One of the standard regularity conditions for such asymptotic results is the exchangeability of the differentiation and integral signs, i.e., the integral $\int f_{\bbtheta}^{(1+\alpha)}(\bb{z})d\bb{z}$ should be differentiable with respect to $\bb{\theta}$ for any $\alpha \in [0, 1]$ and the derivative can be taken under the integral sign. However, this fact follows as a consequence of assumption~\ref{assum:0} for the elliptically symmetric family of distributions. Assumption~\ref{assum:2} makes the calculation simpler, but it is not strictly necessary to establish the asymptotic properties of the proposed estimator. However, it is also known that the set of random matrices with i.i.d. entries with a repeated eigenvalue is negligible~\citep{tao2012topics}. \cite{Kumar_2017} verify similar conclusions for a broader range of distribution of random matrices using numerical simulations. Thus, assumption~\ref{assum:2} holds for almost all positive definite matrices $\bb{\Sigma}^\ast$.

We begin with two generic lemmas describing the quantity $\bbxi_{\bbtheta}$ and $\bb{J}_{\bbtheta}$ as a function of the integral of the model density function and its derivatives. These lemmas are generic; they are applicable in any MDPDE setup, not only in particular to RPCA.

\begin{lemma}\label{lemma:general-xi}
    Let, $c_{\alpha}(\bb{\theta}) = \int f^{(1+\alpha)}_{\bb{\theta}}(\bb{x}) d\bb{x}$. Then under the assumption of thrice differentiability of $f_{\bbtheta}(x)$ and the exchangeability of the differentiation and integral signs,
    \begin{equation*}
        \bbxi_{\bbtheta} = (1+\alpha)^{-1}{c_{\alpha}(\bbtheta)} \fracpartial{}{\bbtheta}\log(c_{\alpha}(\bbtheta)).
    \end{equation*}
\end{lemma}

\begin{lemma}\label{lemma:general-J}
    Under the assumption of thrice differentiability of $f_{\bbtheta}(x)$ and the exchangeability of the differentiation and integral signs, 
    \begin{equation}
        \bb{J}_{\bbtheta} = \dfrac{c_\alpha(\bbtheta)}{(1+\alpha)^2} \left( i^h(\bbtheta) + \left(\dfrac{\partial}{\partial\bbtheta}\log(c_\alpha(\bbtheta))\right)\left(\dfrac{\partial}{\partial\bbtheta}\log(c_\alpha(\bbtheta)) \right)\tr \right)
        \label{eqn:easy-J}
    \end{equation}
    \noindent where $i^h(\bbtheta)$ is the expected Fisher information matrix for a single observation $\bb{x}$ following the density function $h_{\bbtheta}(\bb{x}) = c^{-1}_\alpha(\bbtheta)f^{(1+\alpha)}_{\bbtheta}(\bb{x})$. 
\end{lemma}

Before proceeding with the computation of these quantities $\bbxi_{\bbtheta}, \bb{J}_{\bbtheta}$ and $\bb{K}_{\bbtheta}$ for the particular setup of the rPCAdpd estimator, we recognize that the estimation of the principal components is essentially a two-step procedure. In the first step, we use a consistent robust estimator $\bbhat{\mu}$ to estimate the location parameter $\bb{\mu}^\ast$. In the next step, the rSVDdpd procedure was used to obtain the MDPDE of $\bbtheta$ using the model family densities $f_{\bbtheta}(x)$ as in Eq.~\eqref{eqn:model-family} by replacing $\bb{\mu}^\ast$ with its estimate $\bbhat{\mu}$ from the first step. Therefore, in the following, we compute the quantities $\bbxi_{\bbtheta}, \bb{J}_{\bbtheta}$ and $\bb{K}_{\bbtheta}$ conditional on the value of $\bbhat{\mu}$, which will lead to the conditional asymptotic distribution of $\bbhat{\theta}$ (The proof of which is described in Appendix~\ref{appendix:normality-general-proof}). However, as we shall show later in Theorem~\ref{thm:normality-general}, this conditional distribution turns out to be free of $\bbhat{\mu}$, hence the unconditional asymptotic distribution of $\bbhat{\theta}$ will also remain the same. 

We start by using Lemma~\ref{lemma:general-xi} in combination with Assumption~\ref{assum:2} for our specific use case. To compactly write $\bbxi_{\bbtheta^\ast}$, we introduce the diagonal matrix $\bb{\Gamma}_{p \times p}$ with nonzero entries $\gamma_1^\ast, \dots \gamma_p^\ast$.  
\begin{corollary}
    \label{cor:ellip-xi}
    If $f_{\bbtheta}(\bb{x})$ is a density function belonging to an elliptically symmetric family of distributions with generating function $g(\cdot)$ as given in Eq.~\eqref{eqn:model-family}, then under assumptions~\ref{assum:0}-\ref{assum:3} when the location parameter $\mu^\ast$ is a fixed quantity,
    \begin{equation*}
        \bbxi_{\bbtheta^\ast}
        = \dfrac{ c_{(1+\alpha)g} }{(1+\alpha)(c_g)^{(1+\alpha)} } \prod_{k=1}^p (\gamma_k^\ast)^{-\alpha/2}
        \begin{bmatrix}
            -\dfrac{\alpha}{2} \diag{\bb{\Gamma}^{-1}} \\
            0\\
        \end{bmatrix}.
    \end{equation*}
\end{corollary}
\noindent The quantity $\bb{J}_{\bbtheta^\ast}$ for the current setup can be expressed similarly.
\begin{corollary}\label{cor:ellip-jtheta}
    If $f_{\bbtheta}(\bb{x})$ is a density function belonging to an elliptically symmetric family of distributions with generating function $g(\cdot)$ as given in Eq.~\eqref{eqn:model-family}, then under assumptions~\ref{assum:0}-\ref{assum:3} when the location parameter $\mu^\ast$ is a fixed quantity,
    \begin{equation*}
        \bb{J}_{\bbtheta^\ast} = \dfrac{c_{(1+\alpha)g}}{(1+\alpha)^2 c_{g}^{(1+\alpha)}} \prod_{k=1}^p (\gamma_k^\ast)^{-1/2}
        \begin{bmatrix}
            i^h(\bb{\gamma}, \bb{\gamma}) + \frac{\alpha^2}{4}\left(\diag{\bb{\Gamma}^{-1}} \right)\left( \diag{\bb{\Gamma}^{-1}} \right)\tr & i^h(\bb{\gamma}, \bb{\eta}) \\
            i^h(\bb{\gamma}, \bb{\eta})\tr & i^h(\bb{\eta}, \bb{\eta})
        \end{bmatrix}.
    \end{equation*}
\end{corollary}
\noindent The quantities $i^h(\cdot, \cdot)$ are given by the following formulae
\begin{align*}
    i^h(\bb{\gamma}, \bb{\gamma})
    & = -\dfrac{1}{4}\left(\diag{\bb{\Gamma}^{-1}} \right)\left( \diag{\bb{\Gamma}^{-1}} \right)\tr + \bb{\Gamma}^{-2} \bb{V}\tr A_4((1+\alpha)g) \bb{V} \bb{\Gamma}^{-2},\\
    i^h(\bb{\gamma}, \bb{\eta})
    & = -2 \bb{\Gamma}^{-2} \bb{V}\tr (\bb{I}_p \otimes \bb{\Gamma}^{-1}) A_4((1+\alpha)g) \bb{G}\tr,\\
    i^h(\bb{\eta}, \bb{\eta})
    & = 4 \bb{G} (\bb{I}_p \otimes \bb{\Gamma}^{-1}) A_4((1+\alpha)g) (\bb{I}_p \otimes \bb{\Gamma}^{-1})\tr \bb{G}\tr.
\end{align*}
\noindent where 
\begin{align*}
    Q(\bb{x}) & = \bb{x}\tr \sum_{k=1}^p (\gamma_k^\ast)^{-1} \bb{v}_k^\ast (\bb{v}_k^\ast)\tr \bb{x},\\
    \bb{V}_{p^2 \times p}
    & = \begin{bmatrix}
        \bb{v}_1^\ast & 0 & \dots & 0\\
        0 & \bb{v}_2^\ast & \dots & 0\\
        \vdots & \vdots & \ddots & \vdots\\
        0 & 0 & \dots & \bb{v}_p^\ast
    \end{bmatrix},\\
    \bb{G}_{p(p+1)/2 \times p^2}
    & = \begin{bmatrix}
        \dfrac{\partial \bb{v}_1}{\partial \bbeta}\mid_{\bbeta = \bbeta^\ast} & \dfrac{\partial \bb{v}_2}{\partial \bbeta}\mid_{\bbeta = \bbeta^\ast} & \dots & \dfrac{\partial \bb{v}_p}{\partial \bbeta}\mid_{\bbeta = \bbeta^\ast} \\
    \end{bmatrix}\tr,\\
\end{align*}
and $A_4(g)$ be the $p^2 \times p^2$ matrix comprising of the partitions $A_4(g; \bb{v}_i^\ast, \bb{v}_j^\ast)$ for $i, j = 1, 2, \dots p$, where 
\begin{equation*}
    A_4(g; \bb{u}, \bb{v}) = \int \left(g\dash(Q(\bb{x})) \right)^2 \bb{x} \bb{x}\tr \bb{u}\bb{v}\tr \bb{x}\bb{x}\tr \mathcal{C}_{g}^{-1}\exp(g(Q(\bb{x})) )d\bb{x}.
\end{equation*}

For the particular setup of principal components for the elliptically symmetric family, the assumptions~\ref{assum:0}-\ref{assum:3} indicate all the necessary assumptions (A1)-(A5) of~\cite{basu1998robust}. Thus, we can readily use Theorem 2.2 of the same to establish the asymptotic properties such as consistency and the asymptotic normality of the converged rPCAdpd estimator of the principal components. However, since the quantities $\bbxi_{\bbtheta}, \bb{J}_{\bbtheta}$ are obtained for a fixed value of $\bbhat{\mu}$, the resulting asymptotic normal distribution is also obtained conditional on the values of $\bbhat{\mu}$. However, the conditional asymptotic distribution is independent of $\bbhat{\mu}$, hence the unconditional distribution also turns out to be the same. For the technical details, one may refer to Appendix~\ref{appendix:normality-general-proof}.
\begin{theorem}\label{thm:consistency-general}
    Under the Assumptions~\ref{assum:0}-\ref{assum:3}, for any $\alpha \in [0, 1]$, the converged rPCAdpd estimator $\bbhat{\theta} = (\widehat{\gamma_1}, \dots \widehat{\gamma_p}, \bbhat{\eta})$ as in Eq.~\eqref{eqn:dpd-general-model-4} satisfy the following as the sample size $n \rightarrow \infty$, provided that the location estimator $\bbhat{\mu}$ is consistent for $\mu^\ast$:
    \begin{enumerate}
        \item The estimated eigenvalue $\widehat{\gamma_j}$ is $\sqrt{n}$-consistent for $\gamma_j^\ast$ for $j = 1, 2, \dots p$.
        \item Similarly, the corresponding estimated eigenvector $\bbhat{v}_j$ is also $\sqrt{n}$-consistent for the true eigenvector $\bb{v}_j^\ast$ for $j = 1, 2, \dots p$.
    \end{enumerate}
\end{theorem}
\begin{remark}
    The consistency of $\bbhat{v}_j$ for $\bb{v}_j^\ast$ follows from the fact that $\bbhat{\eta}$ is consistent for $\bbeta^\ast$ and the parameter $\bbeta$ is simply a parametrization of the Stiefel manifold, hence each of $\bb{v}_1, \dots \bb{v}_p$ is a continuous and smooth function of $\bbeta$.
\end{remark}
\begin{theorem}\label{thm:normality-general}
    Under the Assumptions~\ref{assum:0}-\ref{assum:3}, the converged rPCAdpd estimator $\bbhat{\theta} = (\bbhat{\mu}, \widehat{\gamma_1}, \dots \widehat{\gamma_p}, \bbeta)$ as defined in Eq.~\eqref{eqn:dpd-general-model-4} for the general elliptically symmetric family has an asymptotic normal distribution as $n \rightarrow \infty$ after proper centering and scaling, provided that the location estimator $\bbhat{\mu}$ is consistent for $\mu^\ast$. In particular, 
    \begin{equation*}
        \sqrt{n} \bb{J}_{\bbtheta^\ast} \bb{K}_{\bbtheta^\ast}^{-1/2} \left( \bbhat{\theta} - \bbtheta^\ast \right)
    \end{equation*}
    \noindent converges in distribution to a standard normal random variable as $n \rightarrow \infty$. Here,
    \begin{align*}
        \bb{J}_{\bbtheta^\ast}
        & = \dfrac{c_{(1+\alpha)g}}{(1+\alpha)^2 c_g^{(1+\alpha)}}
        \begin{bmatrix}
            \bb{J}_{11} & \bb{J}_{12}\\
            \bb{J}_{12}\tr & \bb{J}_{22}
        \end{bmatrix},\\
        \bb{J}_{11} 
        & = \dfrac{(\alpha^2 - 1)}{4}\left(\diag{\bb{\Gamma}^{-1}}\right)\left(\diag{\bb{\Gamma}^{-1}}\right)\tr + \bb{\Gamma}^{-2}\bb{V}\tr \bb{A}_4((1+\alpha)g) \bb{V}\bb{\Gamma}^{-2},\\
        \bb{J}_{12}
        & = -2\bb{\Gamma}^{-2}\bb{V}\tr \left( \bb{I}_p \otimes \bb{\Gamma}^{-1} \right) A_4((1+\alpha)g)\bb{G}\tr,  \\
        \bb{J}_{22} 
        & =  4 \bb{G} \left( \bb{I}_p \otimes \bb{\Gamma}^{-1} \right) A_4((1+\alpha)g) \left( \bb{I}_p \otimes \bb{\Gamma}^{-1} \right)\tr \bb{G}\tr,
    \end{align*}
    \noindent and
    \begin{align*}
        \bb{K}_{\bbtheta^\ast}
        & = \dfrac{c_{(1+2\alpha)g}}{(1+2\alpha)^2 c_g^{(1+2\alpha)}}
        \begin{bmatrix}
            \bb{K}_{11} & \bb{K}_{12}\\
            \bb{K}_{12}\tr & \bb{K}_{22}\\
        \end{bmatrix} 
        - \dfrac{c^2_{(1+\alpha)g}}{(1+\alpha)^2 c_g^{(2+2\alpha)}}\begin{bmatrix}
            \frac{\alpha^2}{4} \diag{\bb{\Gamma}^{-1}}\diag{\bb{\Gamma}^{-1}}\tr  & 0\\
            0 & 0\\
        \end{bmatrix},\\
        \bb{K}_{11} 
        & = \frac{(4\alpha^2 - 1)}{4}\left(\diag{\bb{\Gamma}^{-1}}\right)\left(\diag{\bb{\Gamma}^{-1}}\right)\tr + \bb{\Gamma}^{-2}\bb{V}\tr \bb{A}_4((1+2\alpha)g) \bb{V}\bb{\Gamma}^{-2}, \\
        \bb{K}_{12}
        & = -2\bb{\Gamma}^{-2}\bb{V}\tr \left( \bb{I}_p \otimes \bb{\Gamma}^{-1} \right) A_4((1+2\alpha)g)\bb{G}\tr,  \\
        \bb{K}_{22} 
        & =  4 \bb{G} \left( \bb{I}_p \otimes \bb{\Gamma}^{-1} \right) A_4((1+2\alpha)g)  \left( \bb{I}_p \otimes \bb{\Gamma}^{-1} \right)\tr \bb{G}\tr.
    \end{align*}
\end{theorem}
One may also be interested in the special case when the underlying elliptically symmetric distribution is assumed to be Gaussian. Formally, if we consider that the sample observations $\bb{X}_1, \dots \bb{X}_n$ are distributed according to a $p$-variate normal distribution with unknown mean $\bb{\mu}^\ast$ and unknown dispersion matrix $\Sigma^\ast = \sum_{k=1}^p \gamma_k^\ast \bb{v}_k^\ast (\bb{v}_k^\ast)\tr$, then it follows that under the same set of assumptions, one can establish the following corollary.

\begin{corollary}\label{thm:normality-normal}
    Under the Assumptions~\ref{assum:0}-\ref{assum:3} and that the location estimator $\bbhat{\mu}$ is consistent for $\mu^\ast$, the converged rPCAdpd estimator $\bbhat{\theta} = (\widehat{\gamma_1}, \dots \widehat{\gamma_p}, \bbeta)$ as in Eq.~\eqref{eqn:dpd-normal-model-4} for the Gaussian model family of distributions, satisfy the following as the sample size $n \rightarrow \infty$:, 
    \begin{enumerate}
    \item The eigenvalues $\widehat{\gamma_j}$ is consistent for $\gamma_j^\ast$ and $\bbhat{v}_j$ is consistent for $\bb{v}_j^\ast$ for $j = 1, 2, \dots p$.
    \item The scaled and centred estimated principal component eigenvalues
    \begin{equation*}
        \sqrt{n}\left( 
        \begin{bmatrix}
        \widehat{\gamma_1}\\
        \dots \\
        \widehat{\gamma_p}
        \end{bmatrix} 
        -
        \begin{bmatrix}
        \gamma_1^\ast\\
        \dots \\
        \gamma_p^\ast
        \end{bmatrix}  
        \right)
    \end{equation*} 
    \noindent has an asymptotic $p$-variate normal distribution with mean $\bb{0}$ and dispersion matrix 
    \begin{equation*}
        \dfrac{(1+\alpha)^{p+4}}{(1+2\alpha)^{p/2}} \bb{M}^{-1} \left( A_1(\alpha) \diag{\bb{\Gamma}^{-1}} \diag{\bb{\Gamma}^{-1}}\tr + \frac{1}{2(1+2\alpha)^2}\bb{\Gamma}^{-2} \right) \bb{M}^{-1},
    \end{equation*}
    \noindent where 
    \begin{align*}
        \bb{M}
        = \left( \frac{\alpha^2}{4} \diag{\bb{\Gamma}^{-1}}\diag{\bb{\Gamma}^{-1}}\tr + \dfrac{1}{2}\bb{\Gamma}^{-2} \right), \
        A_1(\alpha) 
        = \alpha^2\left[ \dfrac{1}{(1+2\alpha)^2} - \dfrac{(1+2\alpha)^{p/2}}{4(1+\alpha)^{p+2}}\right].
    \end{align*}
    \item The scaled and centered estimated $\bbhat{\eta}$ corresponding to the principal component eigenvectors, i.e., $\sqrt{n}(\bbhat{\eta} - \bbeta^\ast)$ has an asymptotic normal distribution with mean $0$ and dispersion matrix 
    \begin{equation*}
        \dfrac{(1+\alpha)^{p+4}}{(1+2\alpha)^{2+p/2}} \left( \sum_{k=1}^p \sum_{l=1}^p \left( 1 - \dfrac{\gamma_k^\ast}{\gamma_l^\ast} \right)\bb{G}_k(\bb{v}_l^\ast) (\bb{v}_k^\ast) \tr \bb{G}_l\tr \right)^{-1},
    \end{equation*}
    \noindent where $\bb{G}_k = \dfrac{\partial \bb{v}_k}{\partial \bbeta}\vert_{\bbeta = \bbeta^\ast}$, the matrix corresponding of the gradients of the eigenvector $\bb{v}_k$ with respect to its natural parametrization $\bbeta$.
    \item The rPCAdpd estimate of the eigenvalues $(\widehat{\gamma}_1, \dots, \widehat{\gamma}_p)$ and estimate of the eigenvectors $(\bbhat{v}_1, \dots \bbhat{v}_p)$ are asymptotically independent.
    \end{enumerate}
\end{corollary}

\begin{remark}
    The independence of the rPCAdpd estimate of eigenvalues and eigenvectors can enable one to create confidence intervals for the eigenvalues and eigenvectors separately. To create the asymptotic confidence interval for the eigenvalues, the knowledge of the corresponding estimates of eigenvalues is sufficient. In contrast, the asymptotic confidence band for eigenvectors require both the eigenvalues and the eigenvectors. 
\end{remark}

\begin{remark}\label{remark:alpha-0-result}
    The density power divergence introduced in~\cite{basu1998robust} becomes the same as the Kullback-Leibler divergence between the true density and the model density $f_{\bbtheta}(\cdot)$ as $\alpha \rightarrow 0$. Thus, for $\alpha \rightarrow 0$, the estimating equations for the MDPDE turn out to be equivalent to the estimating equations corresponding to the log-likelihood. Consequently, the MDPDE coincides with the maximum likelihood estimator as $\alpha \rightarrow 0$. From Corollary~\ref{thm:normality-normal} it then follows that the maximum likelihood estimates (MLE) of the eigenvalues of the covariance matrix under the Gaussian distribution are asymptotically normal with mean $\gamma_j$ and covariance $2\gamma_j^2/n$ and are asymptotically independent. This result has been well established in the literature; see~\cite{girshick1939sampling} for references. A similar result for the asymptotic distribution of the MLE of eigenvectors was derived by~\cite{anderson1963asymptotic}. Results on the asymptotic independence between the MLE of the eigenvalues and eigenvectors were also derived by~\cite{tyler1981asymptotic} for a general setup with repeated eigenvalues. The Corollary~\ref{thm:normality-normal} can be seen as a generalization of these results.
\end{remark}

\begin{remark}
    In contrast to Remark~\ref{remark:alpha-0-result}, for $\alpha = 1$, the form of density power divergence becomes same as the $L_2$ distance between the true density and the model density $f_{\bbtheta}(\bb{x})$. If we denote the minimum $L_2$ distance estimator of the eigenvalues by $\bbtilde{\gamma} = \left(\widetilde{\gamma_1}, \dots \widetilde{\gamma_p}\right)\tr$ and the true eigenvalues by $\bb{\gamma}^\ast = \left( \gamma_1^\ast, \dots \gamma_p^\ast\right)\tr$, then
    \begin{equation*}
        \sqrt{n}(\bbtilde{\gamma} - \bb{\gamma}^\ast) \xrightarrow{d} \normdist_p \left( \bb{0}, \mathcal{V}_2 \right),
    \end{equation*}
    \noindent as $n \rightarrow \infty$. Here, $\xrightarrow{d}$ denotes the convergence in law. The asymptotic variance is given by
    \begin{equation*}
        \mathcal{V}_2 = \dfrac{2^{(p+8)}}{3^{(p/2)}} \bb{M}_1^{-1} \left( \left( \dfrac{1}{9} - \dfrac{3^{(p/2)}}{2^{(p+4)}}\right) \diag{\bb{\Gamma}^{-1}}\diag{\bb{\Gamma}^{-1}}\tr + \dfrac{1}{18}\bb{\Gamma}^{-1} \right) \bb{M}_1^{-1},
    \end{equation*}
    \noindent where $\bb{M}_1 = \left( \diag{\bb{\Gamma}^{-1}}\diag{\bb{\Gamma}^{-1}}\tr + 2\bb{\Gamma}^{-2} \right)$. Since the quantity $\frac{2^{\left(x+4\right)}}{3^{\left(x/2\right)}}\left(\frac{1}{9}-\frac{3^{\left(x/2\right)}}{2^{\left(x+4\right)}}\right)$ increases exponentially fast as $x$ increases, the variance of the minimum $L_2$-distance estimator increases exponentially with increase in the dimension $p$. This shows that by using the highly robust minimum $L_2$ distance estimator to obtain the principal components, one sacrifices considerable efficiency in estimation.
\end{remark}

\subsection{Influence Function Analysis}

The influence function is a local measure of the sensitivity and robustness of an estimator~\citep{hampel2011robust}. In this section, we investigate the influence function of the rPCAdpd estimator for the Gaussian model family of distributions. For this particular choice, the asymptotic independence of the eigenvalues and the eigenvectors as shown in Theorem~\ref{thm:normality-normal} helps in deriving the influence functions quite nicely. Let us assume that instead of the true distribution $\Phi_{\bbtheta^\ast}(\bb{x})$, the observations $\bb{X}_i$s come from a contaminated distribution $G_{\epsilon}(\bb{x}) = (1-\epsilon) \Phi_{\bbtheta^\ast}(\bb{x}) + \epsilon \delta_{\bb{y}}(\bb{x})$, where $\delta_{\bb{y}}(\cdot)$ is the degenerate distribution at $\bb{y} \in \R^p$. Let $\phi_{\bbtheta^\ast}(\bb{x})$ be the density function corresponding to the Gaussian distribution function $\Phi_{\bbtheta^\ast}(\bb{x})$. Then the influence of this contamination on the estimated principal components can be readily obtained from the influence function derived in~\cite{basu1998robust}. Due to the asymptotic independence, the influence functions for the estimators of the eigenvalues and the eigenvectors can be separately obtained along with an application of the chain rule to incorporate the influence of the robust location estimator. It turns out that
\begin{multline*}
    I_{\alpha}(\Phi_{\bbtheta^\ast}, \bb{\gamma}; \bb{y}) 
    = \dfrac{4(1+\alpha)^2}{C_\alpha} \left[ \alpha^2 \diag{\bb{\Gamma}^{-1}} \diag{\bb{\Gamma}^{-1}}\tr + 2\bb{\Gamma}^{-2} \right]^{-1} \\
    \left[ 
    \begin{bmatrix}
        u_{\gamma_1^\ast}(\bb{y})\\
        \vdots\\ 
        u_{\gamma_p^\ast}(\bb{y})
    \end{bmatrix}    
    \phi_{\bbtheta^\ast}^{\alpha}(\bb{y})I(\Phi_{\bbtheta^\ast}, \bbhat{\mu}; \bb{y}) -
    \begin{bmatrix}
        \xi_{\gamma_1^\ast}\\
        \vdots \\
        \xi_{\gamma_p^\ast}
    \end{bmatrix}
    \right],
\end{multline*}
\begin{equation*}
    I_{\alpha}(\Phi_{\bbtheta^\ast}, \bb{\eta}; \bb{y}) = -\dfrac{(1+\alpha)^2}{C_\alpha} \left[\sum_{k=1}^p \dfrac{G_k\bb{\Sigma}^\ast G_k\tr}{\gamma_k^\ast} \right]^{-1} \sum_{k=1}^p \dfrac{G_k}{\gamma_k^\ast} (\bb{y} - \bb{\mu}^\ast)(\bb{y} - \bb{\mu}^\ast)\tr \bb{v}_k^\ast  \phi_{\bbtheta}^{\alpha}(\bb{y}) I(\Phi_{\bbtheta^\ast}, \bbhat{\mu}; \bb{y}).
\end{equation*}
\noindent Here, $u_{\gamma_j^\ast}(\bb{y})$ denotes the score function with respect to the $j$-th eigenvalue $\gamma_j^\ast$ evaluated at the contaminating point $\bb{y}$ and $I(\Phi_{\bbtheta^\ast}, \bbhat{\mu}; \bb{y})$ is the influence function of the location estimator $\bbhat{\mu}$ at $\bb{y}$. We assume that the location estimator $\bbhat{\mu}$ is robust and hence has a bounded influence function, which is true for the $L_1$-median. To show that both the above influence functions are bounded, one may note that the exponential quantity $e^{ -\alpha (\bb{y} - \bb{\mu}^\ast)\tr(\bb{\Sigma}^\ast)^{-1} (\bb{y} - \bb{\mu}^\ast)/2}$ present in the Gaussian density $\phi_{\bbtheta^\ast}(\bb{y})$ is bounded below by $e^{-\alpha \Vert \bb{y} - \bb{\mu}^\ast\Vert^2/2\gamma_{(p)}^\ast}$ and bounded above by $ e^{-\alpha \Vert \bb{y} - \bb{\mu}^\ast\Vert^2/2\gamma_{(1)}^\ast}$, where $\gamma_{(1)}^\ast$ and $\gamma_{(p)}^\ast$ are the largest and the smallest eigenvalues of $\bb{\Sigma}^\ast$ respectively. Now the boundedness of the influence function follows from assumption~\ref{assum:3}, which can be easily verified for $g(x) = -x/2$ corresponding to the Gaussian distribution. Thus, if the location estimator $\bbhat{\mu}$ is B-robust, the rPCAdpd estimator is also B-robust qualifying for one of the primary requirements for a robust estimator.

\subsection{Breakdown Point Analysis}\label{sec:breakdown}

The breakdown point of an estimator is another accepted measure of the robustness of an estimator besides the influence function which measures the highest level of contamination that an estimator can tolerate~\citep{hampel1971general}. Given the true distribution $H$, \cite{ghosh2013robust} consider the asymptotic breakdown point of an MDPDE functional $T$ as the largest value of $\epsilon$ such that there exists a sequence of distributions $\{ K_m \}$ with $\vert T(H_{\epsilon, m}) - T(H)\vert \rightarrow \infty$ as $m \rightarrow \infty$ where 
\begin{equation}
    H_{\epsilon, m} = (1 - \epsilon) H + \epsilon K_m.
    \label{eqn:contam-model}
\end{equation}
\noindent However, such a definition makes sense only for the location estimators. For general estimators, \cite{maronna2019robust} define the breakdown of a functional $T$ for $\epsilon$-level contamination if $T(H_{\epsilon, m}) \rightarrow \theta_{\infty}$ as $m \rightarrow \infty$ where $\theta_{\infty} \in \partial\Theta$, the boundary of parameter space $\Theta$. In the case of the rPCAdpd estimator of eigenvalues and corresponding eigenvectors, the boundary of the parameter space $\bb{\Theta} = (\R^+)^p \times S$ is 
\begin{equation*}
    \partial\bb{\Theta} = \left\{ (\gamma_1, \dots \gamma_p, \bbeta): \bbeta \in S, \text{ and there exists } k \in \{1, \dots p\} \text{ with } \gamma_k \in\{0, \infty\} \right\},
\end{equation*}
\noindent indicating that the breakdown can happen when any of the estimated eigenvalues either explodes to infinity or implodes to $0$.

Since the rPCAdpd algorithm is composed of two steps: location estimation and eigenvalue and eigenvector estimation using the rSVDdpd procedure, the asymptotic breakdown of the entire procedure is the minimum of the asymptotic breakdown of these individual procedures. It is well known that the robust $L_1$-median (used as the location estimator in our entire study) has an asymptotic breakdown point of $1/2$. Also, under fairly general conditions, \cite{roy2023breakdown} showed that the robust MDPDE has a breakdown point at least $\alpha/(1+\alpha)$, where $\alpha$ is the robustness parameter with $\alpha \in [0, 1]$. Hence, the resulting rPCAdpd estimator has an asymptotic breakdown at least $\alpha/(1+\alpha)$, which is also free of the dimension $p$, demonstrating the scalability aspect of the proposed estimator. 

Let, the distributions $H_{\epsilon, m}, H$ and $K_m$ mentioned in the contamination model~\eqref{eqn:contam-model} have densities $h_{\epsilon,m}, h$ and $k_m$ respectively. In~\cite{roy2023breakdown}, the authors derive a lower bound of the breakdown point of the MDPDE in general under the following set of assumptions.
\begin{enumerate}[label = (BP\arabic*), ref = (BP\arabic*)]
    \item\label{assum:BP-0} $\int \min\{ f_{\bbtheta}(x), k_m(x) \} dx \rightarrow 0$ uniformly as $m \rightarrow \infty$ and $\bbtheta$ is bounded away from the boundary $\partial\bb{\Theta}$.
    \item\label{assum:BP-1} $\int \min\{ h(x), f_{\bbtheta_m}(x) \} dx \rightarrow 0$ as $m \rightarrow \infty$ if $\bbtheta_m \rightarrow \bbtheta_\infty$ where $\bbtheta_\infty$ is some point on the boundary $\partial\bb{\Theta}$.
    \item\label{assum:BP-3} $M_{f_{\theta_m}} \geq M_{k_m}$ for all $m \geq M$ for sufficiently large $M$ for any $\bbtheta_m \rightarrow \bbtheta_\infty$ where $\bbtheta_\infty$ is some point on the boundary $\partial\bb{\Theta}$ and $M_f = \int f^{1+\alpha}(x)dx$.
\end{enumerate}
\noindent Assumptions~\ref{assum:BP-0} and~\ref{assum:BP-1} are quite standard assumptions for breakdown analysis. To verify assumption~\ref{assum:BP-3} for our setup, we note that $M_{f_{\bbtheta_m}} = \frac{c_{(1+\alpha)g}}{c_g} \prod_{k=1}^p \gamma_{k,m}^{-\alpha/2}$ where $\{ \gamma_{k,m}\}$ is the sequence of eigenvalues in $\bbtheta_m$. Clearly, when $\{\bbtheta_m \}$ tends to a point on the boundary of the parameter space, for some $k = 1, \dots p$, either $\gamma_{k, m} \rightarrow 0$ or $\gamma_{k, m} \rightarrow \infty$ as $m \rightarrow \infty$. Since $\alpha > 0$, either $M_{f_{\bbtheta_m}} \rightarrow \infty$ or $M_{f_{\bbtheta_m}} \rightarrow 0$ as $m \rightarrow \infty$. When $M_{f_{\bbtheta_m}}$ increases to $\infty$, Assumption (BP3) holds trivially. When $M_{f_{\bbtheta_m}}$ decreases to $0$, Assumption~\ref{assum:BP-3} holds if $M_{k_m}$ decreases to $0$ at a faster rate than $M_{f_{\bbtheta_m}}$. To ensure this, one such particular choice would be to restrict the contaminating distribution to any elliptically symmetric family of distributions with a singular dispersion matrix, implying that the high dimensional data have outlying values not all of the $p$-coordinates. Such outliers are more common when $p$ is large; data where outlyingness occurs in all of the $p$-coordinates rarely show up for almost all practical purposes. Thus, we have the following corollary.
\begin{corollary}\label{cor:rpca-breakdown}
    Under the assumptions~\ref{assum:BP-0}-\ref{assum:BP-3}, if the true density belongs to the model family of elliptically symmetric distributions, then the rPCAdpd estimator has a breakdown point at least as large as $\alpha/(1+\alpha)$ for $\alpha \in [0, 1]$, provided that the robust location estimator used also has an asymptotic breakdown point larger than $\alpha/(1+\alpha)$.
\end{corollary}
\begin{remark}
    Corollary~\ref{cor:rpca-breakdown} shows that by tuning the parameter $\alpha$, one can change the breakdown point of the rPCAdpd estimator irrespective of the dimension $p$ of the data. Also, note that as $\alpha \rightarrow 0$, the lower bound of the breakdown becomes $0$ suggesting a lack of robustness, while for $\alpha = 1$, one would get the highest possible breakdown $1/2$.
\end{remark}

Note that, Corollary~\ref{cor:rpca-breakdown} is in contrast to the breakdown point result obtained by~\cite{maronna1976robust} for an affine equivariant M-estimator, which states that an affine equivariant M-estimator has a breakdown point at most $1/(p+1)$ where $p$ is the dimensionality of the data. As explained in~\cite{basu1998robust}, the MDPDE is a special case of the M-estimator, and also we showed the orthogonal equivariance property of the rPCAdpd estimator in Section~\ref{sec:orthogonal-equivariance}. This discrepancy holds because the classes of the M-estimator differs from the classes of minimum divergence estimators in which MDPDE belongs. In particular, \cite{maronna1976robust} considered the estimators given as the solution to the system of equations
\begin{equation*}
    \begin{split}
        \sum_{i=1}^n u_1\left( (\bb{X}_i - \bb{\mu})\tr \bb{\Sigma}^{-1}(\bb{X}_i - \bb{\mu}) \right) (\bb{X}_i - \bb{\mu}) & = 0,\\
        \sum_{i=1}^n u_2\left( (\bb{X}_i - \bb{\mu})\tr \bb{\Sigma}^{-1}(\bb{X}_i - \bb{\mu}) \right) (\bb{X}_i - \bb{\mu})(\bb{X}_i - \bb{\mu})\tr & = \bb{\Sigma},
    \end{split}
\end{equation*}
\noindent where $u_1(s)$ and $u_2(s)$ are suitable nonincreasing functions for $s \geq 0$. On the other hand, denoting $\bb{\Sigma} = \sum_{k=1}^p \gamma_k \bb{v}_k \bb{v}_k\tr$, the estimating equations for MDPDE turn out to be
\begin{equation*}
    \begin{split}
        \sum_{i=1}^n \exp\left( -0.5\alpha (\bb{X}_i - \bb{\mu})\tr \bb{\Sigma}^{-1}(\bb{X}_i - \bb{\mu}) \right) (\bb{X}_i - \bb{\mu}) & = 0,\\
        \sum_{i=1}^n \exp\left( -0.5\alpha (\bb{X}_i - \bb{\mu})\tr \bb{\Sigma}^{-1}(\bb{X}_i - \bb{\mu}) \right) \left( (\bb{X}_i - \bb{\mu})(\bb{X}_i - \bb{\mu})\tr -\bb{\Sigma} \right) & = 0,
    \end{split}
\end{equation*}
\noindent under the Gaussian model as in Eq.~\eqref{eqn:dpd-normal-model-4}. Therefore, the breakdown point results provided by~\cite{maronna1976robust} do not apply to our proposed rPCAdpd estimator. This independence of the dimension $p$ in the lower bound of the breakdown implies that in contrast to the classical M-estimator~\citep{maronna1976robust}, the rPCAdpd estimator can still remain useful for estimating principal components robustly in arbitrarily high dimensional data.

\section{Simulation Studies}\label{sec:simulation}

In this section, we perform a principal component analysis for data matrices with varying levels of contamination using the existing robust PCA algorithms and our proposed rPCAdpd algorithm. Among the plethora of existing RPCA methods, we take the classical PCA~\citep{jolliffe2002principal}, spherical and elliptical PCA (LOC)~\citep{locantore1999robust}, ROBPCA algorithm by~\cite{hubert2005robpca}, projection pursuit based methods Proj and Grid~\citep{croux2005high}, robust PCA using robust covariance matrix estimation (RobCov)~\citep{todorov2010object}, principal component pursuit (PCP) algorithm by~\cite{candes2011robust} and Gmedian based robust principal component analysis (Gmed) by~\cite{cardot2017fast}, for comparison purposes. We have performed the simulations with several variants of the rPCAdpd algorithm differing only in the location estimator used. Based on empirical performance, we have seen that $L_1$-median as a location estimator provides a desirable balance between robustness, efficiency and computational complexity, hence it is the only variant demonstrated in the results described in this section.

\subsection{Simulation Settings}

In the simulation experiments, we consider a data matrix comprised of i.i.d. rows. The rows $\bb{X}_i$ are generated as $\bb{X}_i = (1 -\delta_i)\bbtilde{X}_i + \delta_i \bb{\epsilon}_i$ for $i = 1, 2, \dots n$. The uncontaminated sample $\bbtilde{X}_i$ is normally distributed with zero mean vector and a dispersion matrix $\bb{\Sigma}$ whose elements are given by $\bb{\Sigma}_{ij} = \min(i,j)/p$ for $i, j = 1, 2, \dots p$. This setup is similar to the one described in~\cite{cardot2017fast} and can be regarded as a discretized version of a Brownian motion within the unit $(0, 1)$ interval. The random variables $\delta_i$ which control the level of contamination are i.i.d. Bernoulli random variable with success probability $\delta$. The contaminating variable $\bb{\epsilon}_i$s are chosen to possess different features compared to $\bbtilde{X}_i$, and in this regard, we feel that the choice of the distribution of outliers as given in~\cite{cardot2017fast} is too restrictive. In comparison, \cite{hubert2005robpca} consider outliers that have changes in both mean and variance components separately, and hence we choose to work with them. In summary, we consider the following simulation scenarios.
\begin{enumerate}
    \item[(S1)] $\delta = 0$, i.e., only pure data is present and there is no contamination. 
    \item[(S2)] Here a proportion of elements are contaminated. The contaminating variable $\bb{\epsilon}_i$s are i.i.d. $p$-variate normal random variables with mean $\mu(f_1)$ and variance $\bb{\Sigma}/f_2$. The mean vector $\mu(f_1)$ is a $p$-length vector where $10\%$ of the entries are equal to $f_1$ while the rest of the entries are equal to $0$.
    \begin{enumerate}
        \item[(S2a)] Here, $f_1 = 3, f_2 = 1$ and $\delta = 0.1$. Therefore, on average $10\%$ of the data will be contaminated.
        \item[(S2b)] Here, $f_1 = 3, f_2 = 1$ and $\delta = 0.2$. Therefore, on average $20\%$ of the data will be contaminated. 
        \item[(S2c)] Similar to (S2a) but with $f_2 = 5$.
        \item[(S2d)] Similar to (S2b) but with $f_2 = 5$.
    \end{enumerate}
    \item[(S3)] This is similar to simulation scenario (S2) but the contaminating variable $\bb{\epsilon}_i$s are i.i.d. $p$-variate $t$-distribution with $5$ degrees of freedom with dispersion matrix $\bb{\Sigma}/f_2$ and a non-centrality parameter $\mu(f_1)$. This is used to understand the behaviour of the PCA algorithms for heavy-tailed contaminating variables.
    \begin{enumerate}
        \item[(S3a)] $f_1 = 3, f_2 = 1$ and $\delta = 0.1$.
        \item[(S3b)] $f_1 = 3, f_2 = 1$ and $\delta = 0.2$.
        \item[(S3c)] $f_1 = 3, f_2 = 5$ and $\delta = 0.1$.
        \item[(S3d)] $f_1 = 3, f_2 = 5$ and $\delta = 0.2$. 
    \end{enumerate}    
\end{enumerate}
In each of the above simulation scenarios, we consider five different situations with the number of samples $n = 50$ but with different dimensions ranging from very small to moderately large ($p = 10, 25, 50, 100, 250$). Based on $1000$ repetitions of each exercise, we obtained an estimate of bias, mean absolute error (MAE) of the estimated eigenvalues as 
\begin{equation*}
        \text{Bias}_k = \dfrac{1}{B} \sum_{b=1}^{B} \widehat{\gamma}_k^{(b)} - \gamma_k, \ 
        \text{MAE}_k = \dfrac{1}{B}\sum_{b=1}^{B} \left\vert \widehat{\gamma}_k^{(b)} - \gamma_k \right\vert,\\
\end{equation*}
\noindent where $\widehat{\gamma}_k^{(b)}, \gamma_k$ respectively denote the estimate and the true $k$-th eigenvalue for the $b$-th sample. Similarly, to measure discrepancy in the estimated eigenvalues we look at the Subspace Recovery Error (SRE) given by
\begin{equation*}
    \text{SRE} = \dfrac{1}{B}\sum_{b=1}^{B} 2\left(r - \trace{\bbhat{P}_b \bb{P}} \right),
\end{equation*}
\noindent where $\bbhat{P}_b = \sum_{k = 1}^{r} \bbhat{v}_k^{(b)} (\bbhat{v}_k^{(b)})\tr$ is the projection matrix onto the span of the estimated eigenvectors corresponding to the largest $r$ eigenvalues from $b$-th sample, and $\bb{P} = \sum_{k=1}^{r} \bb{v}_k\bb{v}_k\tr$ be the corresponding projection matrix from the true eigenvectors. In each of these simulation scenarios, we keep the choice of $r = 5$ fixed, as more than $90\%$ of the variability can be explained by the first $5$ principal components.

\subsection{Simulation Results}

The simulation results from the aforementioned algorithms are demonstrated in Tables~\ref{tbl:sim-S1}-\ref{tbl:sim-S3d}. We denote the rPCAdpd estimator with $L_1$-median as the location estimator in these tables as the DPD method, with the robustness parameter shown in parenthesis. Also, the RobCov algorithm~\citep{todorov2010object} uses MCD-based robust covariance estimation for RPCA. Thus, it is inapplicable when variables outnumber samples ($n \leq p$), and those entries are marked as NA in these tables. 

\begin{table}[t]
    \resizebox*{\textwidth}{!}{
    \begin{tabular}{ccrrrrrrrrrrrr}
        \toprule
        Metric & $p$ & Classical & LOC   & ROBPCA & Proj  & RobCov & Grid  & Gmed  & PCP   & \begin{tabular}[c]{@{}r@{}}DPD\\ (0.25)\end{tabular} & \begin{tabular}[c]{@{}r@{}}DPD\\ (0.5)\end{tabular} & \begin{tabular}[c]{@{}r@{}}DPD\\ (0.75)\end{tabular} & \begin{tabular}[c]{@{}r@{}}DPD\\ (1)\end{tabular} \\ \midrule
        \multirow{5}{*}{Bias} & 10  & 0.059 & 0.723 & 0.194 & 0.229 & 0.431 & 0.416 & 0.043 & 1.066 & 0.06 & 0.062 & 0.065 & 0.068\\ 
        & 25  & 0.019 & 2.175 & 0.227 & 0.362 & 0.336 & 0.807 & 0.079 & 2.45 & 0.017 & 0.013 & 0.01 & 0.008\\ 
        & 50  & 0.031 & 4.572 & 0.519 & 0.467 & NA & 1.414 & 0.177 & 4.729 & 0.026 & 0.017 & 0.007 & 0.017\\ 
        & 100  & 0.194 & 9.366 & 0.944 & 1.058 & NA & 2.827 & 0.314 & 9.387 & 0.201 & 0.216 & 0.233 & 0.254\\ 
        & 250  & 0.154 & 23.76 & 2.847 & 2.239 & NA & 6.906 & 0.644 & 23.301 & 0.184 & 0.236 & 0.295 & 0.359\\ 
        \midrule
        \multirow{5}{*}{MAE} & 10  & 17.919 & 72.334 & 26.756 & 33.798 & 45.663 & 50.391 & 19.122 & 106.477 & 17.921 & 17.936 & 17.981 & 18.054\\ 
        & 25  & 38.106 & 217.462 & 50.069 & 75.426 & 49.757 & 123.166 & 40.445 & 244.951 & 38.25 & 38.485 & 38.814 & 39.252\\ 
        & 50  & 73.085 & 457.212 & 110.189 & 137.425 & NA & 240.293 & 84.614 & 472.875 & 73.126 & 73.225 & 73.489 & 73.803\\ 
        & 100  & 143.086 & 936.571 & 200.658 & 263.141 & NA & 381.432 & 154.736 & 938.685 & 143.426 & 144.012 & 144.595 & 145.234\\ 
        & 250  & 395.183 & 2375.96 & 536.355 & 731.985 & NA & 1010.852 & 434.823 & 2330.092 & 395.893 & 396.863 & 397.827 & 396.641\\ 
        \midrule
        \multirow{5}{*}{SRE} & 10  & 1 & 1.347 & 1.172 & 1.677 & 1.429 & 2.498 & 1.126 & 1.188 & 0.99 & 0.983 & 0.99 & 0.995\\ 
        & 25  & 0.829 & 1.417 & 1.028 & 1.76 & 1.127 & 3.573 & 1.004 & 1.136 & 0.833 & 0.84 & 0.845 & 0.872\\ 
        & 50  & 0.766 & 1.336 & 0.959 & 1.653 & NA & 4.038 & 0.931 & 0.871 & 0.766 & 0.771 & 0.795 & 0.818\\ 
        & 100  & 0.836 & 1.459 & 0.985 & 1.587 & NA & 3.313 & 1.045 & 0.923 & 0.84 & 0.847 & 0.857 & 0.872\\ 
        & 250  & 0.828 & 1.268 & 0.927 & 1.494 & NA & 3.265 & 0.939 & 0.899 & 0.828 & 0.832 & 0.84 & 0.859\\ 
        \bottomrule
    \end{tabular}}
    \caption{Estimated Bias, Mean Absolute Error and Subspace Recovery Error (SRE) for different PCA algorithm for simulation scenario (S1)}
    \label{tbl:sim-S1}
\end{table}

Table~\ref{tbl:sim-S1} presents metrics for various PCA algorithms in setup (S1). For uncontaminated data, classical PCA outperforms all robust methods across all metrics. Gmed and ROBPCA exhibit relatively less efficiency loss. However, the proposed rPCAdpd consistently outperforms both under any $\alpha \in [0, 1]$ and regardless of the location estimator used. Increasing $\alpha$ escalates efficiency loss moderately compared to other methods. Although the $L_1$-median is quite inefficient~\citep{huber2004robust,hampel2011robust}, its strong robustness properties allow rPCAdpd to achieve extremely low MAE.

\begin{table}[th]
    \resizebox*{\textwidth}{!}{
    \begin{tabular}{ccrrrrrrrrrrrr}
        \toprule
        Metric & $p$ & Classical & LOC   & ROBPCA & Proj  & RobCov & Grid  & Gmed  & PCP   & \begin{tabular}[c]{@{}r@{}}DPD\\ (0.25)\end{tabular} & \begin{tabular}[c]{@{}r@{}}DPD\\ (0.5)\end{tabular} & \begin{tabular}[c]{@{}r@{}}DPD\\ (0.75)\end{tabular} & \begin{tabular}[c]{@{}r@{}}DPD\\ (1)\end{tabular} \\ \midrule
        \multirow{5}{*}{ Bias } & 10  & 0.158 & 0.74 & 0.22 & 0.205 & 0.44 & 0.458 & 0.08 & 1.065 & 0.15 & 0.08 & 0.017 & 0.009\\ 
        & 25  & 0.304 & 2.183 & 0.386 & 0.416 & 0.459 & 1.046 & 0.152 & 2.452 & 0.281 & 0.097 & 0.023 & 0.025\\ 
        & 50  & 0.874 & 4.585 & 0.807 & 1.074 & NA & 2.308 & 0.274 & 4.724 & 0.792 & 0.248 & 0.121 & 0.129\\ 
        & 100  & 1.627 & 9.382 & 1.41 & 1.63 & NA & 4.246 & 0.433 & 9.386 & 1.445 & 0.251 & 0.111 & 0.132\\ 
        & 250  & 4.567 & 23.777 & 3.114 & 4.573 & NA & 11.98 & 1.471 & 23.303 & 4.134 & 0.866 & 0.718 & 0.777\\ 
        \midrule
        \multirow{5}{*}{ MAE } & 10  & 29.382 & 74.012 & 30.359 & 37.423 & 47.943 & 59.314 & 24.69 & 106.383 & 30.81 & 24.729 & 18.938 & 18.165\\ 
        & 25  & 62.13 & 218.321 & 62.944 & 83.99 & 61.246 & 141.267 & 56.95 & 245.214 & 63.996 & 47.073 & 39.385 & 39.137\\ 
        & 50  & 138.901 & 458.488 & 124.057 & 163.26 & NA & 305.897 & 111.013 & 472.415 & 145.313 & 95.448 & 82.258 & 82.154\\ 
        & 100  & 258.437 & 938.246 & 213.108 & 296.41 & NA & 495.19 & 211.008 & 938.639 & 268.129 & 155.77 & 140.017 & 139.704\\ 
        & 250  & 693.852 & 2377.669 & 558.396 & 729.947 & NA & 1311.666 & 545.383 & 2330.337 & 709.073 & 398.446 & 380.915 & 383.593\\ 
        \midrule
        \multirow{5}{*}{ SRE } & 10  & 1.779 & 1.875 & 1.056 & 2.016 & 1.405 & 2.697 & 1.843 & 1.171 & 1.787 & 1.46 & 1.063 & 1.005\\ 
        & 25  & 2.135 & 2.322 & 1.063 & 2.243 & 1.076 & 3.774 & 2.197 & 1.152 & 2.137 & 1.261 & 0.872 & 0.852\\ 
        & 50  & 2.185 & 2.43 & 0.998 & 2.2 & NA & 4.395 & 2.263 & 0.924 & 2.172 & 1.145 & 0.847 & 0.871\\ 
        & 100  & 2.251 & 2.482 & 1.084 & 2.3 & NA & 3.544 & 2.351 & 0.986 & 2.228 & 1.075 & 0.898 & 0.901\\ 
        & 250  & 2.231 & 2.504 & 0.991 & 2.229 & NA & 3.599 & 2.317 & 0.912 & 2.196 & 0.936 & 0.869 & 0.882\\ 
        \bottomrule
    \end{tabular}}
    \caption{Estimated Bias, Mean Absolute Error and Subspace Recovery Error (SRE) for different PCA algorithm for simulation scenario (S2a)}
    \label{tbl:sim-S2a}
\end{table}

\begin{table}
    \resizebox*{\textwidth}{!}{
    \begin{tabular}{ccrrrrrrrrrrrr}
        \toprule
        Metric & $p$ & Classical & LOC   & ROBPCA & Proj  & RobCov & Grid  & Gmed  & PCP   & \begin{tabular}[c]{@{}r@{}}DPD\\ (0.25)\end{tabular} & \begin{tabular}[c]{@{}r@{}}DPD\\ (0.5)\end{tabular} & \begin{tabular}[c]{@{}r@{}}DPD\\ (0.75)\end{tabular} & \begin{tabular}[c]{@{}r@{}}DPD\\ (1)\end{tabular} \\ \midrule
        \multirow{5}{*}{ Bias } & 10  & 0.321 & 0.757 & 0.381 & 0.429 & 0.589 & 0.757 & 0.14 & 1.065 & 0.329 & 0.281 & 0.138 & 0.067\\ 
        & 25  & 0.553 & 2.198 & 0.368 & 0.635 & 1.004 & 1.344 & 0.235 & 2.451 & 0.568 & 0.364 & 0.073 & 0.036\\ 
        & 50  & 1.467 & 4.602 & 0.829 & 1.796 & NA & 3.221 & 0.583 & 4.617 & 1.48 & 0.97 & 0.323 & 0.182\\ 
        & 100  & 2.66 & 9.414 & 1.028 & 2.692 & NA & 6.019 & 1.235 & 9.159 & 2.766 & 2.005 & 0.533 & 0.2\\ 
        & 250  & 7.033 & 23.805 & 3.245 & 8.006 & NA & 15.969 & 2.746 & 22.799 & 7.089 & 4.447 & 1.446 & 0.299\\ 
        \midrule
        \multirow{5}{*}{ MAE } & 10  & 41.99 & 75.693 & 43.08 & 52.712 & 60.646 & 82.448 & 30.185 & 106.511 & 45.196 & 42.803 & 29.261 & 22.409\\ 
        & 25  & 85.197 & 219.781 & 63.246 & 93.376 & 112.498 & 165.495 & 65.545 & 245.114 & 90.83 & 74.713 & 45.453 & 41.413\\ 
        & 50  & 194.589 & 460.223 & 130.581 & 236.929 & NA & 406.956 & 144.635 & 462.199 & 209.841 & 172.386 & 110.373 & 96.321\\ 
        & 100  & 364.678 & 941.397 & 221.517 & 400.614 & NA & 665.195 & 267.786 & 916.897 & 394.475 & 317.498 & 173.981 & 142.885\\ 
        & 250  & 957.207 & 2380.505 & 518.404 & 1066.532 & NA & 1696.499 & 658.65 & 2283.607 & 1060.277 & 838.361 & 545.85 & 432.927\\
        \midrule 
        \multirow{5}{*}{SRE} & 10  & 1.812 & 2.049 & 1.109 & 2.346 & 1.424 & 2.886 & 1.889 & 1.197 & 1.811 & 1.774 & 1.405 & 1.111\\ 
        & 25  & 2.14 & 2.422 & 1.021 & 2.645 & 2.212 & 4.19 & 2.26 & 1.276 & 2.152 & 1.832 & 1.111 & 1.03\\ 
        & 50  & 2.219 & 2.472 & 1.02 & 2.828 & NA & 4.985 & 2.314 & 2.265 & 2.24 & 1.819 & 1.201 & 1.049\\ 
        & 100  & 2.227 & 2.453 & 1.043 & 2.868 & NA & 3.86 & 2.326 & 2.272 & 2.242 & 1.868 & 1.153 & 1.007\\ 
        & 250  & 2.249 & 2.549 & 1.066 & 2.976 & NA & 3.901 & 2.362 & 2.302 & 2.262 & 1.767 & 1.16 & 1.007\\ 
        \bottomrule
    \end{tabular}}
    \caption{Estimated Bias, Mean Absolute Error and Subspace Recovery Error (SRE) for different PCA algorithm for simulation scenario (S2b)}
    \label{tbl:sim-S2b}
\end{table}

\begin{table}[th]
    \resizebox*{\textwidth}{!}{
    \begin{tabular}{ccrrrrrrrrrrrr}
        \toprule
        Metric & $p$ & Classical & LOC   & ROBPCA & Proj  & RobCov & Grid  & Gmed  & PCP   & \begin{tabular}[c]{@{}r@{}}DPD\\ (0.25)\end{tabular} & \begin{tabular}[c]{@{}r@{}}DPD\\ (0.5)\end{tabular} & \begin{tabular}[c]{@{}r@{}}DPD\\ (0.75)\end{tabular} & \begin{tabular}[c]{@{}r@{}}DPD\\ (1)\end{tabular} \\ \midrule
        \multirow{5}{*}{ Bias } & 10  & 0.215 & 0.746 & 0.166 & 0.159 & 0.396 & 0.434 & 0.161 & 1.065 & 0.234 & 0.131 & 0.112 & 0.11\\ 
        & 25  & 0.328 & 2.182 & 0.249 & 0.305 & 0.604 & 1.168 & 0.286 & 2.458 & 0.335 & 0.127 & 0.112 & 0.105\\ 
        & 50  & 0.855 & 4.592 & 0.139 & 0.494 & NA & 2.04 & 0.845 & 4.745 & 0.944 & 0.356 & 0.348 & 0.347\\ 
        & 100  & 1.793 & 9.394 & 0.152 & 1.086 & NA & 3.896 & 1.713 & 9.403 & 1.861 & 0.858 & 0.8 & 0.775\\ 
        & 250  & 3.839 & 23.786 & 1.204 & 2.871 & NA & 10.833 & 3.264 & 23.392 & 4.119 & 1.396 & 1.331 & 1.275\\ 
        \midrule
        \multirow{5}{*}{ MAE } & 10  & 29.127 & 74.586 & 28.237 & 32.563 & 44.906 & 57.255 & 25.461 & 106.444 & 31.154 & 22.089 & 19.424 & 19.468\\ 
        & 25  & 60.747 & 218.249 & 56.479 & 76.163 & 87.508 & 156.242 & 60.369 & 245.818 & 64.18 & 44.461 & 42.378 & 42.696\\ 
        & 50  & 129.217 & 459.232 & 99.63 & 149.773 & NA & 314.016 & 127.928 & 474.518 & 140.922 & 83.036 & 82.464 & 81.935\\ 
        & 100  & 253.278 & 939.405 & 195.978 & 280.408 & NA & 510.712 & 251.441 & 940.342 & 266.573 & 163.12 & 158.681 & 158.084\\ 
        & 250  & 633.745 & 2378.584 & 496.981 & 745.615 & NA & 1303.445 & 623.223 & 2339.228 & 676.812 & 398.135 & 394.581 & 395.265\\ 
        \midrule
        \multirow{5}{*}{ SRE } & 10  & 1.815 & 2.014 & 1.099 & 2.118 & 1.485 & 2.823 & 1.87 & 1.194 & 1.821 & 1.159 & 1 & 0.997\\ 
        & 25  & 2.167 & 2.43 & 1.013 & 2.408 & 2.127 & 4.035 & 2.261 & 1.151 & 2.064 & 0.992 & 0.888 & 0.902\\ 
        & 50  & 2.221 & 2.47 & 1.043 & 2.531 & NA & 4.64 & 2.328 & 1.03 & 2.101 & 0.983 & 0.971 & 0.969\\ 
        & 100  & 2.242 & 2.472 & 1.028 & 2.531 & NA & 3.721 & 2.34 & 0.96 & 1.942 & 0.918 & 0.882 & 0.893\\ 
        & 250  & 2.251 & 2.515 & 0.963 & 2.535 & NA & 3.702 & 2.379 & 0.906 & 2.019 & 0.882 & 0.869 & 0.897\\ 
        \bottomrule
    \end{tabular}}
    \caption{Estimated Bias, Mean Absolute Error and Subspace Recovery Error (SRE) for different PCA algorithm for simulation scenario (S2c)}
    \label{tbl:sim-S2c}
\end{table}

\begin{table}[th]
    \resizebox*{\textwidth}{!}{
    \begin{tabular}{ccrrrrrrrrrrrr}
        \toprule
        Metric & $p$ & Classical & LOC   & ROBPCA & Proj  & RobCov & Grid  & Gmed  & PCP   & \begin{tabular}[c]{@{}r@{}}DPD\\ (0.25)\end{tabular} & \begin{tabular}[c]{@{}r@{}}DPD\\ (0.5)\end{tabular} & \begin{tabular}[c]{@{}r@{}}DPD\\ (0.75)\end{tabular} & \begin{tabular}[c]{@{}r@{}}DPD\\ (1)\end{tabular} \\ \midrule
        \multirow{5}{*}{ Bias } & 10  & 0.318 & 0.795 & 0.143 & 0.351 & 0.624 & 0.732 & 0.238 & 1.066 & 0.381 & 0.223 & 0.171 & 0.173\\ 
        & 25  & 0.673 & 2.244 & 0.251 & 0.785 & 0.76 & 1.607 & 0.491 & 2.464 & 0.741 & 0.485 & 0.329 & 0.33\\ 
        & 50  & 1.558 & 4.655 & 0.258 & 1.328 & NA & 3.931 & 0.893 & 4.662 & 1.873 & 1.132 & 0.747 & 0.738\\ 
        & 100  & 3.048 & 9.455 & 0.55 & 2.703 & NA & 7.156 & 1.806 & 9.224 & 3.675 & 2.412 & 1.444 & 1.409\\ 
        & 250  & 7.077 & 23.848 & 0.789 & 7.754 & NA & 18.827 & 4.491 & 22.946 & 8.81 & 5.079 & 3.244 & 3.285\\ 
        \midrule
        \multirow{5}{*}{ MAE } & 10  & 37.458 & 79.459 & 29.663 & 51.391 & 69.393 & 83.281 & 30.899 & 106.478 & 43.167 & 30.213 & 23.255 & 21.96\\ 
        & 25  & 83.302 & 224.442 & 56.003 & 114.552 & 109.815 & 199.736 & 68.858 & 246.374 & 92.75 & 66.028 & 50.217 & 48.364\\ 
        & 50  & 183.396 & 465.533 & 100.959 & 222.596 & NA & 473.89 & 139.529 & 466.556 & 216.876 & 143.9 & 100.715 & 96.639\\ 
        & 100  & 365.033 & 945.488 & 194.688 & 453.592 & NA & 787.951 & 276.815 & 923.643 & 438.872 & 299.393 & 204.371 & 195.749\\ 
        & 250  & 896.782 & 2384.753 & 491.336 & 1154.62 & NA & 2015.285 & 682.064 & 2297.192 & 1077.574 & 670.359 & 488.388 & 482.877\\ 
        \midrule
        \multirow{5}{*}{ SRE } & 10  & 1.882 & 2.106 & 1.159 & 2.529 & 2.221 & 3.09 & 1.991 & 1.205 & 1.886 & 1.491 & 1.214 & 1.174\\ 
        & 25  & 2.136 & 2.387 & 1.094 & 2.75 & 2.318 & 4.552 & 2.209 & 1.261 & 2.132 & 1.539 & 1.164 & 1.125\\ 
        & 50  & 2.233 & 2.506 & 0.978 & 2.915 & NA & 5.408 & 2.323 & 2.272 & 2.274 & 1.523 & 1.09 & 1.035\\ 
        & 100  & 2.225 & 2.532 & 0.982 & 2.926 & NA & 4.051 & 2.33 & 2.274 & 2.25 & 1.514 & 1.066 & 1.002\\ 
        & 250  & 2.256 & 2.479 & 0.933 & 2.907 & NA & 4.143 & 2.344 & 2.3 & 2.265 & 1.346 & 0.972 & 0.937\\ 
        \bottomrule
    \end{tabular}}
    \caption{Estimated Bias, Mean Absolute Error and Subspace Recovery Error (SRE) for different PCA algorithm for simulation scenario (S2d)}
    \label{tbl:sim-S2d}
\end{table}

Tables~\ref{tbl:sim-S2a} and~\ref{tbl:sim-S2b} respectively show results for setups (S2a) and (S2b) differing in contamination level. As the level of contamination increases, classical PCA worsens as expected, spherical PCA~\citep{locantore1999robust} yields biased estimates for large number of variables (large $p$), and the projection pursuit-based methods also perform poorly under the considered simulation scenarios. The ROBPCA algorithm by~\cite{hubert2005robpca} and the Gmedian algorithm by~\cite{cardot2017fast} stand out to be the most promising among the existing methods. However, Gmedian algorithm suits applications where the outlying distribution and the true distribution have the same theoretical mean but a different covariance structure. In contrast, the ROBPCA algorithm works well with significant changes in mean between outlying distribution and true distribution. The proposed rPCAdpd algorithm, suited for similar scenarios with changes in mean, surpasses ROBPCA at high robustness parameter $\alpha$, and is significantly better in high dimensions. The PCP algorithm~\citep{candes2011robust} has consistent results across setups (S1), (S2a), and (S2b). This is due to the fact that the error comes only from the perturbation matrix $\bb{E}$ in Eq.~\eqref{eqn:decomposition}, which is inestimable by the PCP method. Table~\ref{tbl:sim-S2c} and~\ref{tbl:sim-S2d} summarises the results obtained for scenario (S2c) and (S2d). These results closely mirror those in scenarios (S2a) and (S2b) respectively.

\begin{table}[th]
    \resizebox*{\textwidth}{!}{
    \begin{tabular}{ccrrrrrrrrrrrr}
        \toprule
        Metric & $p$ & Classical & LOC   & ROBPCA & Proj  & RobCov & Grid  & Gmed  & PCP   & \begin{tabular}[c]{@{}r@{}}DPD\\ (0.25)\end{tabular} & \begin{tabular}[c]{@{}r@{}}DPD\\ (0.5)\end{tabular} & \begin{tabular}[c]{@{}r@{}}DPD\\ (0.75)\end{tabular} & \begin{tabular}[c]{@{}r@{}}DPD\\ (1)\end{tabular} \\ \midrule
        \multirow{5}{*}{ Bias } & 10  & 0.212 & 0.743 & 0.23 & 0.263 & 0.429 & 0.519 & 0.05 & 1.066 & 0.2 & 0.138 & 0.082 & 0.064\\ 
        & 25  & 0.534 & 2.185 & 0.403 & 0.494 & 0.412 & 1.09 & 0.127 & 2.452 & 0.515 & 0.324 & 0.247 & 0.247\\ 
        & 50  & 1.218 & 4.58 & 1.018 & 0.997 & NA & 2.249 & 0.296 & 4.717 & 1.159 & 0.715 & 0.493 & 0.455\\ 
        & 100  & 2.39 & 9.389 & 1.431 & 1.876 & NA & 4.388 & 0.597 & 9.372 & 2.218 & 1.201 & 0.873 & 0.857\\ 
        & 250  & 5.09 & 23.783 & 2.675 & 2.997 & NA & 9.718 & 1.631 & 23.309 & 4.503 & 1.83 & 1.424 & 1.451\\ 
        \midrule
        \multirow{5}{*}{ MAE } & 10  & 32.656 & 74.327 & 30.411 & 42.304 & 48.105 & 66.025 & 24.609 & 106.512 & 34.086 & 29.286 & 23.804 & 21.794\\ 
        & 25  & 76.302 & 218.539 & 58.718 & 82.371 & 62.299 & 144.168 & 53.57 & 245.173 & 78.951 & 61.355 & 54.034 & 54.426\\ 
        & 50  & 160.631 & 458.022 & 131.667 & 166.613 & NA & 302.974 & 113.246 & 471.68 & 166.973 & 127.601 & 105.154 & 100.407\\ 
        & 100  & 303.042 & 938.894 & 218.99 & 303.707 & NA & 503.095 & 218.692 & 937.193 & 314.138 & 220.16 & 188.221 & 189.425\\ 
        & 250  & 815.964 & 2378.289 & 589.99 & 858.143 & NA & 1279.957 & 662.041 & 2330.966 & 817.858 & 559.381 & 515.87 & 516.502\\ 
        \midrule
        \multirow{5}{*}{ SRE } & 10  & 1.822 & 1.882 & 1.152 & 1.945 & 1.537 & 2.698 & 1.822 & 1.155 & 1.818 & 1.636 & 1.292 & 1.158\\ 
        & 25  & 2.154 & 2.272 & 1.003 & 2.128 & 1.048 & 3.757 & 2.185 & 1.142 & 2.155 & 1.4 & 1.054 & 1.053\\ 
        & 50  & 2.204 & 2.368 & 1.017 & 2.152 & NA & 4.395 & 2.287 & 0.978 & 2.206 & 1.423 & 0.961 & 0.919\\ 
        & 100  & 2.251 & 2.475 & 1.02 & 2.279 & NA & 3.587 & 2.311 & 1.01 & 2.23 & 1.282 & 0.951 & 0.955\\ 
        & 250  & 2.264 & 2.521 & 1.072 & 2.202 & NA & 3.525 & 2.343 & 1.015 & 2.16 & 1.165 & 0.98 & 0.98\\ 
        \bottomrule
    \end{tabular}}
    \caption{Estimated Bias, Mean Absolute Error and Subspace Recovery Error (SRE) for different PCA algorithm for simulation scenario (S3a)}
    \label{tbl:sim-S3a}
\end{table}

\begin{table}[th]
    \resizebox*{\textwidth}{!}{
    \begin{tabular}{ccrrrrrrrrrrrr}
        \toprule
        Metric & $p$ & Classical & LOC   & ROBPCA & Proj  & RobCov & Grid  & Gmed  & PCP   & \begin{tabular}[c]{@{}r@{}}DPD\\ (0.25)\end{tabular} & \begin{tabular}[c]{@{}r@{}}DPD\\ (0.5)\end{tabular} & \begin{tabular}[c]{@{}r@{}}DPD\\ (0.75)\end{tabular} & \begin{tabular}[c]{@{}r@{}}DPD\\ (1)\end{tabular} \\ \midrule
        \multirow{5}{*}{ Bias } & 10  & 0.454 & 0.756 & 0.376 & 0.415 & 0.509 & 0.749 & 0.172 & 1.065 & 0.478 & 0.399 & 0.283 & 0.207\\ 
        & 25  & 0.957 & 2.198 & 0.529 & 0.828 & 1.049 & 1.596 & 0.363 & 2.448 & 0.968 & 0.817 & 0.523 & 0.416\\ 
        & 50  & 2.053 & 4.603 & 0.829 & 1.571 & NA & 3.221 & 0.729 & 4.613 & 2.108 & 1.738 & 0.952 & 0.674\\ 
        & 100  & 4.085 & 9.41 & 1.438 & 2.712 & NA & 5.838 & 1.174 & 9.136 & 4.235 & 3.121 & 1.585 & 1.015\\ 
        & 250  & 10.553 & 23.793 & 5.036 & 8.705 & NA & 17.022 & 4.318 & 22.72 & 10.683 & 8.138 & 5.083 & 3.947\\ 
        \midrule
        \multirow{5}{*}{ MAE } & 10  & 52.663 & 75.648 & 40.059 & 50.066 & 53.751 & 81.22 & 29.654 & 106.464 & 56.791 & 50.076 & 39.234 & 32.751\\ 
        & 25  & 113.553 & 219.799 & 73.592 & 109.254 & 119.102 & 191.308 & 75.088 & 244.825 & 119.018 & 106.796 & 78.32 & 67.627\\ 
        & 50  & 236.518 & 460.337 & 122.695 & 206.081 & NA & 411.252 & 144.926 & 461.708 & 249.263 & 218.761 & 141.867 & 114.571\\ 
        & 100  & 492.217 & 941.042 & 244.944 & 399.289 & NA & 647.709 & 288.144 & 914.552 & 534.963 & 445.369 & 293.533 & 237.476\\ 
        & 250  & 1175.092 & 2379.306 & 659.666 & 1087.267 & NA & 1791.531 & 725.623 & 2273.85 & 1231.25 & 1015.697 & 716.547 & 604.369\\ 
        \midrule
        \multirow{5}{*}{ SRE } & 10  & 1.839 & 1.993 & 1.134 & 2.326 & 1.465 & 2.825 & 1.886 & 1.198 & 1.85 & 1.797 & 1.556 & 1.351\\ 
        & 25  & 2.22 & 2.414 & 1.071 & 2.784 & 2.217 & 4.183 & 2.249 & 1.234 & 2.237 & 2.026 & 1.356 & 1.176\\ 
        & 50  & 2.304 & 2.528 & 0.97 & 2.888 & NA & 4.909 & 2.358 & 2.287 & 2.312 & 2.149 & 1.43 & 1.196\\ 
        & 100  & 2.286 & 2.491 & 1.034 & 2.838 & NA & 3.809 & 2.307 & 2.288 & 2.309 & 1.993 & 1.223 & 0.995\\ 
        & 250  & 2.307 & 2.481 & 0.952 & 2.83 & NA & 3.847 & 2.34 & 2.308 & 2.317 & 1.997 & 1.348 & 1.172\\
        \bottomrule
    \end{tabular}}
    \caption{Estimated Bias, Mean Absolute Error and Subspace Recovery Error (SRE) for different PCA algorithm for simulation scenario (S3b)}
    \label{tbl:sim-S3b}
\end{table}

\begin{table}[th]
    \resizebox*{\textwidth}{!}{
    \begin{tabular}{ccrrrrrrrrrrrr}
        \toprule
        Metric & $p$ & Classical & LOC   & ROBPCA & Proj  & RobCov & Grid  & Gmed  & PCP   & \begin{tabular}[c]{@{}r@{}}DPD\\ (0.25)\end{tabular} & \begin{tabular}[c]{@{}r@{}}DPD\\ (0.5)\end{tabular} & \begin{tabular}[c]{@{}r@{}}DPD\\ (0.75)\end{tabular} & \begin{tabular}[c]{@{}r@{}}DPD\\ (1)\end{tabular} \\ \midrule
        \multirow{5}{*}{ Bias } & 10  & 0.149 & 0.746 & 0.228 & 0.181 & 0.464 & 0.505 & 0.123 & 1.065 & 0.163 & 0.081 & 0.043 & 0.041\\ 
        & 25  & 0.357 & 2.188 & 0.258 & 0.402 & 0.543 & 1.057 & 0.381 & 2.457 & 0.382 & 0.174 & 0.148 & 0.143\\ 
        & 50  & 0.777 & 4.59 & 0.33 & 0.455 & NA & 2.011 & 0.676 & 4.738 & 0.823 & 0.296 & 0.265 & 0.253\\ 
        & 100  & 1.517 & 9.387 & 0.667 & 1.276 & NA & 4.291 & 1.144 & 9.402 & 1.495 & 0.442 & 0.392 & 0.372\\ 
        & 250  & 3.886 & 23.785 & 1.012 & 2.727 & NA & 9.358 & 3.513 & 23.383 & 3.883 & 1.464 & 1.386 & 1.359\\ 
        \midrule
        \multirow{5}{*}{ MAE } & 10  & 25.443 & 74.609 & 32.294 & 37.131 & 50.803 & 64.494 & 22.821 & 106.453 & 26.557 & 19.682 & 15.649 & 15.393\\ 
        & 25  & 59.571 & 218.815 & 53.687 & 74.807 & 76.628 & 147.007 & 60.024 & 245.654 & 65.782 & 44.871 & 42.176 & 41.994\\ 
        & 50  & 133.973 & 459.029 & 108.565 & 159.38 & NA & 319.77 & 123.649 & 473.841 & 141.108 & 87.353 & 84.297 & 84.503\\ 
        & 100  & 256.543 & 938.692 & 193.825 & 284.691 & NA & 496.1 & 225.919 & 940.223 & 263.121 & 159.77 & 154.23 & 154.083\\ 
        & 250  & 676.504 & 2378.474 & 524.335 & 711.74 & NA & 1230.442 & 649.349 & 2338.346 & 678.719 & 435.267 & 427.714 & 430.439\\ 
        \midrule
        \multirow{5}{*}{ SRE } & 10  & 1.796 & 2.006 & 1.052 & 2.079 & 1.468 & 2.768 & 1.872 & 1.144 & 1.788 & 1.286 & 0.952 & 0.958\\ 
        & 25  & 2.141 & 2.342 & 0.981 & 2.397 & 1.775 & 3.969 & 2.211 & 1.186 & 2.057 & 0.994 & 0.861 & 0.87\\ 
        & 50  & 2.179 & 2.418 & 0.954 & 2.389 & NA & 4.572 & 2.289 & 0.858 & 2.038 & 0.908 & 0.841 & 0.851\\ 
        & 100  & 2.23 & 2.487 & 1.053 & 2.476 & NA & 3.66 & 2.315 & 0.929 & 1.983 & 0.899 & 0.849 & 0.883\\ 
        & 250  & 2.254 & 2.555 & 0.999 & 2.479 & NA & 3.719 & 2.364 & 0.927 & 1.908 & 0.841 & 0.826 & 0.834\\ 
        \bottomrule
    \end{tabular}}
    \caption{Estimated Bias, Mean Absolute Error and Subspace Recovery Error (SRE) for different PCA algorithm for simulation scenario (S3c)}
    \label{tbl:sim-S3c}
\end{table}

\begin{table}[th]
    \resizebox*{\textwidth}{!}{
    \begin{tabular}{ccrrrrrrrrrrrr}
        \toprule
        Metric & $p$ & Classical & LOC   & ROBPCA & Proj  & RobCov & Grid  & Gmed  & PCP   & \begin{tabular}[c]{@{}r@{}}DPD\\ (0.25)\end{tabular} & \begin{tabular}[c]{@{}r@{}}DPD\\ (0.5)\end{tabular} & \begin{tabular}[c]{@{}r@{}}DPD\\ (0.75)\end{tabular} & \begin{tabular}[c]{@{}r@{}}DPD\\ (1)\end{tabular} \\ \midrule
        \multirow{5}{*}{ Bias } & 10  & 0.268 & 0.79 & 0.139 & 0.347 & 0.633 & 0.724 & 0.183 & 1.066 & 0.324 & 0.217 & 0.115 & 0.113\\ 
        & 25  & 0.673 & 2.231 & 0.243 & 0.71 & 0.725 & 1.615 & 0.486 & 2.461 & 0.748 & 0.458 & 0.304 & 0.297\\ 
        & 50  & 1.325 & 4.641 & 0.13 & 1.474 & NA & 3.602 & 0.709 & 4.649 & 1.669 & 0.934 & 0.522 & 0.481\\ 
        & 100  & 2.772 & 9.436 & 0.101 & 2.798 & NA & 6.713 & 1.462 & 9.206 & 3.428 & 2.046 & 1.142 & 1.098\\ 
        & 250  & 6.861 & 23.832 & 0.34 & 6.809 & NA & 17.565 & 3.587 & 22.922 & 8.555 & 5.192 & 3.156 & 2.813\\ 
        \midrule
        \multirow{5}{*}{ MAE } & 10  & 35.048 & 79.01 & 31.374 & 49.556 & 67.262 & 81.659 & 27.681 & 106.519 & 39.974 & 32.114 & 20.677 & 19.517\\ 
        & 25  & 83.555 & 223.086 & 58.178 & 105.35 & 112.409 & 201.416 & 70.421 & 246.107 & 91.476 & 65.052 & 49.229 & 46.971\\ 
        & 50  & 179.148 & 464.061 & 112.288 & 227.737 & NA & 458.559 & 129.521 & 465.585 & 210.767 & 138.549 & 96.856 & 89.979\\ 
        & 100  & 370.199 & 943.586 & 216.485 & 440.313 & NA & 738.531 & 272.546 & 921.936 & 430.631 & 287.974 & 191.931 & 187.771\\ 
        & 250  & 908.806 & 2383.166 & 566.649 & 1118.667 & NA & 1918.17 & 688.634 & 2294.489 & 1054.348 & 709.767 & 492.689 & 450.457\\ 
        \midrule
        \multirow{5}{*}{ SRE } & 10  & 1.849 & 2.045 & 1.084 & 2.42 & 2.053 & 3.117 & 1.955 & 1.026 & 1.851 & 1.608 & 1.225 & 1.16\\ 
        & 25  & 2.145 & 2.383 & 0.948 & 2.836 & 2.283 & 4.553 & 2.233 & 1.077 & 2.134 & 1.409 & 1.04 & 0.965\\ 
        & 50  & 2.235 & 2.495 & 0.983 & 2.951 & NA & 5.331 & 2.34 & 2.274 & 2.286 & 1.496 & 1.112 & 1.022\\ 
        & 100  & 2.269 & 2.529 & 0.985 & 2.928 & NA & 3.971 & 2.351 & 2.311 & 2.296 & 1.503 & 1.01 & 0.953\\ 
        & 250  & 2.28 & 2.506 & 1.039 & 3.009 & NA & 4.132 & 2.388 & 2.326 & 2.295 & 1.551 & 1.176 & 1.076\\ 
        \bottomrule
    \end{tabular}}
    \caption{Estimated Bias, Mean Absolute Error and Subspace Recovery Error (SRE) for different PCA algorithm for simulation scenario (S3d)}
    \label{tbl:sim-S3d}
\end{table}

In scenarios (S3a)-(S3d), the contaminating distribution changes to $t$-distribution with $5$ degrees of freedom with a heavy tail. In these scenarios, ROBPCA~\citep{hubert2005robpca}, Gmedian~\citep{cardot2017fast} algorithm and the proposed rPCAdpd methods perform closely. In (S3a), the rPCAdpd method excels for large values of $\alpha$. As the contamination rises to $20\%$, as shown in Table~\ref{tbl:sim-S3b}, all of the chosen algorithms show a significant increase in MAE. However, the proposed estimator maintains a low bias for all components even for large $p$ relative to $n$, consistent with its theoretical breakdown point behaviour as pointed out in Section~\ref{sec:breakdown}. 

In essence, the proposed rPCAdpd algorithm excels at detecting and removing low-variance, different-location contaminating components, compared to the primary data distribution component. In all other cases, its performance is closely comparable to the existing algorithms. Also, across all of the simulation setups considered, the proposed rPCAdpd algorithm yields significantly better estimates of principal components than the existing algorithms when the dimension of the data $p$ is large, which is also theoretically justified by its dimension-independent asymptotic breakdown point.

\section{Real Data Analysis}\label{sec:real-data-analysis}

In this section, we demonstrate applications of the proposed rPCAdpd estimator on three real-life datasets. The first two datasets, namely the 
Car dataset and the Octane dataset are popular benchmark datasets used to compare performances of different RPCA algorithms (see~\cite{hubert2005robpca} for details). We also consider a novel Credit Card Fraud Detection dataset to demonstrate how the proposed robust PCA estimator can serve as a preliminary preprocessing step to identify fraudulent transactions using credit cards before applying binary classification algorithms.
    
\subsection{Car Dataset}\label{sec:car-data}

The Car dataset comprises $n = 111$ observations of cars with $p = 11$ variables, including the length, width, and height of the car. This dataset has served as a benchmark for various robust PCA methods~\citep{hubert2005robpca,croux2007algorithms}. We utilize it to assess the performance of our proposed rPCAdpd method on outlier detection. For visual evaluation, we adopt orthogonal and score distances as diagnostic metrics~\citep{hubert2005robpca}.

\begin{figure}[t]
    \centering
    \begin{subfigure}{0.35\linewidth}
    \includegraphics[width=\textwidth]{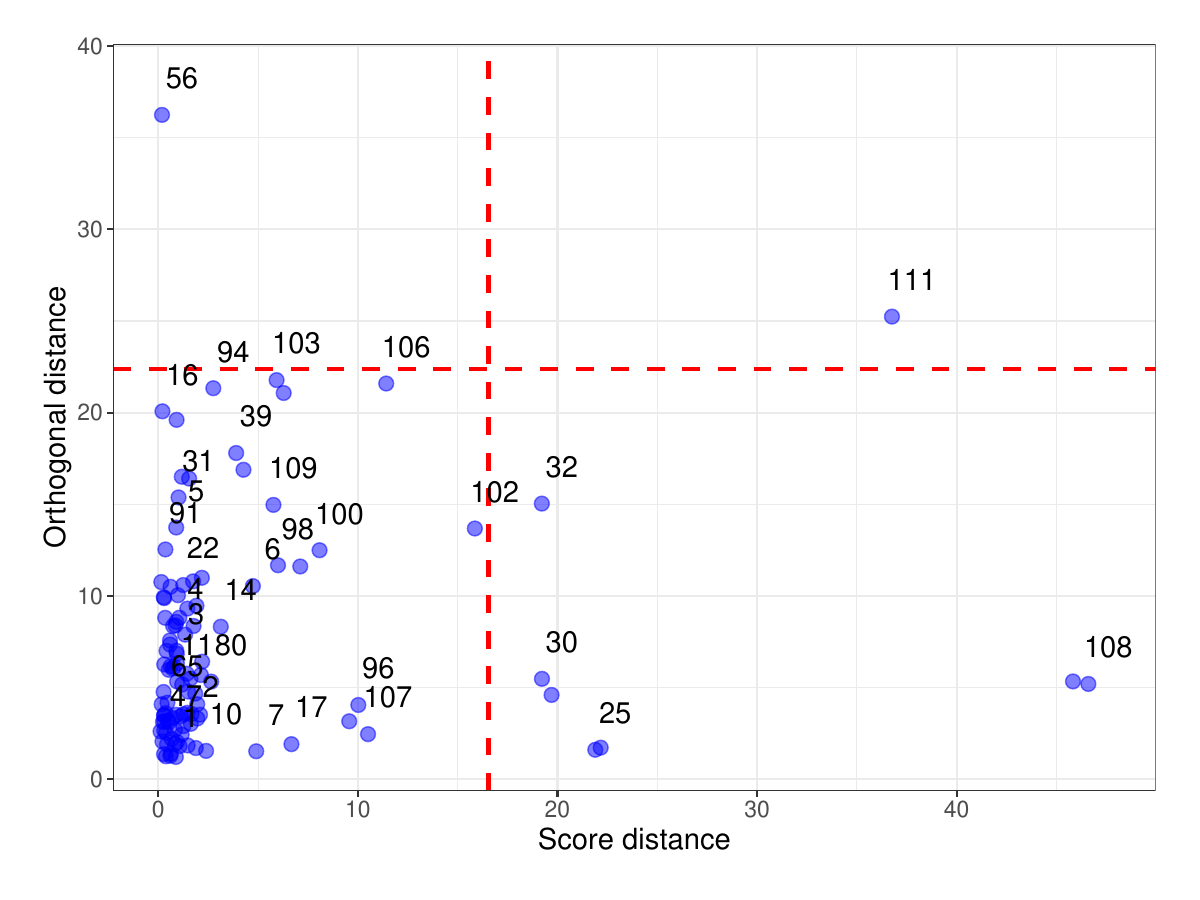}
    \caption{Classical PCA}
    \end{subfigure}
    \begin{subfigure}{0.35\linewidth}
    \includegraphics[width=\textwidth]{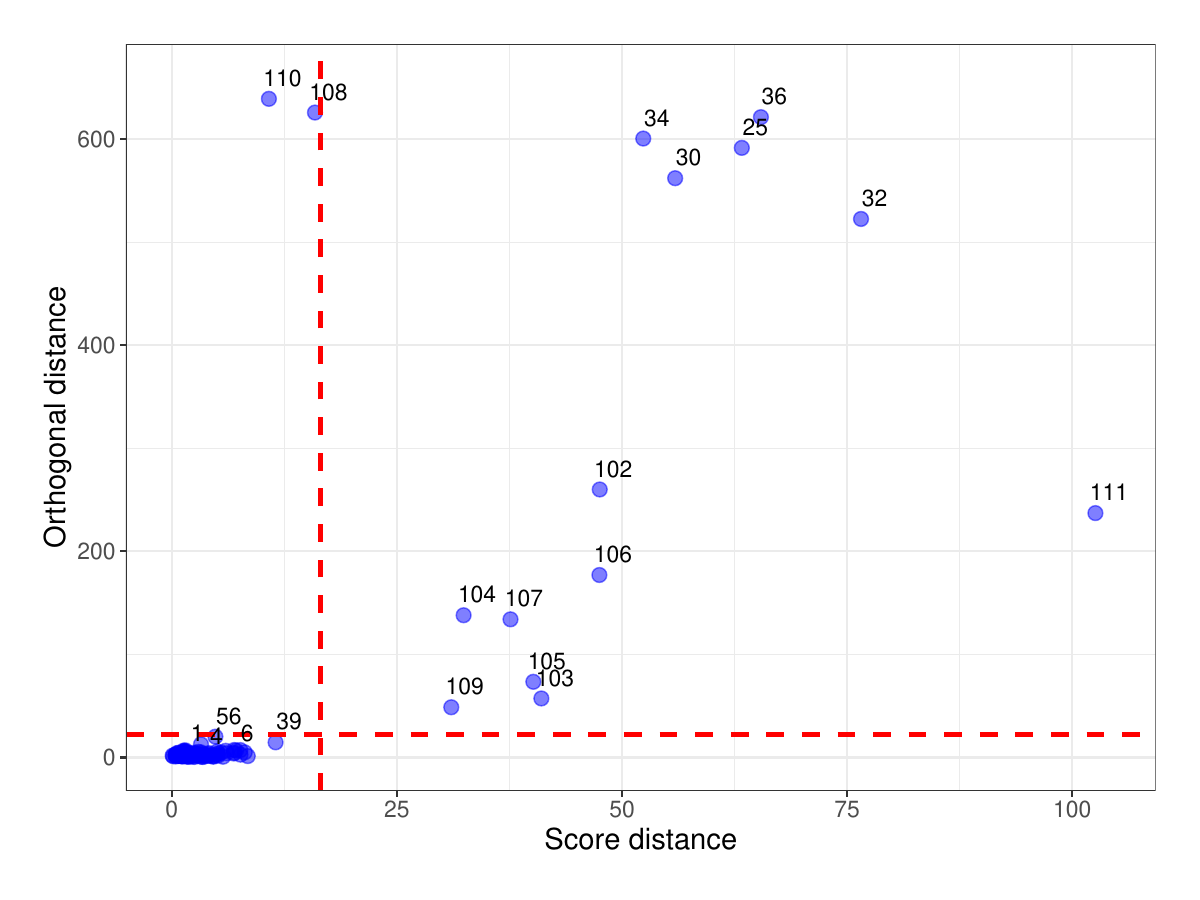}
    \caption{rPCAdpd}
    \end{subfigure}
    \begin{subfigure}{0.35\linewidth}
        \includegraphics[width=\textwidth]{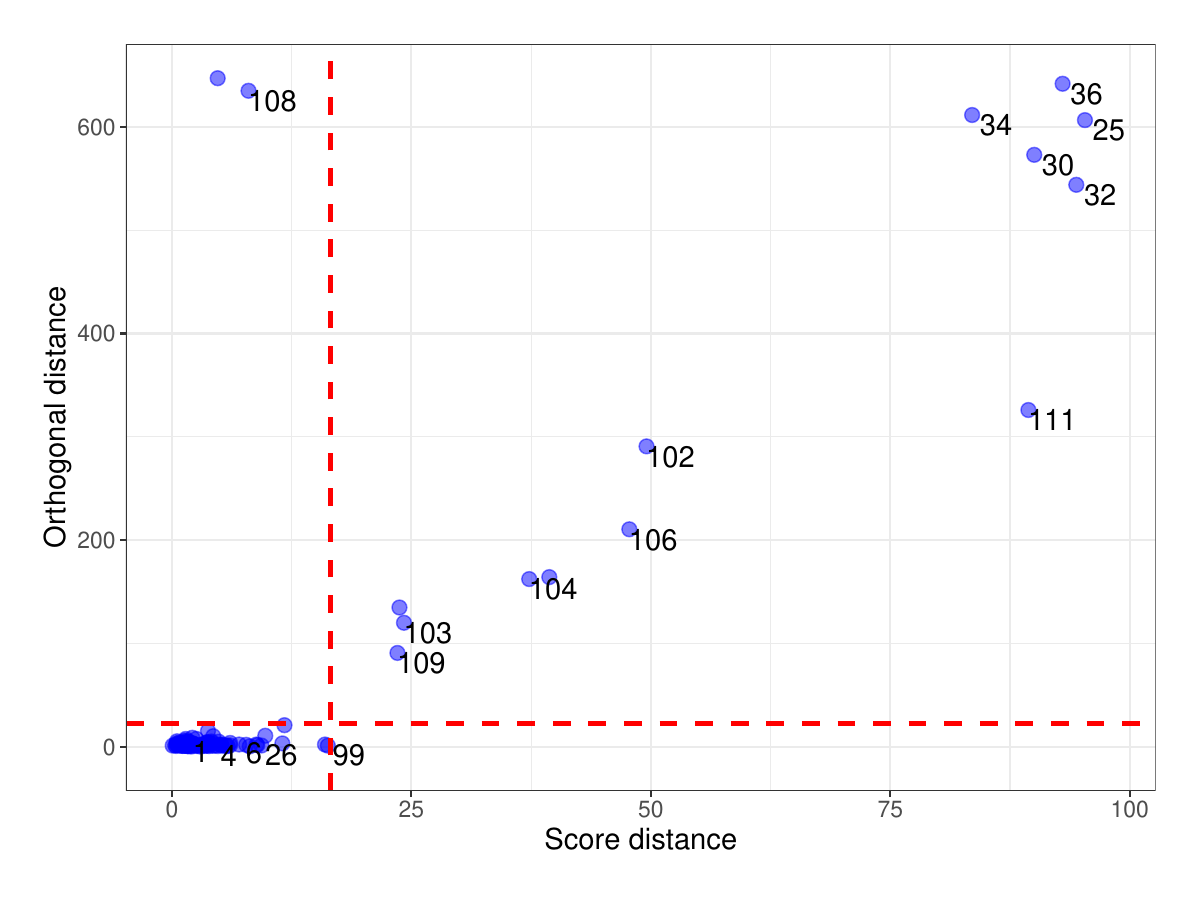}
        \caption{ROBPCA}
    \end{subfigure}
    \begin{subfigure}{0.35\linewidth}
        \includegraphics[width=\textwidth]{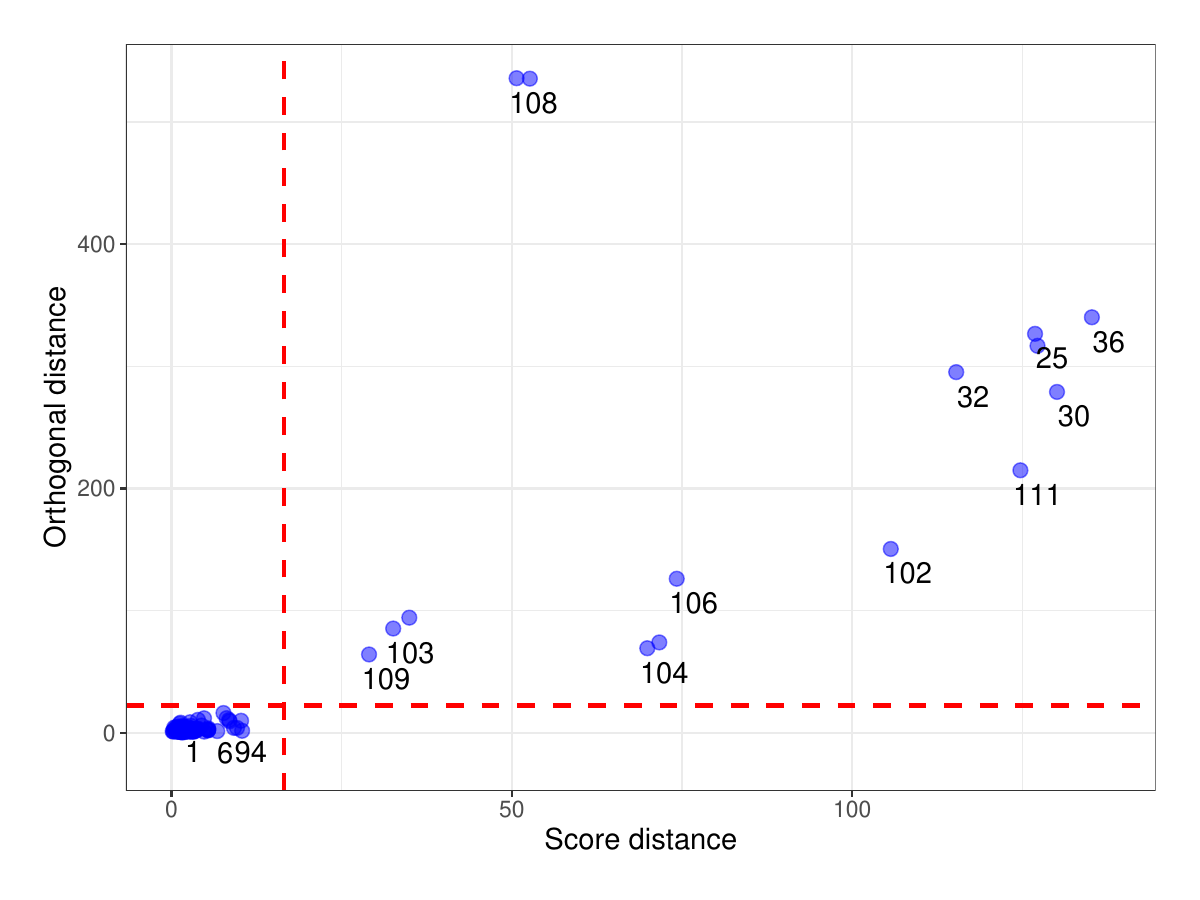}
        \caption{Gmedian}
    \end{subfigure}    
    \caption{Diagnostic plots for the Car dataset}
    \label{fig:cardata-plot}
\end{figure}

Analyzing screeplots for both rPCAdpd and the classical PCA for the Car dataset reveals that the first four principal components capture more than $90\%$ of the variance. We thus apply both algorithms to extract these components and compute orthogonal and score distances for each observation. Figure~\ref{fig:cardata-plot} illustrates this diagnostic analysis. Classical PCA identifies a cluster of influential points (observations $25, 30, 32, 34$, and $36$), which are also detected by rPCAdpd estimator. These points share a value of $(-2)$ for $4$ of the $11$ variables: Rear.Hd, Rear.Seat, Rear.Shld, and Luggage. However, classical PCA assigns low orthogonal distances to these outliers, indicating their good fit, thus inflating distances for most points. Conversely, rPCAdpd assigns high orthogonal distances to these outliers. Additionally, rPCAdpd identifies a different set of outliers (observations $102-107, 109$), unnoticed by classical PCA, consistent with findings in~\cite{hubert2005robpca}. As demonstrated in Figure~\ref{fig:cardata-plot}, ROBPCA and Gmedian algorithms also spot such outliers.

\subsection{Octane Data}

The Octane dataset, sourced from~\cite{esbensen2002multivariate}, features spectroscopic data with octane numbers derived from near-infrared (NIR) absorbance spectra of $39$ gasoline samples. Measurements span $226$ electromagnetic radiation wavelengths ($1102$ nm to $1552$ nm), each of which gives rise to a feature. With $39$ observations and $226$ features, principal component analysis (PCA) is pivotal for dimension reduction and subsequent analysis. Six samples ($25$, $26$, and $36-39$) contain additional alcohol, making them distinct~\citep{hubert2005robpca}. 

\begin{figure}[t]
    \centering
    \begin{subfigure}{0.35\linewidth}
    \includegraphics[width=\textwidth]{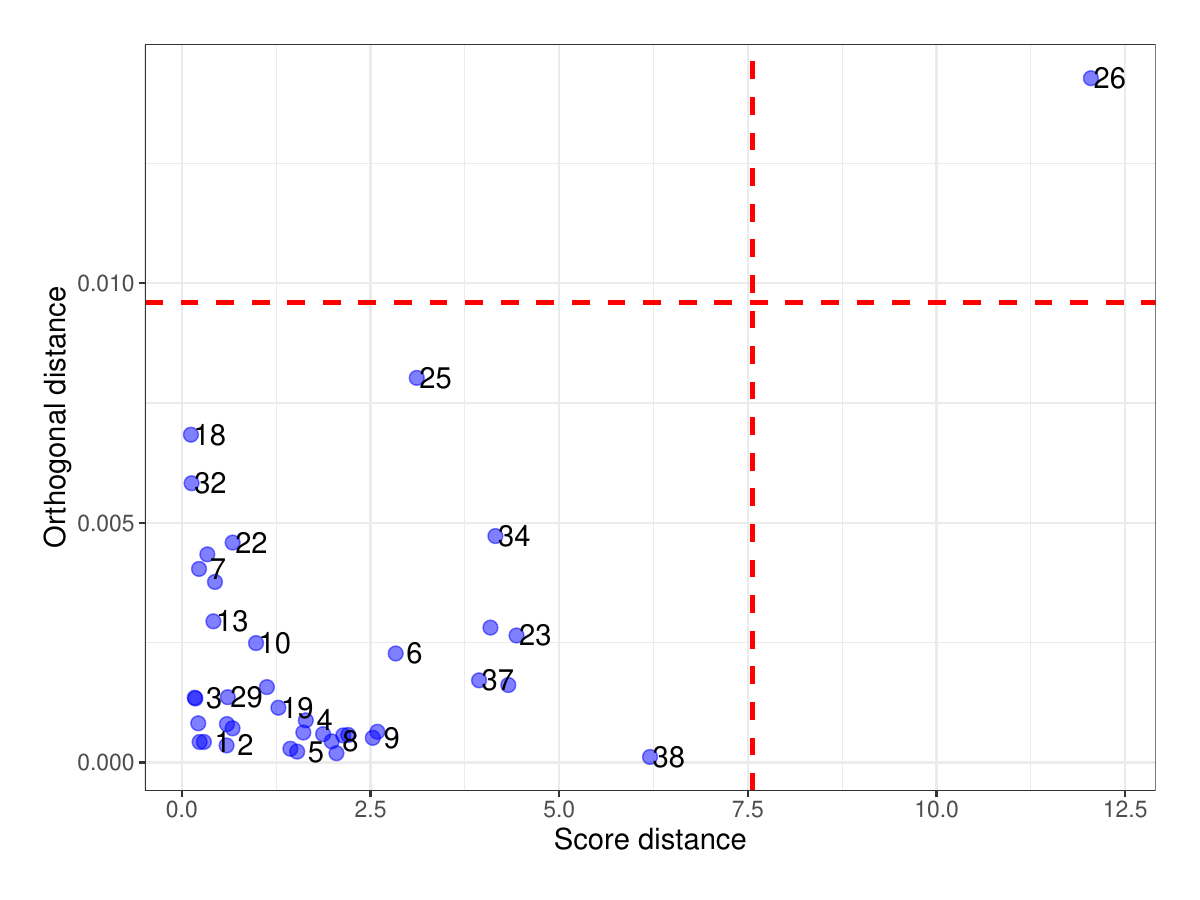}
    \caption{Classical PCA}
    \end{subfigure}
    \begin{subfigure}{0.35\linewidth}
    \includegraphics[width=\textwidth]{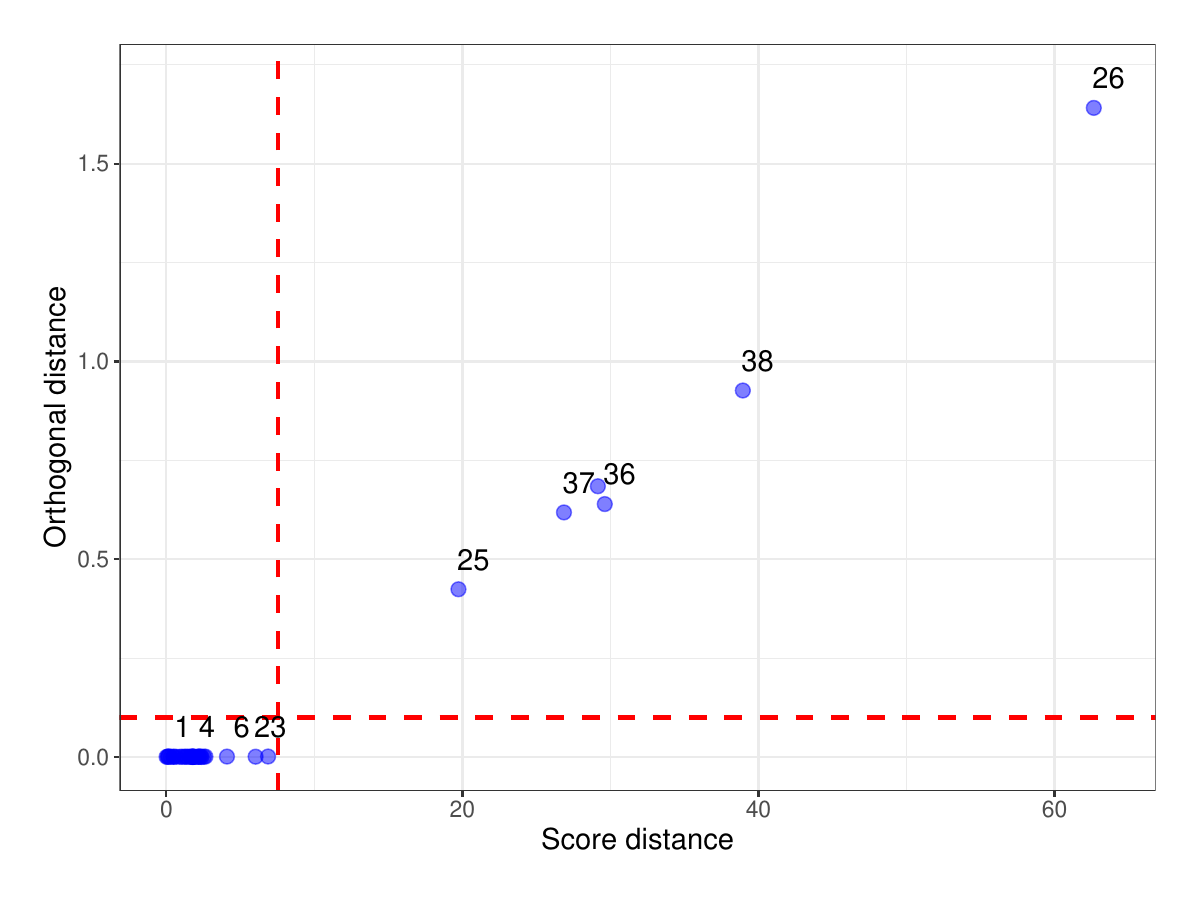}
    \caption{rPCAdpd}
    \end{subfigure}
    \begin{subfigure}{0.35\linewidth}
        \includegraphics[width=\textwidth]{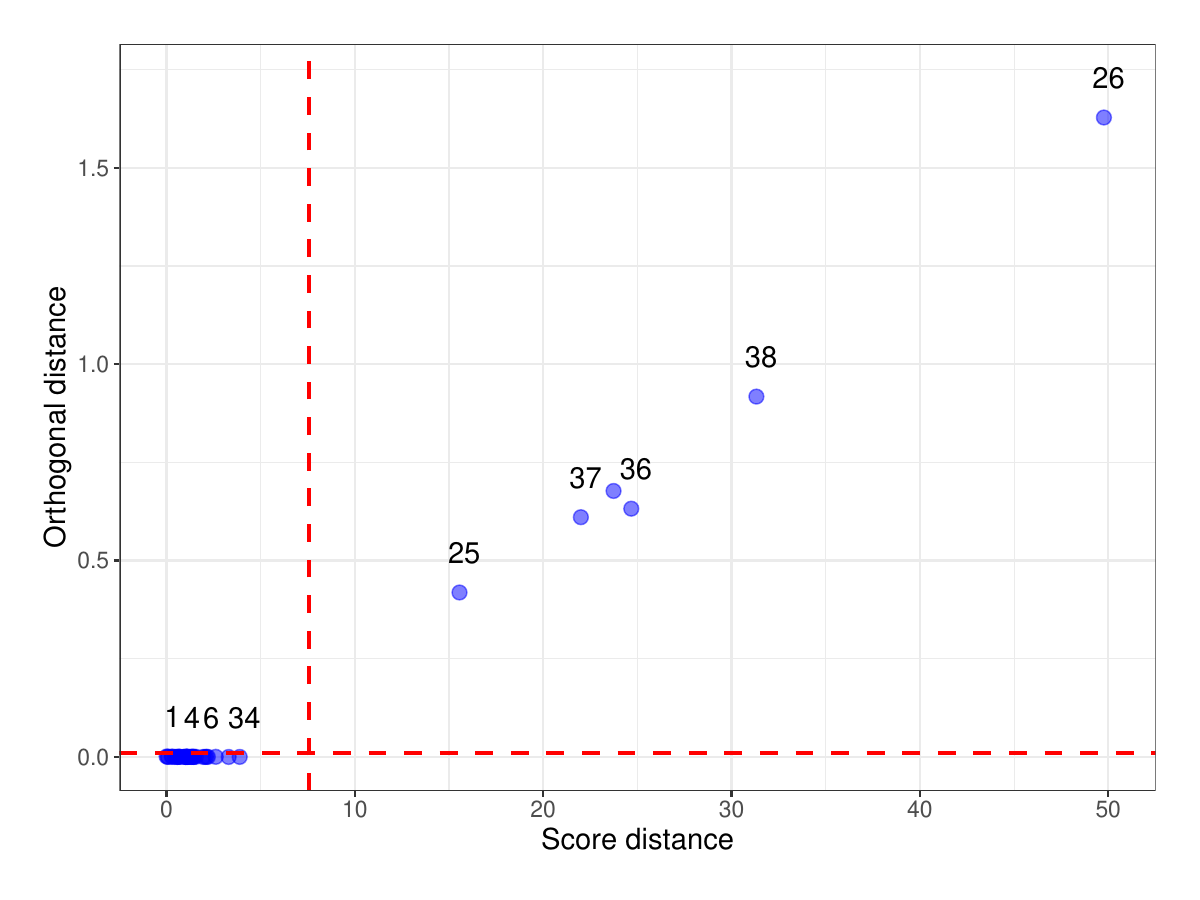}
        \caption{ROBPCA}
    \end{subfigure}
    \begin{subfigure}{0.35\linewidth}
        \includegraphics[width=\textwidth]{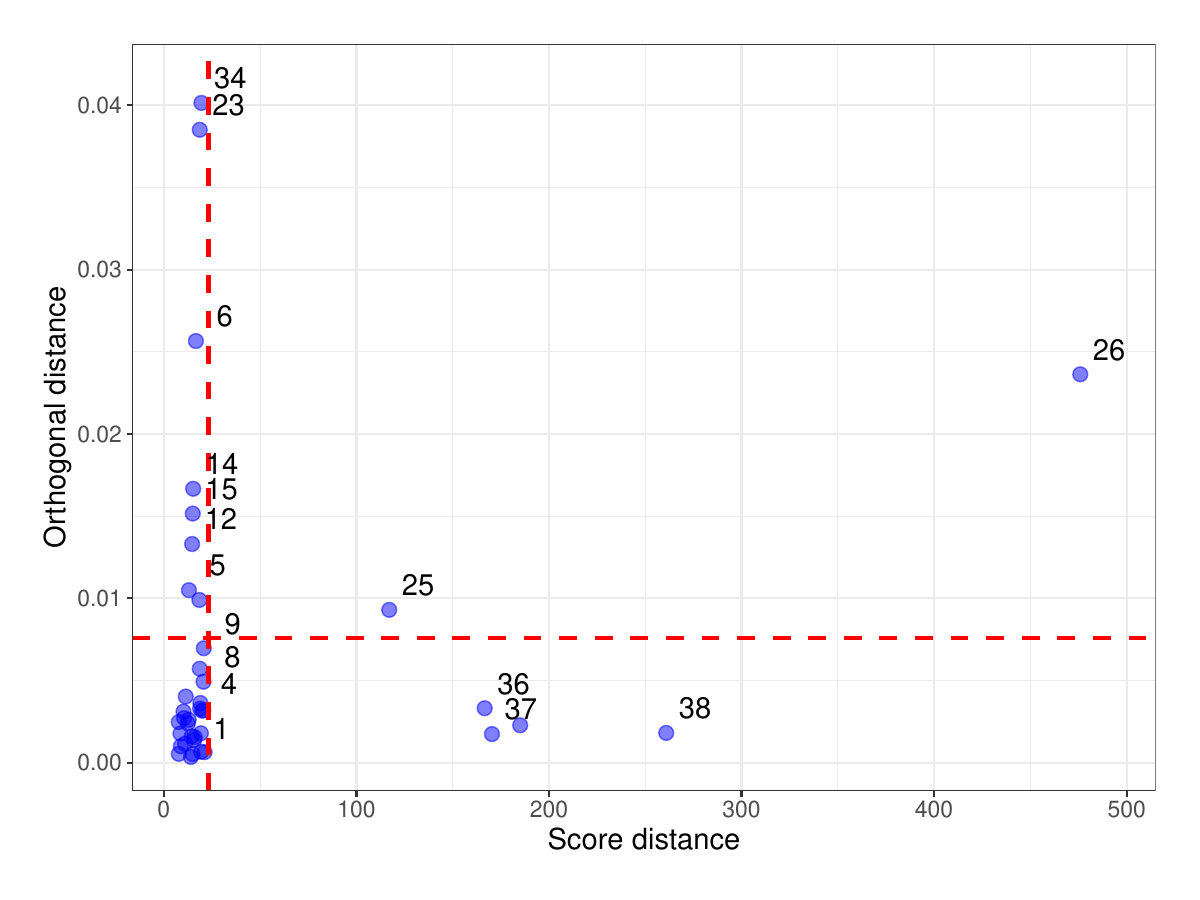}
        \caption{GMedian}
    \end{subfigure}
    \caption{Diagnostic plots for the Octane dataset}
    \label{fig:octane-plot}
\end{figure}

Similar to the Car dataset, a screenplot analysis reveals that there are only $2$ significant principal components present in the Octane dataset. However, the first principal value estimated by the classical PCA ($0.132$) is several magnitudes higher than the first principal value estimated by rPCAdpd ($0.01075$), which aligns with the estimates obtained from existing robust PCA algorithms~\citep{hubert2005robpca}. Diagnostic plots in Figure~\ref{fig:octane-plot} demonstrate classical PCA's failure to detect outliers, except observation $26$, while rPCAdpd identifies alcohol-mixed gasoline samples accurately. The ROBPCA algorithm also detects these outliers, with a similar score and orthogonal distances. However, the Gmedian algorithm labels most of these points as orthogonal outliers only.

\subsection{Credit Card Fraud Detection}

Credit card fraud detection is a very challenging problem because of the specific nature of transaction data and the labelling process. Most of the practical transaction data is highly imbalanced, and the number of fraudulent transactions is far too less compared to the extremely large number of valid transactions made on a day-to-day basis. There are primarily two kinds of strategies to detect such fraudulent transactions: the first one models the situation as a binary classification problem with some sampling procedures to counter class imbalance, and the second approach assumes that the fraudulent transactions are outliers in the data and applies an outlier detection algorithm. Many existing supervised and unsupervised machine learning algorithms~\citep{carcillo2018,carcillo2019} employ outlier detection to spot such fraudulent transactions. These methods often begin with a principal component analysis (PCA) to reduce dimensions and training time for real-time application.

\begin{figure}[t]
    \centering
    \begin{subfigure}{0.35\linewidth}
    \includegraphics[width=\textwidth]{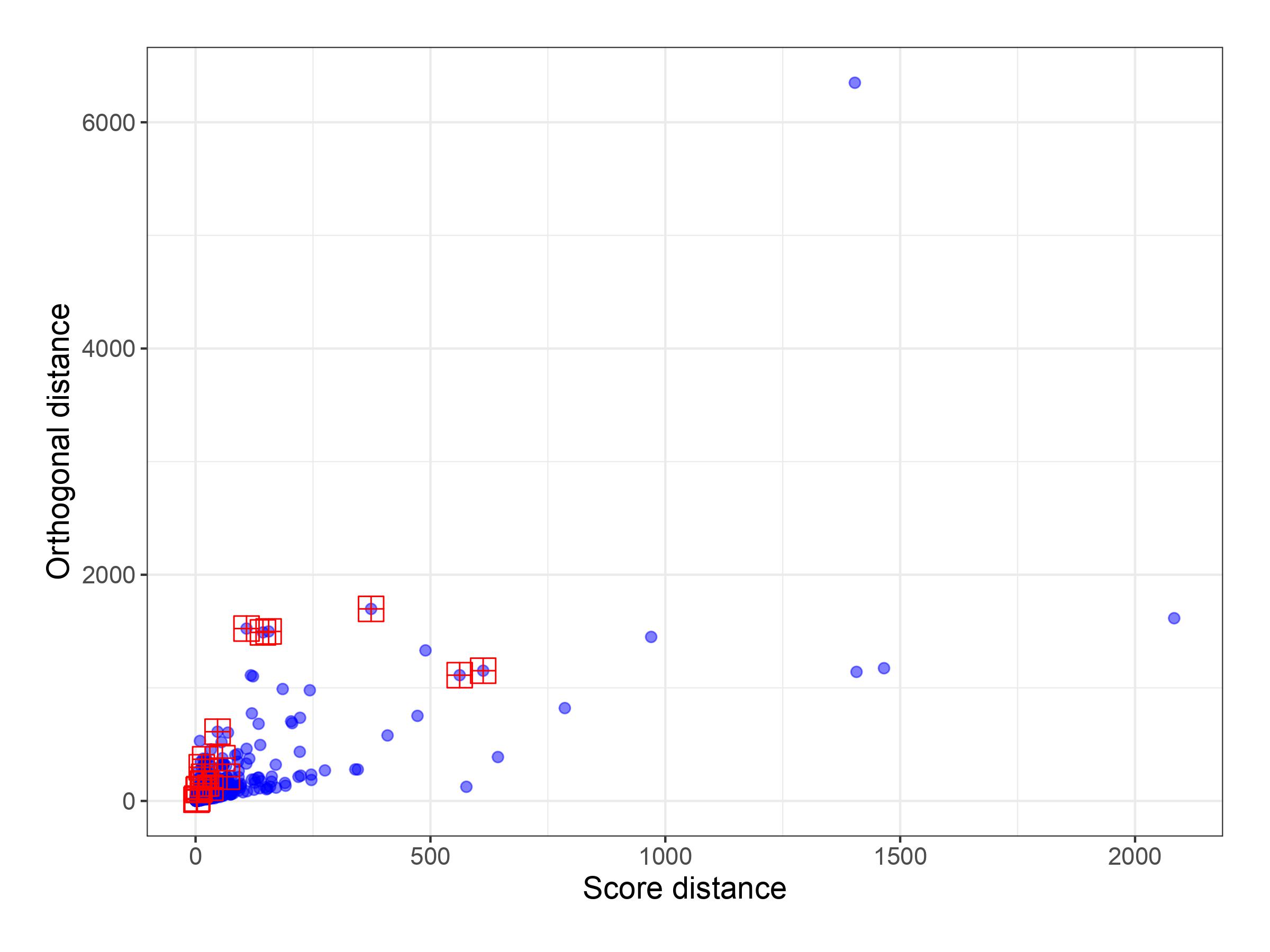}
    \caption{Classical PCA}
    \end{subfigure}
    \begin{subfigure}{0.35\linewidth}
    \includegraphics[width=\textwidth]{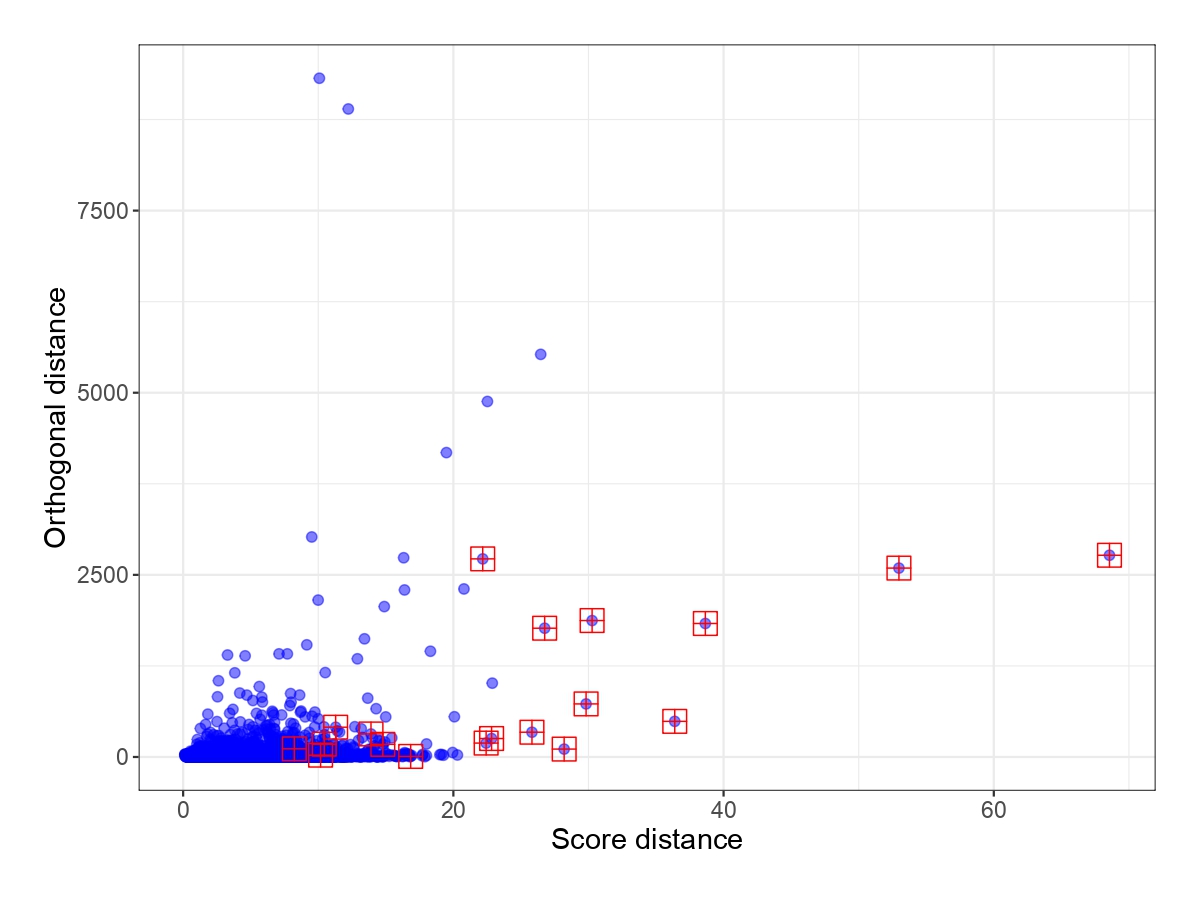}
    \caption{rPCAdpd}
    \end{subfigure}
    \begin{subfigure}{0.35\linewidth}
        \includegraphics[width=\textwidth]{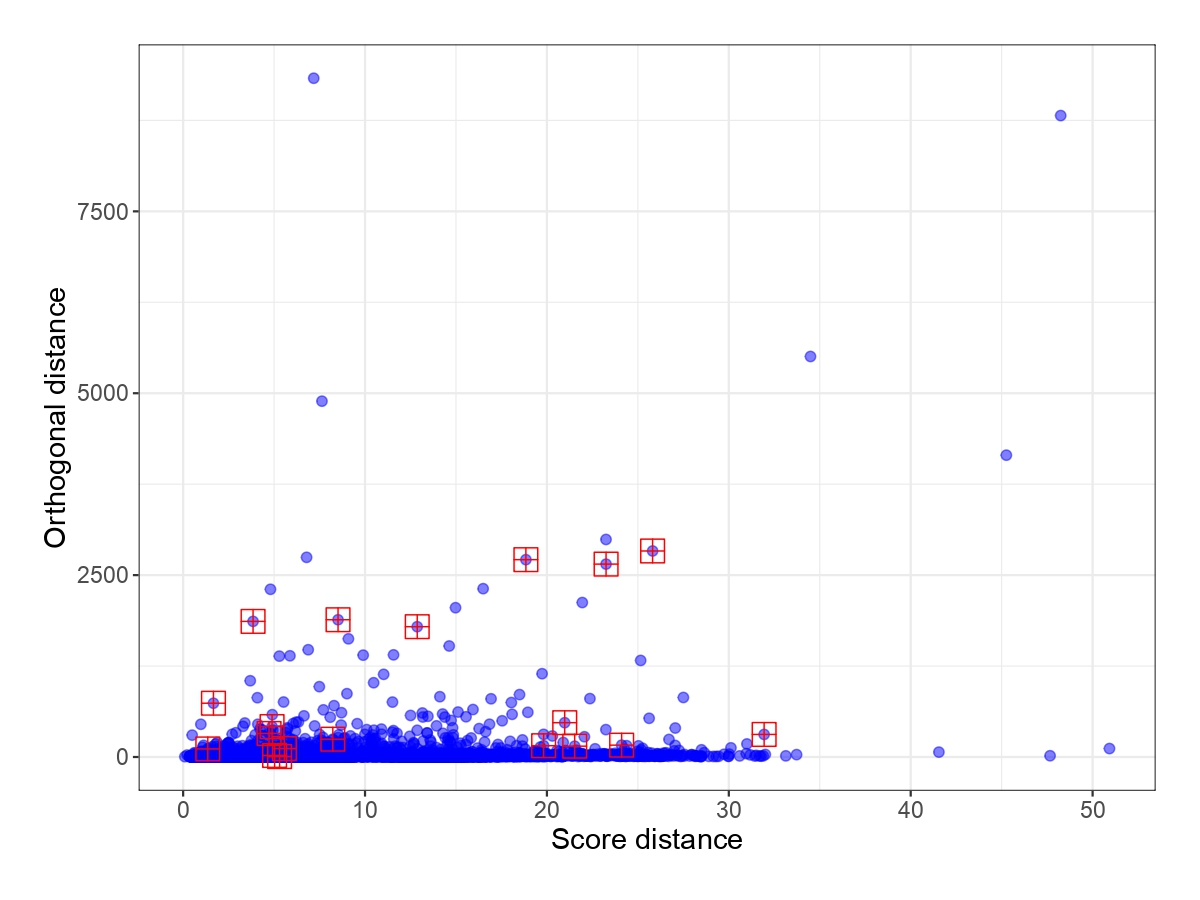}
        \caption{ROBPCA}
    \end{subfigure}
    \begin{subfigure}{0.35\linewidth}
        \includegraphics[width=\textwidth]{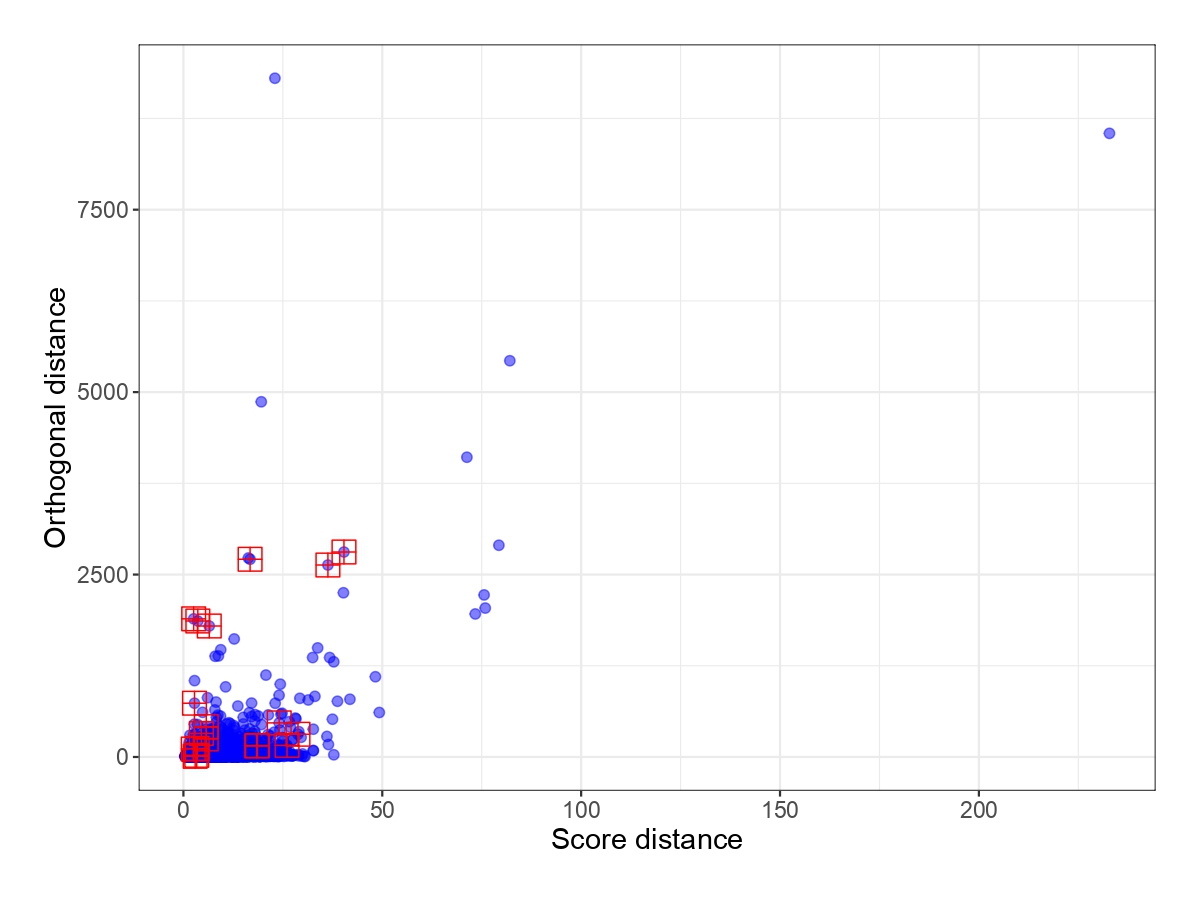}
        \caption{GMedian}
        \end{subfigure}
    \caption{Diagnostic plots for the Credit Card dataset for different robust PCA methods}
    \label{fig:ccard-plot}
\end{figure}

To this end, we anticipate that the proposed robust PCA algorithm will outperform classical PCA in dimensionality reduction and provide reliable principal component estimates. We demonstrate this using the Credit Card Fraud Detection Dataset from~\cite{le2004machine}. The dataset encompasses $28$ anonymized features over $284807$ transactions, with only $0.1\%$ ($492$) being fraudulent. For demonstration, we randomly sample $5\%$ of the dataset, including $19$ fraudulent transactions. The first $5$ principal components, explaining over $80\%$ of variation, are retained for both classical and rPCAdpd algorithms. Diagnostic plots in Figure~\ref{fig:ccard-plot} portray the outcomes, with red squares denoting fraudulent transactions. As shown in Figure~\ref{fig:ccard-plot}, the classical PCA method fails to separate most of the fraudulent transactions, correctly identifying only $5$ (in red). In contrast, the rPCAdpd algorithm separates out $13$ out of $19$ outliers. Existing robust PCA methods such as ROBPCA and GMedian spot $7$ and $6$ outliers respectively, which are better than classical PCA but at the cost of many false positives (outliers without red squares). Thus, substituting classical PCA with the robust rPCAdpd algorithm in the preprocessing or dimensionality reduction step of this analysis can greatly enhance the results of the existing machine learning algorithms. By doing so, valuable insights about fraudulent transactions can assist the existing outlier detection and classification algorithms on the transformed, lower-dimensional data.

\section{Conclusion}

As described in Section~\ref{sec:intro}, a plethora of algorithms from an extensive range of disciplines use principal component analysis. Unfortunately, with the emergence of the era of big data, it has become increasingly difficult to check or validate the authenticity, trustworthiness and overall correctness of the data. As a result, most of the input data to these algorithms are highly susceptible of being contaminated by various forms of noise and outlying observations. Since classical PCA is heavily affected by such outliers, several robust PCA algorithms have been proposed in the last two decades. Many of these are not both fast and scalable. M-estimation based techniques are computationally efficient to obtain, but their breakdown point decays rapidly with the increase in dimension making it unacceptable for being used for high dimensional data. On the other hand, MVE, MCD and other projection pursuit based methods are highly scalable, but they are either computationally extremely intensive or lack proper theoretical guarantees of consistency, asymptotic normality or bounded influence function along with high breakdown. We believe that this paper will help to fill this gap by providing a robust, scalable and efficient PCA estimator with the help of the popular density power divergence. We demonstrate its various desirable theoretical properties in the present work. It also has a dimension-free breakdown point making it attractive to be used in arbitrarily high dimensional data analysis. Also, the robustness parameter $\alpha$ in rPCAdpd can be tuned to provide a smooth bridge between efficiency in estimation and robustness capabilities. 

In all the datasets used to describe the practical applicability of the rPCAdpd, we estimate the significant number of principal components to be extracted based on thresholding the proportion of the variation explained by the first few principal components. However, such a procedure would require the estimation of all principal components first and then computing the proportion. From a computational point of view, it is highly beneficial to estimate the rank of the low-rank matrix $\bb{L}$ first, and then proceed with the estimation of principal components. We will investigate this direction in a future study.


\clearpage

\begin{appendices}

    \section{Proofs of the Results}

    \subsection{Normalization constant of Elliptically Symmetric Families of Distributions}\label{appendix:normal-constant}
    Here we show that the normalizing constant for the elliptically symmetric family of densities is of the form $c_g \det(\bb{\Sigma})^{1/2}$. To see this, we note that it can be expressed as
    \begin{equation*}
        \mathcal{C}_g = \int_{\R^p} \exp\left[ g\left( (\bb{x} - \bb{\mu})\tr \sum_{k=1}^p \gamma_k^{-1}\bb{v}_k\bb{v}_k\tr (\bb{x} - \bb{\mu}) \right) \right] d\bb{x}.
    \end{equation*}
    \noindent Let $\bb{P}$ be the $p \times p$ orthogonal matrix whose rows are the vectors $\bb{v}_k\tr$ for $k = 1, 2, \dots p$. Then, applying a change of variable $\bb{z} = \bb{P}\tr(\bb{x} - \bb{\mu})$, we can rewrite the integral as 
    \begin{equation*}
        \mathcal{C}_g = \int_{\R^p} \exp\left[ g\left( \sum_{k=1}^p \gamma_k^{-1}z_k^2 \right) \right] d\bb{z},
    \end{equation*}
    \noindent where $\bb{z} = (z_1, z_2, \dots z_p)\tr$. Finally, another change of variable with $u_k = z_k/\sqrt{\gamma_k}$ for $k = 1, 2, \dots p$ yields,
    \begin{equation*}
        \mathcal{C}_g = \int_{\R^p} 2^{-p}\prod_{k=1}^p \gamma_k^{1/2} \prod_{k=1}^{p}u_k^{-1/2} \exp\left[ g\left( \sum_{k=1}^p u_k^2 \right) \right] du_1 du_2 \dots du_p = \det(\bb{\Sigma})^{1/2} c_g,
    \end{equation*}
    \noindent where the constant $c_g$ is the integral it is replacing. Clearly, the term $c_g$ is free of the mean $\bb{\mu}$ and the dispersion $\bb{\Sigma}$ matrix, and hence is a constant depending only on the $g$ function.

    \subsection{Proof of Theorem~\ref{thm:existence}}
    First note that the eigenvectors $\bb{v}_k$ lie in the Stiefel manifold of order $p$, which is a closed and bounded subset of $\R^p$, hence is compact.
    
    Also, since $g(x)$ is a continuous decreasing function, $\lim_{x \rightarrow \infty} e^{g(x)} = 0$. Otherwise if $\lim_{x \rightarrow \infty} e^{g(x)} = \epsilon > 0$, it implies that the integral $\int_{0}^{\infty}e^{g(x)}$ diverges by comparison test, contradicting the existence of the elliptically symmetric probability density function. 
    
    Fixing $\bb{\mu} \in \R^p$, let us now observe how the objective function $Q$ behaves for extreme values of the eigenvalues $\gamma_1, \dots \gamma_p$. If $\gamma_1 \rightarrow 0$, then it follows that 
    \begin{equation*}
        \lim_{\gamma_1 \rightarrow 0} Q(\gamma_1, \dots, \gamma_p, \bbeta) = \lim_{\gamma_1 \rightarrow 0} \gamma_1^{-1/2} \left[ \dfrac{c_{(1+\alpha)g}}{c_g} - \lim_{x \rightarrow \infty} e^{g(x)} \right] \geq 0,
    \end{equation*}
    \noindent since $c_g > 0$ for any choice of $g$ function by definition. On the other hand, if $\gamma_1 \rightarrow \infty$, the quadratic form 
    \begin{equation*}
        (\bb{X}_i - \bb{\mu})\tr\sum_{k=1}^p \gamma_k^{-1}\bb{v}_k(\bbeta)\bb{v}_k(\bbeta)\tr (\bb{X}_i - \bb{\mu}) \rightarrow (\bb{X}_i - \bb{\mu})\tr\sum_{k=2}^p \gamma_k^{-1}\bb{v}_k(\bbeta)\bb{v}_k(\bbeta)\tr (\bb{X}_i - \bb{\mu}).
    \end{equation*}
    Then by the strong law of large numbers, it follows that for sufficiently large $n$, with probability $1$,
    \begin{align*}
        & \dfrac{1}{n}\sum_{i=1}^n \exp\left\{ \alpha g\left( (\bb{X}_i - \bb{\mu})\tr\sum_{k=2}^p \gamma_k^{-1}\bb{v}_k(\bbeta)\bb{v}_k(\bbeta)\tr (\bb{X}_i - \bb{\mu})  \right) \right\}\\
        \rightarrow \quad & \E\left[ \exp\left\{ \alpha g\left( (\bb{X} - \bb{\mu})\tr\sum_{k=2}^p \gamma_k^{-1}\bb{v}_k(\bbeta)\bb{v}_k(\bbeta)\tr (\bb{X} - \bb{\mu})  \right) \right\} \right]\\
        \geq \quad & \E\left[ \exp\left\{ \alpha g\left( (\bb{X} - \bb{\mu})\tr\sum_{k=1}^p \gamma_k^{-1}\bb{v}_k(\bbeta)\bb{v}_k(\bbeta)\tr (\bb{X} - \bb{\mu})  \right) \right\} \right]\\
        = \quad & \int_{R^p} \exp\left\{ (1 +\alpha) g\left( (\bb{x} - \bb{\mu})\tr\sum_{k=1}^p \gamma_k^{-1}\bb{v}_k(\bbeta)\bb{v}_k(\bbeta)\tr (\bb{x} - \bb{\mu})  \right) \right\} d\bb{x}\\
        = \quad & \dfrac{c_{(1+\alpha)g}}{c_g},
    \end{align*}
    \noindent where the inequality uses the fact that $g$ is monotonically decreasing. Therefore, for sufficiently large $n$, with probability $1$, $Q(\gamma_1, \dots \gamma_p, \bbeta)$ increases to $0$ as $\gamma_1$ increases to $\infty$. Therefore, for any given $\epsilon > 0$, there exists $0 < a_1 < b_1 < \infty$ such that $Q(\gamma_1, \gamma_2, \dots \gamma_p) > (-\epsilon)$ for any $\gamma_1 \not\in [a_1, b_1]$. Note that, since $\gamma_1$ is chosen arbitrarily, the same conclusion also holds for all other eigenvalues, possibly with different choices of $a_k$ and $b_k$ for $k = 1, 2, \dots p$. Letting, $\epsilon = -\inf Q(\gamma_1, \dots \gamma_p, \bbeta)/2$ (which is finite by continuity of $Q$ and the limiting behaviour described above) and considering the set $K = \prod_{k=1}^p [a_k, b_k] \times S$, we note that the infimum of $Q$ must exist within the set $K$. Since $K$ is a compact subset of $\R^p$, it follows by the Extreme Value Theorem that the infimum must be attained. This proves the existence of the rPCAdpd estimator for any arbitrary value of $\mu$, including the location estimate $\widehat{\mu}$.

    \subsection{Proof of Theorem~\ref{thm:orthogonal-equivariance}}
    
    Let $\bbhat{\mu}_Y$ and $\bbhat{\mu}_X$ be the robust estimates of the location based on the sample $\bb{Y}_1, \dots \bb{Y}_n$ and $\bb{X}_1, \dots \bb{X}_n$ respectively. Then by the orthogonal equivariance of the location estimator, we have that $\bbhat{\mu}_Y = a \bb{P}\bbhat{\mu}_X + \bb{b}$. The equivariance property for the estimated eigenvalues and eigenvectors by the rPCAdpd algorithm then follows from the observation that the quadratic form of the transformed data can be expressed as
    \begin{align*}
        (\bb{Y}_i - \bbhat{\mu}_Y)\tr \left( \sum_{k=1}^p \gamma_k^{-1}\bb{v}_k \bb{v}_k\tr \right) (\bb{Y}_i - \bbhat{\mu}_Y)
        & = a(\bb{X}_i -  \bbhat{\mu}_X)\tr \bb{P}\tr \left( \sum_{k=1}^p \gamma_k^{-1}\bb{v}_k \bb{v}_k\tr \right) (\bb{X}_i -  \bbhat{\mu}_X)\bb{P} a\\
        & = (\bb{X}_i -  \bbhat{\mu}_X)\tr \left( \sum_{k=1}^p (\gamma_k/a^2)^{-1} \bb{P}\tr\bb{v}_k \bb{v}_k\tr \bb{P} \right) (\bb{X}_i -  \bbhat{\mu}_X).
    \end{align*}
    \noindent It shows that, if rPCAdpd estimates of the eigenvalues for the sample $\bb{X}_1, \dots \bb{X}_p$ are $\gamma_k^\ast$ for $k = 1, 2,\dots p$ respectively, then the rPCAdpd estimate of the same for the transformed sample would be $a^2\gamma_k^\ast$. A similar conclusion can be drawn for the rPCAdpd estimate of eigenvectors as well. 
    
    \subsection{Proof of Lemma~\ref{lemma:general-xi}}
    
    Let, $h_{\bbtheta}(\bb{x}) = c^{-1}_\alpha(\bbtheta)f^{(1+\alpha)}_{\bbtheta}(\bb{x})$ be another density function. Note that 
    \begin{equation*}
        \dfrac{\partial}{\partial\bbtheta}\log(h_{\bbtheta}(\bb{x})) = -\dfrac{\partial}{\partial\bbtheta}\log(c_\alpha(\bbtheta)) + (1+\alpha) u_{\bbtheta}(\bb{x}),
        \label{eqn:easy-xi}
    \end{equation*}
    \noindent where $u_\theta(\bb{x})$ is the score function corresponding to $f_{\bbtheta}(\bb{x})$. Under the standard regularity conditions, one can exchange the differentiation and the integral sign to obtain that the expectation of the score function is equal to $0$. Therefore, 
    $$
        0 = \int \dfrac{\partial}{\partial\bbtheta}\log(h_{\bbtheta}(\bb{x})) h_{\bbtheta}(\bb{x})d\bb{x} = -\dfrac{\partial}{\partial\bbtheta}\log(c_\alpha(\bbtheta)) + \dfrac{(1+\alpha)}{c_\alpha(\bbtheta)} \xi_{\bbtheta}.
    $$
    \noindent Interchanging the sides and solving for $\xi_{\bbtheta}$ yields the result.

    \subsection{Proof of Lemma~\ref{lemma:general-J}}
    
    Starting with the decomposition
    $$
        u^h_{\bbtheta}(\bb{x}) = \dfrac{\partial}{\partial\bbtheta}\log(h_{\bbtheta}(\bb{x})) = -\dfrac{\partial}{\partial\bbtheta}\log(c_\alpha(\bbtheta)) + (1+\alpha) u_{\bbtheta}(\bb{x}),
    $$
    \noindent it follows that 
    \begin{equation*}
        \begin{split}
            \left( u^h_{\bbtheta}(\bb{x}) \right)\left( u^h_{\bbtheta}(\bb{x}) \right)\tr 
            & = \left(\dfrac{\partial}{\partial\bbtheta}\log(c_\alpha(\bbtheta))\right)\left(\dfrac{\partial}{\partial\bbtheta}\log(c_\alpha(\bbtheta)) \right)\tr \\
            & + 2(1+\alpha)u_{\bbtheta}(\bb{x})\left(\dfrac{\partial}{\partial\bbtheta}\log(c_\alpha(\bbtheta))\right)\tr + (1+\alpha)^2 u_{\bbtheta}(\bb{x})u\tr_{\bbtheta}(\bb{x}).
        \end{split}
    \end{equation*} 
    \noindent Multiplying both sides with $h_{\bbtheta}(\bb{x})$ and integrating with respect to $\bb{x}$ yields
    \begin{equation*}
        i^h(\bbtheta) = \left( \dfrac{\nabla_{\bbtheta} c_\alpha(\bbtheta) }{ c_\alpha(\bbtheta) } \right)\left( \dfrac{\nabla_{\bbtheta} c_\alpha(\bbtheta) }{ c_\alpha(\bbtheta) } \right)\tr - 2\left( \dfrac{\nabla_{\bbtheta} c_\alpha(\bbtheta) }{ c_\alpha(\bbtheta) } \right)\left( \dfrac{\nabla_{\bbtheta} c_\alpha(\bbtheta) }{ c_\alpha(\bbtheta) } \right)\tr + \dfrac{(1+\alpha)^2}{c_\alpha(\bbtheta)}J_{\bbtheta},
    \end{equation*}
    \noindent where $\nabla_{\bbtheta} c_\alpha(\bbtheta) = \fracpartial{c_\alpha(\bbtheta)}{\bbtheta}$. Solving for $J_{\bbtheta}$ yields Eq.~\eqref{eqn:easy-J}.

    \subsection{Proof of Corollary~\ref{cor:ellip-xi}}
    Since the normalized density $c_{\alpha}^{-1}(\bbtheta)f_{\bbtheta}^{(1+\alpha)}$ also belongs to an elliptically symmetric class of densities, it follows that
    \begin{equation*}
        c_\alpha(\bbtheta) = c_{(1 + \alpha)g} \prod_{k=1}^p (\gamma_k)^{1/2} c_g^{-(1+\alpha)} \prod_{k=1}^p (\gamma_k)^{-(1+\alpha)/2}.
    \end{equation*}
    \noindent Putting this value and its derivative with respect to $\bbtheta$ into Lemma~\ref{lemma:general-xi} yields Corollary~\ref{cor:ellip-xi}.

    \subsection{Proof of Corollary~\ref{cor:ellip-jtheta}}
    We start by defining a few notations as follows:
    \begin{align*}
        Q(\bb{x}) & = (\bb{x} - \bb{\mu}^\ast)\tr \left(\sum_{k=1}^p \gamma_k^{-1} \bb{v}_k\bb{v}_k\tr \right) (\bb{x} - \bb{\mu}^\ast),\\
        A_2(g) & = \int g\dash(Q(\bb{x})) (\bb{x} - \bb{\mu}^\ast)(\bb{x} - \bb{\mu}^\ast)\tr c_0(\bbtheta)^{-1} \exp(g(Q(\bb{x})))d\bb{x},\\
        A_4(g; \bb{u}, \bb{v}) & = \int \left(g\dash(Q(\bb{x})) \right)^2 (\bb{x} - \bb{\mu}^\ast)(\bb{x} - \bb{\mu}^\ast)\tr \bb{u}\bb{v}\tr (\bb{x} - \bb{\mu}^\ast)(\bb{x} - \bb{\mu}^\ast)\tr c_0(\bbtheta)^{-1}\exp(g(Q(\bb{x})) )d\bb{x}.
    \end{align*}
    \noindent Here, $c_0(\bbtheta)$ is the normalizing constant the normalizing constant for the elliptically symmetric density proportional to $\exp(g(Q(\bb{x}))$. Clearly, $c_0(\bbtheta) = c_g \prod_{k=1}^p \gamma_k^{1/2}$. All of these quantities are well defined due to the Assumptions~\ref{assum:0} and~\ref{assum:3}. Also, let $\bb{G}_k = \fracpartial{\bb{v}_k}{\bbeta}$ denote the $p(p+1)/2 \times p$ matrix whose columns are the gradients of the entries $v_{kj}$ for $j = 1, 2, \dots p$, of $\bb{v}_k$ with respect to the parameter $\bbeta$. One important aspect is to note the quantities $A_2(g)$ and $A_4(g; \bb{u}, \bb{v})$ are free of $\bb{\mu}^\ast$, which can be verified by a simple substitution in the integral. 
    
    Starting with the identity 
    \begin{equation*}
        c_0(\bbtheta) = \int \exp\left( g(Q(\bb{x})) \right)d\bb{x},
    \end{equation*}
    \noindent and differentiating both sides by $\gamma_k$ and $\bbeta$ respectively, we obtain the identities
    \begin{equation}
        \gamma_k^{-2} \bb{v}_k\tr A_2(g) \bb{v}_k = -\dfrac{1}{2\gamma_k}, \ 
        \sum_{k=1}^p \gamma_k^{-1} \bb{G}_k A_2(g) \bb{v}_k = 0,
        \label{eqn:a2-relations}
    \end{equation}
    both of which will be used later in the proof.
    
    Let $h_{\bbtheta}(\bb{x}) = c_{(1+\alpha)}(\bbtheta)^{-1} e^{(1+\alpha)g(Q(\bb{x}))}$ be a density belonging to the same elliptically symmetric family. Then, the score function $u_{\bbtheta}^h(\bb{x})$ corresponding to $h_{\bbtheta}$ can be expressed as
    \begin{equation}
        u_{\bbtheta}^h(\bb{x})
        = \begin{bmatrix}
            \frac{1}{2}\diag{\bb{\Gamma}^{-1}} - (1+\alpha) g\dash(Q(\bb{x})) \bb{\Gamma}^{-2} \bb{V}\tr (\bb{I}_p \otimes (\bb{x} - \bb{\mu})(\bb{x} - \bb{\mu})\tr ) \bb{V}\\
            2(1+\alpha)g\dash(Q(\bb{x})) \bb{G} (\bb{\Gamma}^{-1} \otimes (\bb{x} - \bb{\mu})(\bb{x} - \bb{\mu})\tr ) \bb{V}\bb{1}_p
        \end{bmatrix}.
        \label{eqn:score-h}
    \end{equation}
    \noindent Using the expression for $u_{\bbtheta}^h(\bb{x})$, we can further differentiate this with respect to the entries of $\bbtheta$ and take expectation. This leads to the Fisher Information matrix in the partitioned form as follows,
    \begin{equation*}
        i^h(\bbtheta) = \begin{bmatrix}
            i^h(\gamma_1, \gamma_1) & \dots & i^h(\gamma_1, \gamma_p) & i^h(\gamma_1, \bbeta) \\
            \vdots & \ddots & \vdots & \vdots \\
            i^h(\gamma_p, \gamma_1) & \dots & i^h(\gamma_p, \gamma_p) & i^h(\gamma_p, \bbeta) \\
            i^h(\gamma_1, \bbeta)\tr & \dots & i^h(\gamma_p, \bbeta)\tr & i^h(\bbeta, \bbeta)
        \end{bmatrix},
    \end{equation*}
    \noindent where, 
    \begin{align*}
        i^h(\gamma_k, \gamma_l)
        & = \left( \dfrac{\partial q_{(1+\alpha)g}}{\partial \gamma_k} \right)\left( \dfrac{\partial q_{(1+\alpha)g}}{\partial \gamma_l} \right) + \left( \dfrac{\partial q_{(1+\alpha)g}}{\partial \gamma_k} \right)\gamma_l^{-2} \bb{v}_l\tr A_2((1+\alpha)g) \bb{v}_l \\
        & \qquad + \left( \dfrac{\partial q_{(1+\alpha)g}}{\partial \gamma_l} \right)\gamma_k^{-2} \bb{v}_k\tr A_2((1+\alpha)g) \bb{v}_k + \dfrac{\bb{v}_k\tr A_4((1+\alpha)g; \bb{v}_k, \bb{v}_l) \bb{v}_l}{\gamma_k^2\gamma_l^2}\\
        & = -\left( \dfrac{\partial q_{(1+\alpha)g}}{\partial \gamma_k} \right)\left( \dfrac{\partial q_{(1+\alpha)g}}{\partial \gamma_l} \right) + \dfrac{\bb{v}_k\tr A_4((1+\alpha)g; \bb{v}_k, \bb{v}_l) \bb{v}_l}{\gamma_k^2\gamma_l^2},\ k, l = 1, 2, \dots p\\
        i^h(\gamma_k, \bbeta)
        & = -2\left( \dfrac{\partial q_{(1+\alpha)g}}{\partial \gamma_k} \right)\sum_{k=1}^p \gamma_k^{-1}\bb{G}_k A_2((1+\alpha)g)\bb{v}_k - \dfrac{2}{\gamma_k^2}\sum_{l=1}^p \gamma_l^{-1}\bb{v}_k\tr A_4((1+\alpha)g; \bb{v}_k, \bb{v}_l) \bb{G}_l\tr\\
        & = - \dfrac{2}{\gamma_k^2}\sum_{l=1}^p \gamma_l^{-1}\bb{v}_k\tr A_4((1+\alpha)g; \bb{v}_k, \bb{v}_l) G_l\tr, \ k = 1, \dots p\\
        i^h(\bbeta, \bbeta)
        & = 4 \sum_{k=1}^p \sum_{l=1}^p \gamma_k^{-1}\gamma_l^{-1}\bb{G}_k A_4((1+\alpha)g; \bb{v}_k, \bb{v}_l) \bb{G}_l\tr,
    \end{align*}
    \noindent where we use the identities~\eqref{eqn:a2-relations}. In all of the above expressions, the quantity $q_g$ denoted the logarithm of the normalizing constant, i.e., $q_g = \log(c_0(\bbtheta))$ and $q_{(1+\alpha)g} = \log(c_\alpha(\bbtheta))$. Finally, Corollary~\ref{cor:ellip-jtheta} follows from using Lemma~\ref{lemma:general-J} and the expression of $\bb{\xi}_{\bbtheta}$ given in Corollary~\ref{cor:ellip-xi}.

    \subsection{Proof of the Theorem~\ref{thm:normality-general}}\label{appendix:normality-general-proof}
    
    The proof of the Theorem~\ref{thm:normality-general} closely resembles the proof of Theorem 3.1 of~\cite{ghosh2013robust}. For brevity, we shall only indicate the modifications pertinent to the special scenario of principal components. Given the location estimator $\bbhat{\mu}$, using the same notation as in~\cite{ghosh2013robust}, we define
    \begin{equation*}
        V(\bb{X}, \bbtheta) = \prod_{k=1}^p \gamma_k^{-\alpha/2} \left[ \dfrac{c_{(1+\alpha)g}}{c_g} - \left(1 + \dfrac{1}{\alpha} \right) e^{ \alpha g\left( (\bb{X} - \bbhat{\mu})\tr \sum_{k=1}^p \gamma_k^{-1} \bb{v}_k(\bbeta)\bb{v}_k(\bbeta)\tr (\bb{X} - \bbhat{\mu}) \right)} \right]
    \end{equation*}
    \noindent which are the summands in the objective function in Eq.~\eqref{eqn:Q-function}. Now, conditional on $\bbhat{\mu}$, by an application of the Law of Large Numbers, we have 
    \begin{equation*}
        \dfrac{1}{n}\sum_{i=1}^n \nabla V(\bb{X}_i, \bbtheta^\ast) \mid \bbhat{\mu} \xrightarrow{P} 0, \
        \text{and, } \
        \dfrac{1}{n} \sum_{i=1}^n \nabla^2 V(\bb{X}_i, \bbtheta^\ast) \mid \bbhat{\mu} \xrightarrow{P} \bb{J}_{\bbtheta^\ast}
    \end{equation*}
    \noindent where $\bbtheta^\ast$ is the true value of the parameters. Now, since the right-hand sides of both of these are continuous functions of $\bbhat{\mu}$ and as $\bbhat{\mu} \xrightarrow{P} \bb{\mu}^\ast$ (the true location parameter) due to the consistency of the location estimator, it follows that the unconditional random variables also converges in probability to the same value. As the support of the elliptically symmetric family of distributions is assumed to be the entire space $\R^p$, $\bb{J}_{\bbtheta^\ast}$ becomes free of the choice of location which makes this convergence possible. Now, one can replicate the proof for consistency to show that the rPCAdpd estimator is consistent.
    
    To prove the asymptotic normality, we need to show that $T_n = \frac{1}{\sqrt{n}}\sum_{i=1}^n \nabla^2 V(\bb{X}_i, \bbtheta^\ast)$ converges in distribution to a random variable $\bb{Z}$ following a multivariate normal distribution with mean $0$ and variance $\bb{K}_{\bbtheta^\ast}$. Due to Portmanteau's theorem, it is enough to show that for any bounded continuous function $h$, $\vert \E(h(T_n)) - \E(h(\bb{Z}))\vert \rightarrow 0$ as $n \rightarrow \infty$. An application of Lindeberg-Levy Central Limit Theorem and Portmanteau's theorem yields that as $n \rightarrow \infty$,
    \begin{equation*}
        \vert \E(h(T_n) \mid \bbhat{\mu}) - \E(h(\bb{Z})) \vert \rightarrow 0.
    \end{equation*}
    \noindent Since $\E(h(T_n) \mid \bbhat{\mu})$ is also a bounded and continuous function of $\bbhat{\mu}$, it follows that
    \begin{equation*}
        \E(h(T_n)) = \E\left[ \E(h(T_n) \mid \bbhat{\mu}) \right] \rightarrow \E\left[ \E(h(\bb{Z}) \mid \bb{\mu}^\ast ) \right] = \E(h(\bb{Z})), \text{ as } n \rightarrow \infty,
    \end{equation*}
    \noindent where the last equality follows due to the fact that both mean and the variance $\bb{K}_{\bbtheta^\ast}$ of $\bb{Z}$ is free of the choice of location $\bb{\mu}^\ast$. The rest of the proof follows as in~\cite{ghosh2013robust}.

    \subsection{Proof of the Corollary~\ref{thm:normality-normal}}
    
    The generating function for the Gaussian distribution in the elliptically symmetric family of distributions is $g(x) = (-x/2)$. It follows that $g\dash(x) = -1/2$ and the normalizing constant $\mathcal{C}_g = (2\pi)^{p/2} \prod_{k=1}^p \gamma_k^{1/2}$. For ease of notation, we also define 
    \begin{equation*}
        c_\alpha = \dfrac{\mathcal{C}_{(1+\alpha)g}}{\mathcal{C}_g} = (2\pi)^{-\alpha p/2} (1+\alpha)^{-p/2} \prod_{k=1}^p (\gamma_k^\ast)^{-\alpha/2}.
    \end{equation*}
    \noindent Now, some standard calculation using properties of normal distribution and its quadratic forms~\citep{petersen2008matrix} reveals that $A_2((1+\alpha)g) = (1+\alpha)\bb{\Sigma}^\ast/4$, and 
    \begin{equation*}
        A_4((1+\alpha)g; \bb{u}, \bb{v})
        = \dfrac{1}{4}\left[\bb{\Sigma}^\ast \left( \bb{u}\bb{v}\tr + \bb{v}\bb{u}\tr \right)\Sigma^\ast + \trace{\bb{u}\bb{v}\tr\bb{\Sigma}^\ast}\bb{\Sigma}^\ast \right].
    \end{equation*}
    \noindent In particular, for any $k, l = 1, 2, \dots p$,
    \begin{align*}
        A_4((1+\alpha)g; \bb{v}_k^\ast, \bb{v}_l^\ast)
        & = \dfrac{1}{4} \left[\Sigma^\ast \left( (\bb{v}_k^\ast)(\bb{v}_l^\ast)\tr + (\bb{v}_l^\ast)(\bb{v}_k^\ast)\tr \right)\Sigma^\ast + \trace{(\bb{v}_k^\ast)(\bb{v}_l^\ast)\tr\bb{\Sigma}^\ast}\bb{\Sigma}^\ast \right]\\
        & = \dfrac{1}{4} \left[ \gamma_k^\ast\gamma_l^\ast \left( (\bb{v}_k^\ast)(\bb{v}_l^\ast)\tr + (\bb{v}_l^\ast)(\bb{v}_k^\ast)\tr \right) + \bb{1}_{\{k=l\}}\gamma_l^\ast\bb{\Sigma}^\ast \right],
    \end{align*}
    \noindent where we use the fact that $\bb{v}_k^\ast$ is an eigenvector of $\Sigma^\ast$ corresponding to the eigenvalue $\gamma_k^\ast$. Thus, it turns out that $j^h(\bb{\mu}^\ast,\bb{\mu}^\ast) = \frac{c_\alpha}{(1+\alpha)} (\Sigma^\ast)^{-1}$, and
    \begin{equation*}
        j^{h}(\gamma_k^\ast, \gamma_l^\ast)
        = \dfrac{c_\alpha}{4(1+\alpha)^2\gamma_k^\ast \gamma_l^\ast} \left( \alpha^2 + 2\bb{1}_{\{ k = l\}} \right) 
    \end{equation*}
    \noindent and $j^{h}(\gamma_k^\ast, \bb{\eta}^\ast) = 0$, where we use the fact that $\bb{G}_k \bb{v}_k^\ast = 0$. This equality follows from differentiating both sides of the identity $(\bb{v}_k^\ast)\tr (\bb{v}_k^\ast) = 1$ with respect to the parameter $\bb{\eta}$ at $\bb{\eta} = \bb{\eta}^\ast$. Similarly, differentiating the identity $(\bb{v}_k^\ast)\tr (\bb{v}_l^\ast) = 0$ for $k \neq l$ with respect to $\bb{\eta}$ yields that $\bb{G}_k \bb{v}_l^\ast + \bb{G}_l \bb{v}_k^\ast = 0$. Some lengthy calculation and an application of this identity allows us to obtain 
    \begin{equation*}
        j^h(\bb{\eta}^\ast, \bb{\eta}^\ast)
        = \dfrac{c_\alpha}{(1+\alpha)^2} \left( \sum_{k=1}^p \sum_{l=1}^p \left( 1 - \dfrac{\gamma_k^\ast}{\gamma_l^\ast} \right)\bb{G}_k(\bb{v}_l^\ast) (\bb{v}_k^\ast) \tr \bb{G}_l\tr \right).
    \end{equation*}
    \noindent A similar calculation may be performed to determine the entries of $K_{\bbtheta^\ast}$. This completes the proof of the corollary, with a direct application of Theorem~\ref{thm:normality-general}.
    
    \section{Performance Metrics for Assessmentof Principal\\ Components}
    
    Letting $\bbhat{P}$ denote the estimated principal component matrix where we stack each principal component vector as columns, the quantity $\bbhat{X} = \bb{X}\bbhat{P}\bbhat{P}\tr$ becomes the projection of the samples $\bb{X}_1, \dots \bb{X}_n$ (here $\bb{X}_i$ is the $i$-th row of $\bb{X}$) onto the principal component space (i.e., the vector space spanned by the eigenvectors corresponding to the first $r$ eigenvalues). Then the orthogonal distance for the $i$-th datapoint is calculated as the Euclidean distance between $\bb{X}_i$ and $\bbhat{X}_i$ where $\bbhat{X}_i$ denotes the $i$-th row of $\bbhat{X}$. On the other hand, the score distance of the $i$-th datapoint would be given by
    \begin{equation}
        \text{Score distance}_i = \sum_{k=1}^{r} \dfrac{\widehat{X}_{ik}^2}{\widehat{\gamma}_k},
    \end{equation}
    \noindent where $\widehat{\gamma}_k$ is the $k$-th eigenvalue and $\widehat{X}_{ik}$ is the $k$-th element of the vector $\bbhat{X}_i$.

\end{appendices}


\clearpage
\bibliography{references}
\end{document}